%% file: main.tex
\documentclass[acmsmall] {acmart}



\setcopyright{rightsretained}
\acmDOI{10.1145/3704864}
\acmYear{2025}
\acmJournal{PACMPL}
\acmVolume{9}
\acmNumber{POPL}
\acmArticle{28}
\acmMonth{1}
\received{2024-07-09}
\received[accepted]{2024-11-07}

\usepackage[euler]{textgreek}
\usepackage{mathtools}
\usepackage{subcaption}
\usepackage{graphicx} 
\usepackage{hyperref}
\hypersetup{
    colorlinks=true,
    linkcolor=blue,
    filecolor=magenta,      
    urlcolor=cyan
}
\usepackage{amsmath}
\usepackage{amsthm}
\usepackage{mathpartir}
\usepackage{xcolor}
\usepackage{tikz}
\usetikzlibrary{calc,shadows,shapes,positioning,arrows.meta,bending}
\usepackage{tikz-cd}
\usepackage{tcolorbox}
\usepackage{soul}
\usepackage{listings}
\usepackage{framed}
\usepackage{tabularx,booktabs}
\usepackage{xspace}
\usepackage{etoolbox}
\usepackage{accents}
\usepackage[normalem]{ulem}
\usepackage{natbib}
\usepackage{tikz-qtree}
\usepackage{array}
\usepackage{calc}
\usepackage{multirow}
\usepackage{wrapfig}

\newcommand{\zerodisplayskips}{%
  \setlength{\abovedisplayskip}{0.5em}%
  \setlength{\belowdisplayskip}{1em}%
  \setlength{\abovedisplayshortskip}{0pt}%
  \setlength{\belowdisplayshortskip}{0pt}}
\appto{\normalsize}{\zerodisplayskips}
\appto{\small}{\zerodisplayskips}
\appto{\footnotesize}{\zerodisplayskips}

\newcolumntype{R}{>{$}r<{$}}
\newcolumntype{L}{>{$}l<{$}}
\newcolumntype{C}{>{$}c<{$}}

\newtheorem{definition}{Definition}
\newtheorem{theorem}{Theorem}
\newtheorem{lemma}{Lemma}
\newtheorem{property}{Property}
\newtheorem{example}{Example}
\newtheorem{innercustomthm}{Theorem}
\newenvironment{customthm}[1] 
  {\renewcommand\theinnercustomthm{#1}\innercustomthm}
  {\endinnercustomthm}

\newcommand{\lblkbrbrak}[0]{\langle}
\newcommand{\rblkbrbrak}[0]{\rangle}

\newcommand{\includepantographgraphics}[2][0.25]{\includegraphics[scale=#1]{#2}}

\newcommand{\codeblockeditsequencetwosteps}[2]{
    \centerline{\begin{tikzpicture}
        \node[draw,anchor=north west] (step0) at (0,0) {#1};
        \node[draw,anchor=north west] (step1) at ($(step0.south west) + (2,-0.1)$) {#2};
        \draw[-{[flex,sep]>},thick]
            ($(step0.south west) + (1,0)$)
            .. controls ($(step0.south west) + (1,-1)$) and ($(step1.west) + (-1,0)$) ..
            (step1.west);
    \end{tikzpicture}}
}

\input{syntax}

\title{Pantograph: A Fluid and Typed Structure Editor}
\date{October 2023}

\begin{document}
\author{Jacob Prinz}
\orcid{0000-0002-2702-9319}
\affiliation{%
  \institution{University of Maryland}
  \city{College Park}
  \country{USA}
}
\email{jprinz@umd.edu}

\author{Henry Blanchette}
\orcid{0000-0002-9415-0944}
\affiliation{%
  \institution{University of Maryland}
  \city{College Park}
  \country{USA}
}
\email{blancheh@umd.edu}

\author{Leonidas Lampropoulos}
\orcid{0000-0003-0269-9815}
\affiliation{%
  \institution{University of Maryland}
  \city{College Park}
  \country{USA}
}
\email{leonidas@umd.edu}

\begin{abstract}
Structure editors operate directly on a program's syntactic tree structure.
At first glance, this allows for the exciting possibility that such an editor could enforce correctness properties:  programs could be well-formed and sometimes even well-typed by construction.
Unfortunately, traditional approaches to structure editing that attempt to rigidly enforce these properties face a seemingly fundamental problem, known in the literature as {\em viscosity}.
Making changes to existing programs often requires temporarily breaking program structure---but disallowing such changes makes it difficult to edit programs!

In this paper, we present a scheme for structure editing which always maintains a valid program structure without sacrificing the fluidity necessary to freely edit programs.
Two key pieces help solve this puzzle: first, we develop a novel generalization of {\em selection} for tree-based structures that properly generalizes text-based selection and editing, allowing users to freely rearrange pieces of code by cutting and pasting one-hole contexts;
second, we type these one-hole contexts with a category of {\em type diffs} and explore the metatheory of the system  that arises for maintaining well-typedness systematically.
We implement our approach as an editor called \href{\link}{{\em Pantograph}}, and we conduct a study in which we successfully taught students to program with Pantograph and compare their performance against a traditional text editor.
\end{abstract}

\maketitle

\input{introduction}
\input{zipper-editing}
\input{typed-editing}
\input{simplediffmath}

\input{fullmath}
\input{proofs}

\input{userstudy-v2}

\input{limitations}
\input{related_work}
\input{conclusion}

\section*{Data Availability Statement}

An artifact \cite{artifact} containing the source code for Pantograph, as well as the web application for Pantograph and the user study, is available. The web application can also be found at \href{\link}{\link}.

\begin{acks}
We thank Sankha Narayan Guria, David Thrane Christiansen, our colleagues at the Programming Languages Research group at the University of Maryland, and the
anonymous reviewers for their
helpful comments.


We also thank all those who helped us test the user interface and the user study participants.

This work was supported by NSF award \#2107206,
Efficient and Trustworthy Proof Engineering (any opinions, findings and
conclusions or recommendations expressed in this material are those of
the authors and do not necessarily reflect the views of the NSF).
\end{acks}

\nocite{towardsuserfriendly}
\nocite{amyko1}

\bibliography{main}

\appendix
\input{smallsteprules2}
\input{metatheoryAppendix}
\input{diffappendix}

\end{document}

%% file: syntax.tex
\newcommand{\codefamily}[0]{\sffamily}
\newcommand{\code}[1]{\ensuremath{\mathsf{#1}}}

\lstdefinelanguage{pg}{
    morekeywords={let,in,match,with},
    morecomment=[l]{//},
    commentstyle=\color{gray}
}
\newcommand{\CC}{\lstinline}

\newcommand{\ttt}[1]{\texttt{#1}}
\newcommand{\pipe}{\;\;\;|\;\;\;}
\newcommand{\s}[1]{\{#1\}}

\definecolor{textselection}{RGB}{186,214,251}
\definecolor{diffhl}{RGB}{252, 194, 3}
\definecolor{plusdiffhl}{RGB}{130,255,130}
\definecolor{minusdiffhl}{RGB}{255,150,150}
\definecolor{ghosthl}{RGB}{0,0,0}

\newcommand{\textselection}[1]{{\sethlcolor{textselection}\hl{#1}}}

\newif\ifhighlight
\highlighttrue

\newsavebox{\tempbox}
\newcommand{\myhl}[2][colorA]{
\ifhighlight%
  \ifmmode\savebox{\tempbox}{$#2$}%
  \else\savebox{\tempbox}{#2}%
  \fi
  \leavevmode\rlap{\color{#1}\rule[-\dp\tempbox]{\wd\tempbox}{\dimexpr \ht\tempbox + \dp\tempbox}}%
  \usebox{\tempbox}%
\else
  #2%
\fi}

\newcommand{\hole}[0]{\square}
\newcommand{\holety}[1]{\hole_{#1}}

\newcommand{\diffhl}[1]{\myhl[diffhl!30]{\ensuremath{\displaystyle #1}}} 
\newcommand{\plusdiffhl}[1]{\myhl[plusdiffhl!30]{\ensuremath{\displaystyle #1}}} 
\newcommand{\minusdiffhl}[1]{\myhl[minusdiffhl!30]{\ensuremath{\displaystyle #1}}}
\newcommand{\ghostdiffhl}[1]{\myhl[ghosthl!10]{\ensuremath{\displaystyle #1}}}

\newcommand{\clasp}[1]{\lblkbrbrak #1 \rblkbrbrak}
\newcommand{\ohchole}[1]{[#1]}
\newcommand{\diff} [3] { \diffhl{\lblkbrbrak #1 \lblkbrbrak} #2 \diffhl{\rblkbrbrak #3 \rblkbrbrak} }
\newcommand{\plusdiff} [3] { \plusdiffhl{{}^{+} \! \lblkbrbrak #1 \lblkbrbrak} #2 \plusdiffhl{\rblkbrbrak #3 \rblkbrbrak} }
\newcommand{\minusdiff} [3] { \minusdiffhl{{}^{-} \! \lblkbrbrak #1 \lblkbrbrak} #2 \minusdiffhl{\rblkbrbrak #3 \rblkbrbrak} }
\newcommand{\replacediff}[2]{\minusdiffhl{#1 ~} / \plusdiffhl{~ #2} }

\newcommand{\plusdiffs} [3] { {}^{+} \! \lblkbrbrak #1 \lblkbrbrak #2 \rblkbrbrak #3 \rblkbrbrak}
\newcommand{\minusdiffs} [3] {{}^{-} \! \lblkbrbrak #1 \lblkbrbrak #2 \rblkbrbrak #3 \rblkbrbrak}
\newcommand{\replacediffs}[2]{{#1 ~} / {~ #2}}

\newcommand{\step}[0]{\;\;\leadsto\;\;}
\newcommand{\diffsto}[3]{
    #1 \xRightarrow{\mbox{\normalsize\ensuremath{#2}}} #3
}

\newcommand{\kw}[1]{\text{\bfseries\codefamily #1}}
\newcommand{\fun}[3]{\kw{\textlambda} #1~\kw{:}~#2~\kw{.}~#3}
\newcommand{\llet}[4]{{\kw{let}~#1~\kw{:}~#2~\kw{=}~#3~\kw{in}~#4}}
\newcommand{\matchTwo}[5]{{\kw{match}~#1~\kw{with}~#2~\Rightarrow~#3~\kw{;}~#4~\Rightarrow~#5}}
\newcommand{\rawboundary}[1]{ \{ #1 \} }

\newcommand{\upboundary}[2]{ \s{#2}^{\uparrow}_{{#1}} }
\newcommand{\downboundary}[2]{ \s{#2}^{\downarrow}_{#1} }

\newcommand{\updownboundary}[2]{ \s{#2}^{\updownarrow}_{#1} }

\newcommand{\errorboundary}[3]{ \rawboundary{#3}^{!}_{ #1 / #2} }

\newcommand\reduwave{\bgroup\markoverwith{\textcolor{red}{\rule[-0.5ex]{2pt}{0.4pt}}}\ULon}
\newcommand{\ghostsymbol}{{//}}
\newcommand{\ghostapp}[2]{ \ghostdiffhl{{}^\ghostsymbol \!\lblkbrbrak \lblkbrbrak} #1 \ghostdiffhl{\rblkbrbrak ~ #2 \rblkbrbrak} }
\newcommand{\ghostvar}[2]{ \ghostdiffhl{{}^\ghostsymbol \! #1_{#2}} }

\newcommand{\downa}[2]{ \downboundary{#1}{#2} }
\newcommand{\upa}[2]{ \upboundary{#1}{#2} }
\newcommand{\down}[2]{ \downboundary{#1}{#2} }
\newcommand{\up}[2]{ \upboundary{#1}{#2} }

\newcommand{\cnil}[0]{\code{nil}}
\newcommand{\ccons}[0]{\code{cons}}
\newcommand{\cInt}[0]{\code{Int}}
\newcommand{\cBool}[0]{\code{Bool}}
\newcommand{\cList}[1]{\code{List}~#1}

\newcommand{\metatheorystyle}[1]{\textsf{#1}}

\newcommand{\mdistance}{\metatheorystyle{distance}}
\newcommand{\mcount}{\metatheorystyle{count}}

\newcommand{\ohc}{one-hole context}

\newcommand{\ohcs}{one-hole contexts}

\newcommand{\link}{https://pantographeditor.github.io/Pantograph/}

%% file: introduction.tex
\section{Introduction}
\label{sec:intro}

\newcommand*{\naive}{traditional}

Structure editors allow a programmer to view and operate on the tree structure of their source code. For decades, many structure editors such as
Alfa~\cite{alfa}, Scratch~\cite{scratch}, MPS~\cite{mps}, or Hazel~\cite{hazelnut},
have allowed for partial programs by leaving the nodes of the program that are yet to be written as (usually typed) holes, which can be filled by any construct of the correct type.

For a concrete example, consider the following expression in some such structure editor, which filters out the negative numbers from 
the result of appending a list \CC{l1} to a term hole of type \CC{List  Int}:
%
\begin{center} \includepantographgraphics{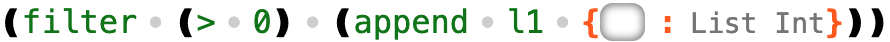} \end{center}
The programmer may then place their cursor in the term hole and fill it with a value \CC{l2} in scope:
%
\begin{center} \includepantographgraphics{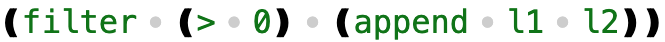} \end{center}
In this manner one can fill in the entire program, and never be permitted to make a syntax or type error.
Unsurprisingly, this quite rigid approach very quickly breaks down: how can one edit existing code? 

\begin{figure}
\begin{subfigure}{\textwidth}
\begin{center}
\begin{tikzpicture}[scale=0.9]
\node[draw] (step0) at (-1,0) {\includepantographgraphics{figures/fill-hole-1.png}};
\node[draw] (step10) at (-6,-1.3) {\includepantographgraphics{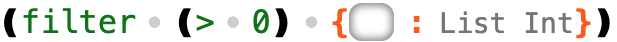}};
\node[draw] (step11) at (0,-1.3) {\includepantographgraphics{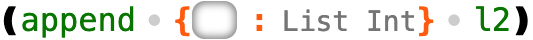}};
\node[draw] (step12) at (4,-1.3) {\includepantographgraphics{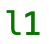}};
\node[draw] (step2) at (-1,-2.7) {\includepantographgraphics{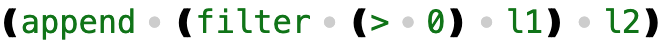}};
\draw[-{[flex,sep]>},thick]
    ($(step0.south) + (-2,0)$)
    .. controls ($(step0.south) + (-2,-0.5)$) and ($(step10.north) + (0,1)$) ..
    ($(step10.north)$);
\draw[->,thick]
    ($(step0.south) + (0.4,0)$)
    .. controls ($(step0.south) + (0.4,-0.5)$) and ($(step11.north) + (0,1)$) ..
    ($(step11.north)$);
\draw[->,thick]
    ($(step0.south) + (1.6,0)$)
    .. controls ($(step0.south) + (1.6,-0.5)$) and ($(step12.north) + (0,1)$) ..
    ($(step12.north)$);
\draw[->,thick]
    ($(step10.south)$)
    .. controls ($(step10.south) + (0,-0.5)$) and ($(step2.north) + (-0.5,1)$) ..
    ($(step2.north) + (-0.5,0)$);
\draw[->,thick]
    ($(step11.south)$)
    .. controls ($(step11.south) + (0,-0.5)$) and ($(step2.north) + (-2,1)$) ..
    ($(step2.north) + (-2,0)$);
\draw[->,thick]
    ($(step12.south)$)
    .. controls ($(step12.south) + (0,-0.5)$) and ($(step2.north) + (1.6,1)$) ..
    ($(step2.north) + (1.6,0)$);
\end{tikzpicture}
\end{center}
\caption{Structured editing traditionally requires breaking a term into a forest of syntactically valid components.}
\label{subfig:structured:rearr}
\vspace*{1em}
\end{subfigure}

\begin{subfigure}{\textwidth}
\begin{small}
\begin{center}
\begin{tabular}{r|l|l}
\hline
\textbf{edit} & \textbf{state} & \textbf{problems}
\\
\hline\hline
select &
\lstinline|(filter (> 0) $\textsf{\textselection{(append }}$l1 l2))| &
\\
cut &
\lstinline|(filter (> 0) l1 l2))| &
ill-formed, ill-typed
\\
paste & 
\lstinline|$\textsf{\textselection{(append }}$(filter (> 0) l1 l2))| &
ill-typed
\\
select & 
\lstinline|(append (filter (> 0) l1$\textsf{\textselection{ l2)}}$)| &
ill-typed
\\
cut & 
\lstinline|(append (filter (> 0) l1)| &
ill-formed, ill-typed
\\
paste & 
\lstinline|(append (filter (> 0) l1)$\textsf{\textselection{ l2)}}$|
\end{tabular}
\end{center}
\end{small}
\caption {Traditional text editing permits convenient grammar- and type-breaking edits. }
\label{subfig:text-rearr}
\end{subfigure}
\vspace*{-1em}
\caption{Structured vs Text Editing Example}
\label{fig:diagramintro}
\vspace*{-1em}
\end{figure}

Suppose that the programmer realizes that \CC{l2} never contains negative numbers, and decides to optimize the expression by moving the execution of the \CC{append} operation to after the \CC{filter}. For this, the mechanism of filling typed holes won't help. In fact, the very constraint that  programmers must work with entire typed terms becomes burdensome. The details depend on the editor, but as Figure~\ref{subfig:structured:rearr} shows, rearranging expressions like this is difficult when you can only operate on entire terms: one essentially has to break down the tree structure into a forest of small sub-trees---in our example \CC{filter (> 0) $\hole{}$}, \CC{append $\hole{}$ l2}, and \CC{l1}---and recombine them after the fact.
This difficulty to edit existing code is known as viscosity in the literature~\cite{BLACKWELL2003103}.

By contrast, this edit is strikingly easy to make in a text editor, which allows users to break the grammatical  and semantic structure of the program.
In particular, one can make this edit with two cut and paste operations as shown in Figure~\ref{subfig:text-rearr}.
Each cut operation temporarily breaks the grammatical structure, and the types don't work out until the very end.
Still, programmers perform such operations all the time.

One approach to solving the viscosity problem is to introduce specific useful actions that make particular structured program edits.
For example, in Alfa~\cite{alfa}, the user edits the program though a menu of language-specific actions.
This contrasts with text editing, which has a simple core interface which allows any edit on any language, on top of which language specific features may be added.

Another approach to solving the viscosity problem, explored by recent work in the structure editor literature, is to allow the user to make text-like edits to the linear projection of a structured program:
rather than rigidly operating on a program only as an intrinsically typed term, such editors allow the user to break the program into syntax which can afterwards be interpreted into typed terms.
In particular, Tylr~\cite{tylr,teentylr} allows a user to move delimiters (such as parentheses), temporarily breaking the grammar of the program, but enforces that these delimiters are eventually placed in a valid location.
Similarly, recent versions of Hazel (as described in \cite{typeerrorlocalization}) allow the user to operate over an untyped grammar, and then mark the program with error forms after the user's edit.
Still, the question remains:
\begin{center}
\emph{Could there be a general fluid editor which operates on strictly structured and typed programs?}
\end{center}

Such an editor would never reduce terms to untyped or ill-formed syntax, and would be characterized by never needing to re-parse or re-typecheck the program.
But at the same time, unlike traditional attempts at such editors, it would not weigh down the programmer when editing existing code.
Further, it would not require the programmer to learn various ad-hoc actions for specific edits, but rather have a small consistent language-generic interface.
If indeed possible, what advantages would such an editor facilitate?

In this paper, we take on this challenge by demonstrating the feasibility and exploring the benefits of such an intrinsically typed editing scheme.
A key observation that enables this fluidity is that existing structure editors lack a proper generalization of text-based selection and related operations.
We develop this single language-generic abstraction based on the well-known notion of one-hole contexts, and it allows for a wide range of structure-preserving operations.
We imbue these operations with a typing structure and explore their metatheory, rooting our design in firm formal foundations.

\begin{figure}
\begin{subfigure}{0.4\textwidth}
    \begin{center}
        \includegraphics[scale=0.25]{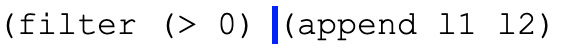}
    \end{center}
    \caption{Text cursor}
    \label{subfig:text-cursor}
    \vspace*{1em}
\end{subfigure}
\quad\quad
\begin{subfigure}{0.4\textwidth}
    \begin{center}
        \includegraphics[scale=0.25]{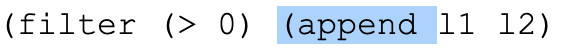}
    \end{center}
    \caption{Text selection}
    \label{subfig:text-select}
    \vspace*{1em}
\end{subfigure}

\begin{subfigure}{0.4\textwidth}
    \begin{center}
        \includepantographgraphics{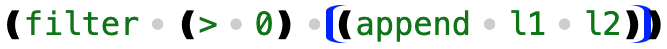}
    \end{center}
    \caption{Tree cursor}
    \label{subfig:tree-cursor}
\end{subfigure}
\quad\quad
\begin{subfigure}{0.4\textwidth}
    \begin{center}
        \includepantographgraphics{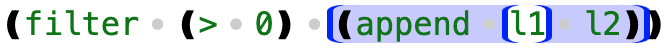}
    \end{center}
    \caption{Tree selection}
    \label{subfig:tree-select}
\end{subfigure}
\caption{Text- and Tree-based Cursors and Selection}
\label{fig:text-and-tree-cursor-select}
\end{figure}

\paragraph{One-Hole Contexts}
Standard text editors allow two different mechanisms of editing a program: a cursor and a selection.
In a text editor, cursors exist between characters, while selections
correspond to the area between two text based cursors. Going back to our earlier example, the text-based
cursor in Figure~\ref{subfig:text-cursor} is before "\CC{(append}", while the text-based selection 
of the same string (Figure~\ref{subfig:text-select}) ranges between two cursor locations and (roughly) corresponds to an application of append to a value.
Traditional structure editors offer users a tree-based analogue for cursors:
where text-based cursors live before characters, tree-based cursors live on top of tree nodes. For example,
the tree-based cursor corresponding to the text one ranges over an entire inner subtree (Figure~\ref{subfig:tree-cursor}).

However, traditional structure editors have no analogue for selection.
Generalizing text-based selection to trees would mean that 
structure editor users should be able to select and edit what lies between two tree-based cursors;
they should be able to select and edit the part of the program that lies between one node and one of its descendants;
they should be able to select and edit expressions with a single subexpression missing;
they should be able to select and edit {\em one-hole contexts}.

One of the contributions of this paper is to introduce such a generalization of structured editing: 
when a one-hole context is cut, 
the expression in its hole takes its place; when a one-hole context is pasted onto an expression, that expression
fills in its hole. For example, our scheme allows the following manipulation with a single cut and paste of a 
one hole context:

\begin{figure}[H]
    \centering
    \vspace*{-1em}
    \begin{tikzpicture}[scale=0.9] 
        \node[anchor=north west] (step0) at (0,0) {\includepantographgraphics{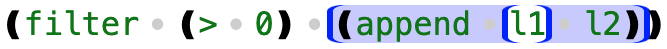}};
        \node[anchor=north west] (step1) at ($(step0.south west) + (0,0.25)$) {\includepantographgraphics{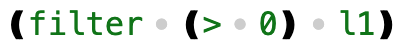}};
        \node[anchor=north west] (step2) at ($(step1.south west) + (0,0.25)$) {\includepantographgraphics{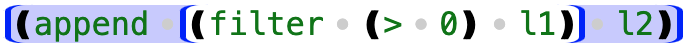}};
        \draw[-{[flex,sep]>},thick]
            ($(step0.west) + (0,0)$)
            .. controls ($(step0.west) + (-0.5,0)$) and ($(step1.west) + (-0.5,0)$) ..
            node[left]{cut}
            ($(step1.west) + (0,0)$);
        \draw[-{[flex,sep]>},thick]
            ($(step1.west) + (0,-0.0)$)
            .. controls ($(step1.west) + (-0.5,0)$) and ($(step2.west) + (-0.5,0)$) ..
            node[left]{paste}
            ($(step2.west) + (0,0)$);
    \end{tikzpicture}
    \vspace*{-1em}
\end{figure}

\noindent
Such selection-based manipulation of trees properly generalizes selection from traditional text-based editing, allowing for fluid editing while preserving the well-formedness of programs at all times. In Section~\ref{sec:zipper-editing}, we provide additional intuition about the edits that this approach enables through examples.
On its own, however, it does nothing to preserve the well-typedness of programs. 


\paragraph{Typed Edits}

For an editor to respect not only syntactic structure but also types, edits must be able to affect multiple locations in a program at once.
Consider, for instance, the following simple recursive program which applies the \CC{not} function to every element of 
an input list \CC{l}:
\begin{center}
    \includepantographgraphics{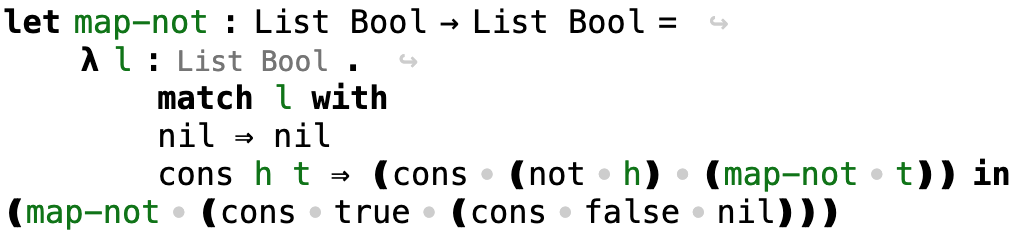}
\end{center}


\noindent
Naturally, a functional programmer will want to abstract away the recursion pattern (here, \CC{map}) from the particular function being applied
(here, \CC{not}). This seemingly requires five distinct edits: an argument must be added to the signature, a lambda to the definition, and an application at both function calls, including the recursive call, and a call to \CC{f} (replacing the call to \CC{not}). This amounts to inserting five different \ohcs{} into the program: 

\label{fig:generalize-map-untyped}
\begin{figure}[H]
    \includepantographgraphics{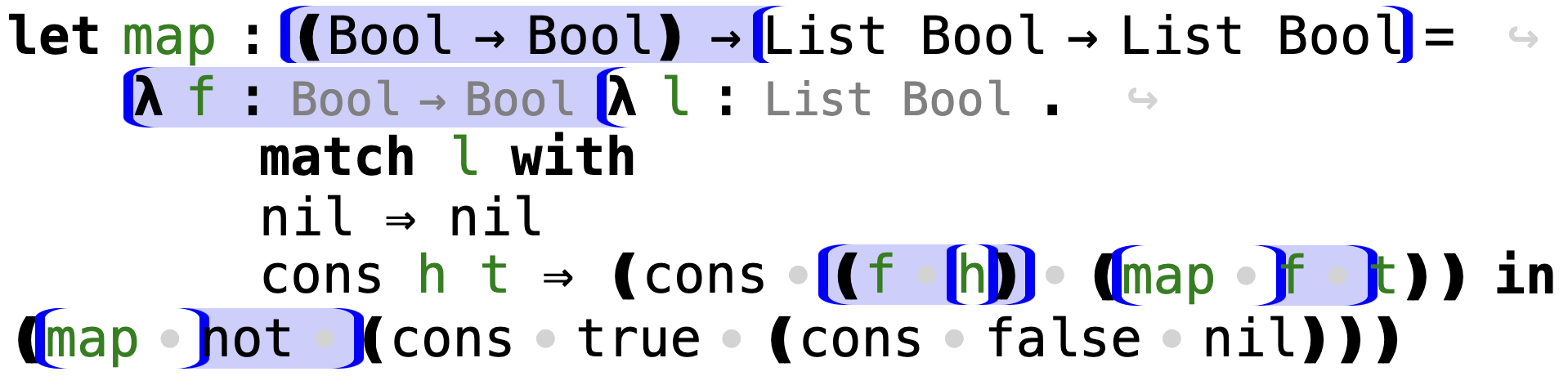}
\end{figure}

Any of these five edits made individually results in an intermediate ill-typed state.
Our second main contribution is a system of typed edits where each one-hole context is typed by a {\em type diff}, corresponding
to the difference between its inner and its outer type.
Armed with this abstraction, programmers need only to insert a 
lambda at the body of \CC{map}, and the extra parameter in the type and applications and holes at the two call sites are automatically inserted by the system. 


This intrinsically typed editing system facilitates useful behaviors that are not possible in other editing systems.
Here, every edit at any term or type results in a well-typed refactoring operation that updates corresponding parts of the program.
In some cases these updates can insert intrinsic type error forms in order to encode type errors into the program itself. However, this approach has access to more information than a traditional type checker because it inputs the edit rather than merely the untyped state after the edit.
This allows it to automatically update corresponding parts of the program, including by placing errors, in ways that are impossible for a traditional type checking or inference algorithm.



We present the following contributions:
\begin{itemize}
    \item We present a new structure editing paradigm that we call "zipper editing" which generalizes text selection to tree-based structures and provides standard edit operations which always maintain the well-formedness of the source program (Section \ref{sec:zipper-editing}).
    \item We define a category of diffs for typing these edits, in which the edits act as refactoring operations on the source program which preserve well-typedness---modulo sometimes inserting explicit type error boundaries
    (Section \ref{sec:typed-editing}).
    We model such refactoring operations as propagating typed edits through a program in a core language with explicit typed edits (Section~\ref{sec:simplemath}); and we explore the metatheory of this language, proving normalization, type-preservation, and confluence (Section~\ref{sec:fullmath}).
    \item We implement our approach in a structure editor we call \href{\link}{Pantograph}.
    To demonstrate the feasibility of our editing paradigm, we conducted a user study in which participants solved functional programming tasks in both our prototype implementation and a 
    traditional text editor (Section~\ref{sec:userstudy}).
\end{itemize}

\noindent
We discuss limitations of our editing paradigm in Section~\ref{sec:limitations}, related work in Section~\ref{sec:relatedwork}, and finally conclude with future work in Section~\ref{sec:conclusion}.

%% file: zipper-editing.tex
\section{Zipper Editing}
\label{sec:zipper-editing}

\begin{wrapfigure}{r}{0.45\textwidth}
  \vspace*{-2em}
  \begin{center}
    \includegraphics[scale=0.33]{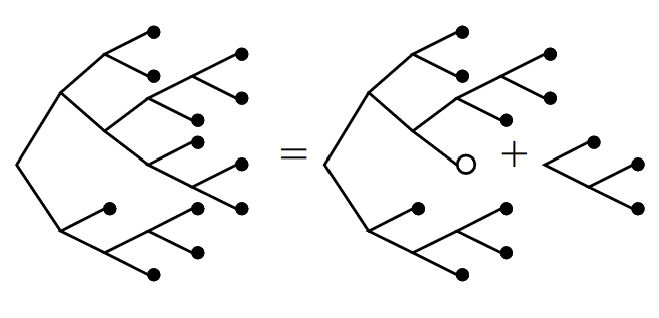}
  \end{center}
  \vspace*{-2em}
  \label{fig:OHC}
\end{wrapfigure}

\begin{figure}[h]
    \centering

    \begin{tikzpicture}[scale=0.9]  
        \node[anchor=north west] (step0) at (0,0) {\includegraphics[scale=0.25]{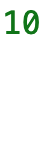}};
        \node[anchor=north west] (step1) at (2,0) {\includegraphics[scale=0.25]{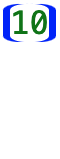}};
        \node[anchor=north west] (step2) at (5.5,0) {\includegraphics[scale=0.25]{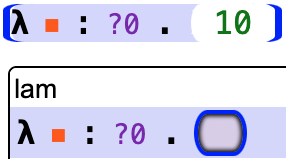}};
        \node[anchor=north west] (step3) at (10,0) {\includegraphics[scale=0.25]{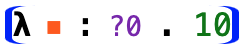}};
        \draw[->,thick] ($(step0.north east) - (0,1em)$) to node[above,text width=5em,align=center]{move} 
                        ($(step1.north west) - (0,1em)$);
        \draw[->,thick] ($(step1.north east) - (0,1em)$) to node[above,text width=5em,align=center]{type ``lam''} 
                        ($(step2.north west) - (0,1em)$);
        \draw[->,thick] ($(step2.north east) - (0,1em)$) to node[above,text width=5em,align=center]{submit} 
                        ($(step3.north west) - (0,1em)$);
    \end{tikzpicture}

    \caption{Inserting a \kw{\textlambda} around a term of the program: Users query the name of the construction that they want to insert, and Pantograph wraps it around the term at the cursor; $?0$ represents an unknown type.\vspace*{-1em}}
    \label{fig:zipper-edit-insert-lambda}
\end{figure}

Zippers, originally proposed by \citet{huet-zipper}, allow trees to be decomposed into a one-hole context and a ``subtree of interest'', as in the figure on the right found in \citet{mcbride-derive-ohc}.
%
This decomposition allows for efficient local editing of tree-like data structures. 
Naturally, it is well known that this notion can represent cursors in both text and structure editors.
In text editors---ignoring their 2D layout aspect---the cursor position could be seen as a pair of strings:
the one hole context that precedes the cursor, and the text that follows. In structure editors, the cursor position
is the one-hole context that surrounds a program node, together with the node under focus. In this representation,
local program operations such as inserting a construct into the program or navigating the cursor
are easily and efficiently expressed~\cite{huet-zipper}.
For example, Figure~\ref{fig:zipper-edit-insert-lambda} shows how easy it is to insert a lambda around a term at the cursor in Pantograph.

For concreteness, consider the following simple functional language core, consisting of $\lambda$ abstractions, applications, variables, \CC{let} expressions, and term holes.
\begin{align*}
    t ~&::=~
        \llet{x}{T}{t}{t} \pipe
        \fun{x}{T}{t} \pipe
        t \; t \pipe
        x
    \\
    C ~&::=~
        \llet{x}{T}{C}{t} \pipe
        \llet{x}{T}{t}{C} \pipe
        \fun{x}{T}{C} \pipe
        C \; t \pipe 
        t \; C
        \pipe \clasp{}
\end{align*}
The corresponding grammar of one-hole contexts allows for exactly one subterm to be missing at the hole $\clasp{}$.
%
%
Given a one hole context $C$ and a term $t$, we write $C[t]$ to represent filling in the hole in $C$ with $t$ to get a single term. 
For instance, if the term in Figure~\ref{fig:zipper-edit-insert-lambda} was the focus of some bigger program $C[\code{10}]$,
inserting that $\kw{\textlambda}$ expression at the cursor results in the new program  $C[\lambda x. 10]$.

However, text editors wouldn't be very useful without the ability to select text. But what is text selection? Following the same intuition as before, it's a triple of strings: the text before the selection, the selection itself, and the text after it. How does that notion generalize to trees?

According to this analogy, selection in trees can be defined as a triple consisting of two one hole contexts $C_1$ and $C_2$ and one term $t$, arranged as $C_1 \ohchole{C_2 \ohchole{t}}$, where $C_2$ is the selected region. In our example from the introduction, the selection of \CC{append} and \CC{l2} corresponds to the following decomposition of the whole
program:


\begin{center}
\begin{small}
\begin{tabular}{ p{13em} p{1em} c c c }
\Tree [.$\code{App}$ $\code{filter}$ [.$\code{(> 0)}$ ] [.$\code{App}$ $\code{append}$ l1 $\code{l2}$ ]]
&
=
&
\Tree [.$\code{App}$ $\code{filter}$ [.$\code{(> 0)}$ ] $\clasp{}$ ]
+
\Tree [.$\code{App}$ $\code{append}$ $\clasp{}$ $\code{l2}$ ]
+
\Tree [.l1 ]
\end{tabular}
\end{small}
\end{center}

\noindent
This decomposition allows for very quick selection-based operations similar to text editing. For example, by reordering the one hole contexts, we can quickly rearrange this expression:

\begin{center}
\begin{small}
\begin{tabular}{ p{16em} p{0em} c c c }
\Tree [.$\code{App}$ $\code{append}$ [.$\code{App}$ $\code{filter}$ [.$\code{(> 0)}$ ] l1 ] $\code{l2}$ ]
&
=
&
\Tree [.$\code{App}$ $\code{append}$ $\clasp{}$ $\code{l2}$ ]
+
\Tree [.$\code{App}$ $\code{filter}$ [.$\code{(> 0)}$ ] $\clasp{}$ ]
+
\Tree [.l1 ]
\end{tabular}
\end{small}
\end{center}


\begin{figure}[b]
    \centering
    \begin{tikzpicture}[scale=0.9]
        \node (step0) at (0,0) {\includegraphics[scale=0.25]{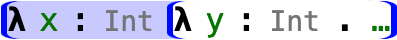}};
        \node (step1) at (3.7,0) {\includegraphics[scale=0.25]{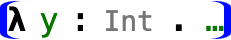}};
        \node (step2) at (6.5,0) {\includegraphics[scale=0.25]{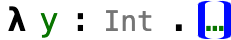}};
        \node (step3) at (10,0) {\includegraphics[scale=0.25]{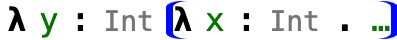}};
        \draw[->,thick] (step0.east) to node[above,yshift=0.5em]{cut} (step1.west);
        \draw[->,thick] (step1.east) to node[above,yshift=0.5em]{move} (step2.west);
        \draw[->,thick] (step2.east) to node[above,yshift=0.5em]{paste} (step3.west);
    \end{tikzpicture}
    
    \caption{Cutting and pasting a selection into a different location in Pantograph. Zipper editing lends itself to a nice user interface similar to text editing. The user may make a selection with a familiar click and drag motion. The user may also cut selections into a clipboard, and paste them later at the cursor.}
    \label{fig:zipper-edit-select-lambda}
\end{figure}
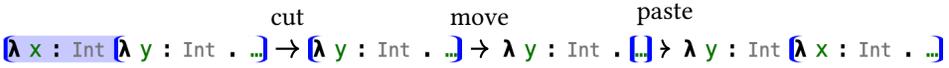

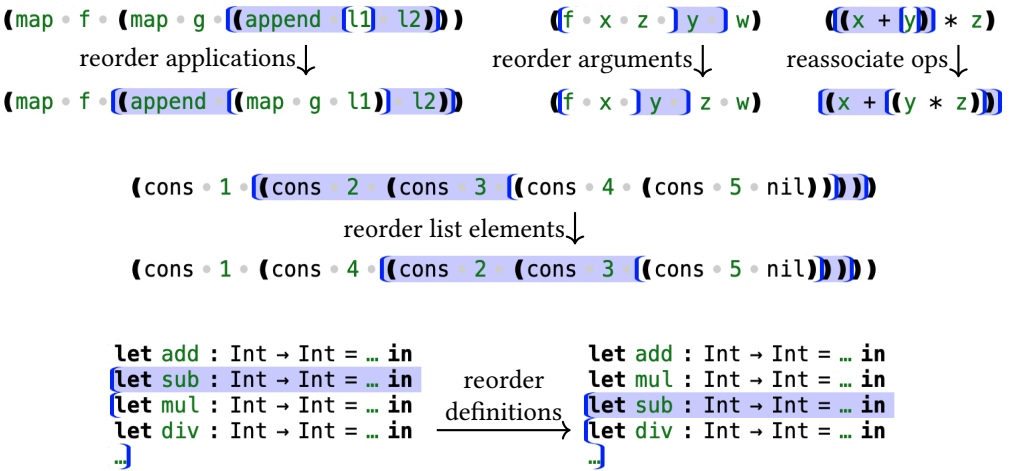
\begin{figure}[b]
\begin{center}
    \begin{tabular}{c}
        \begin{tikzpicture}[scale=0.9] 
            \node (step0) at (0,0) {\includegraphics[scale=0.25]{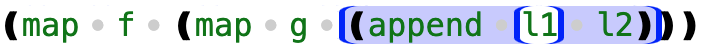}};
            \node (step1) at (0,-1.2) {\includegraphics[scale=0.25]{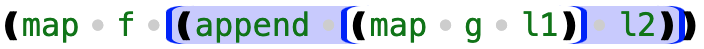}};
            \draw[->,thick] ($(step0.south) + (3em,0)$) to node[left]{reorder applications} ($(step1.north) + (3em,0)$);
        \end{tikzpicture}
        \begin{tikzpicture}[scale=0.9]
            \node (step0) at (0,0) {\includegraphics[scale=0.25]{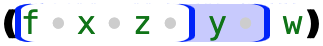}};
            \node (step1) at (0,-1.2) {\includegraphics[scale=0.25]{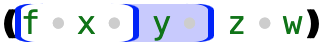}};
            \draw[->,thick] ($(step0.south) + (2em,0)$) to node[left]{reorder arguments} ($(step1.north) + (2em,0)$);
        \end{tikzpicture}
        \begin{tikzpicture}[scale=0.9]   
            \node (step0) at (0,0) {\includegraphics[scale=0.25]{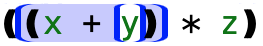}};
            \node (step1) at (0,-1.2) {\includegraphics[scale=0.25]{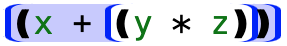}};
            \draw[->,thick] ($(step0.south) + (2em,0)$) to node[left]{reassociate ops} ($(step1.north) + (2em,0)$);
        \end{tikzpicture}
        \\[1em]
        \begin{tikzpicture}[scale=0.9]    
            \node (step0) at (0,0) {\includegraphics[scale=0.25]{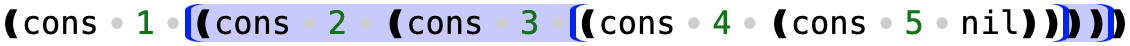}};
            \node (step1) at (0,-1.2) {\includegraphics[scale=0.25]{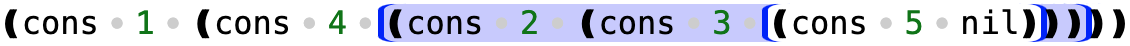}};
            \draw[->,thick] ($(step0.south) + (3em,0)$) to node[left]{reorder list elements} ($(step1.north) + (3em,0)$);
        \end{tikzpicture}
        \\[1em]
        \begin{tikzpicture}[scale=0.9]    
            \node (step0) at (0,0) {\includegraphics[scale=0.25]{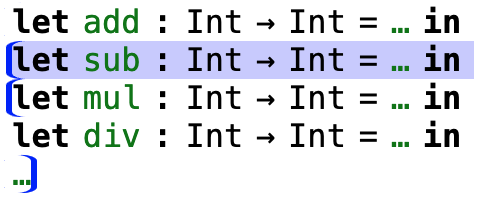}};
            \node (step1) at (7,0) {\includegraphics[scale=0.25]{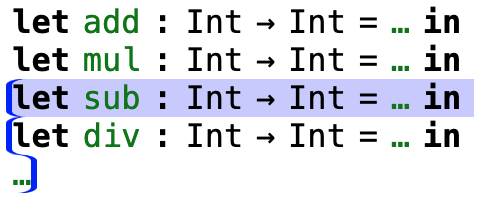}};
            \draw[->,thick] ($(step0.east) - (0,1em)$) to node[above,text width=5em,align=center]{reorder definitions} ($(step1.west) - (0,1em)$);
        \end{tikzpicture}
    \end{tabular}

    \caption{Various edits each performed with a single cut and paste of a \ohc~selection. The first selection in each example is made by the user, but the second selection is only illustrative.}
    \label{fig:simple-zipper-edits}
\end{center}
\end{figure}


Any number of features that a structure editor could have would allow this particular edit.
But, we take inspiration from the design of the standard text editor, where a few simple operations enable arbitrary text manipulation. One-hole context selection is similarly easy to use, as shown in Figure~\ref{fig:zipper-edit-select-lambda}.
It also is similarly versatile; Figure~\ref{fig:simple-zipper-edits}  shows a wide variety of edits that can be made with these selections.
In each example, the user performs a single cut and paste operation.




%% file: typed-editing.tex
\section{Typed Editing}
\label{sec:typed-editing}

Zipper editing, like the traditional structure editing it extends, preserves the syntactic well-formedness of programs.
However, it does not necessarily preserve well-typedness.
Let's add a few types to our core language of the previous
section (integers, booleans, and lists) that will enable us to encode the \CC{map} example of the introduction, and define a 
standard simple type system with recursive $\code{let}$, pattern matching on lists, and typed holes $\holety{T}$, as shown in Figure~\ref{fig:typing-rules}.

\begin{figure}
    \centering
    \begin{align*}
        T ~&::=~
            \cInt \pipe 
            \cBool \pipe
            T \to T \pipe
            \cList{T}
        \hspace{4em}
        \begin{array}{cc}
             \\
            \hline
            \Gamma \vdash \holety{T} : T
        \end{array}
    \end{align*}
    \vspace*{-1em}
    \begin{align*}\begin{array}{c}
        \begin{array}{c} i \in \mathbb{Z} \\ \hline \Gamma \vdash i : \cInt \end{array}
        \hspace{1em}
        \begin{array}{c} b \in \mathbb{B} \\ \hline \Gamma \vdash b : \cBool \end{array}
        \hspace{1em}
        \begin{array}{c} \\ \hline \Gamma \vdash \cnil : \cList{T} \end{array}
        \hspace{1em}
        \begin{array}{c}
            \\ \hline
            \Gamma \vdash \ccons~h~t : T \to \cList{T} \to \cList{T}
        \end{array}
        \\[1.5em]
        \begin{array}{c}
            \\ \hline
            \Gamma, x : T \vdash x : T
        \end{array}
        \hspace{1em}
        \begin{array}{c}
            \Gamma, x : T_1 \vdash b : T_2
            \\ \hline
            \Gamma \vdash \fun{x}{T_1}{b} : T_1 \to T_2
        \end{array}
        \hspace{1em}
        \begin{array}{c}
            \Gamma \vdash f : T_1 \to T_2 \hspace{2em}
            \Gamma \vdash a : T_1
            \\ \hline 
            \Gamma \vdash f~a : T_2
        \end{array}
        \\[2em]
        \begin{array}{c}
            \Gamma, x : T_1 \vdash a : T_1 \\
            \Gamma, x : T_1 \vdash b : T_2
            \\ \hline 
            \Gamma \vdash \llet{x}{T_1}{a}{b} : T_2
        \end{array}
        \hspace{1em}
        \begin{array}{c}
            \Gamma \vdash a : \cList{T_1}
            \hspace{2em}
            \Gamma \vdash b : T_2 \\ 
            \Gamma, h : T_1, t : \cList{T_1} \vdash c : T_2
            \\ \hline
            \Gamma \vdash \matchTwo{a}{\cnil}{b}{\ccons~h~t}{c} : T_2
        \end{array}
    \end{array}\end{align*}

    \caption{A typed core language}
    \vspace*{-1em}
    \label{fig:typing-rules}
\end{figure}

Recall the edits required to add a parameter to the \CC{map} function (Figure~\ref{fig:generalize-map-untyped}). While the result of performing all four of the edits yields a well-typed program, performing just the first edit of inserting 
\code{\fun{f}{\cBool \to \cBool}{\clasp{}}}
leaves the program in an intermediary ill-typed state since type annotation of \code{map} would still be \code{\cList\cBool \to \cList\cBool}.

Any system that aims to operate on intrinsically typed terms needs to account for how such an edit changes types in the program. To that end, we will introduce a grammar of \emph{type diffs}, encoding precisely how a type is transformed into a new type after an edit. We will use $\delta$ to range over these diffs and write $T_1 \xRightarrow{\delta} T_2$ for a diff $\delta$ which changes type $T_1$ to $T_2$. The grammar of diffs for 
our core language is shown below; we will systematically generalize
to other language features in Section~\ref{sec:fullmath}:

\begin{align*}
    \label{fig:type_diff_intro_syntax}
    \delta ~&::=~
        \cInt \pipe \cBool \pipe
        \delta \to \delta \pipe 
        \cList{\delta} \pipe 
        \plusdiff{T \to}{\delta}{} \pipe 
        \minusdiff{T \to}{\delta}{}
        \pipe \replacediff{T}{T}
\end{align*}


\vspace{1em}
The first two of these constructors are identity diffs which don't alter the type:

\begin{align*}
    \inferrule{}{
        \diffsto{\cInt}{\code{Int}}{\cInt}
    }
    \hspace{3em}
    \inferrule{}{
         \diffsto{\cBool}{\cBool}{\cBool}
    }
\end{align*}
The next two constructors represent diffs which preserve a top level constructor of a type, and apply diffs to the child types:

\begin{align*}
    \inferrule{
    \diffsto{T_1}{\delta_1}{T_1'}
    \\
    \diffsto{T_2}{\delta_2}{T_2'}
    }{
    \diffsto{T_1 \to T_2}{\delta_1 \to \delta_2}{T_1' \to T_2'}
    }
    \hspace{3em}
    \inferrule{
        \diffsto{T}{\delta}{T'}
    }{
        \diffsto{\cList{T}}{\cList{\delta}}{\cList{T'}}
    }
\end{align*}

The final three constructors represent diffs which actually alter types: the first two insert or remove a function type respectively, and then (potentially) change the output type of the function by a diff; the latter simply replaces one type with another:

\begin{align*}
\inferrule{
    \diffsto{B_1}{\delta}{B_2}
}{
    \diffsto{B_1}{\plusdiff{A \to}{\delta}{}}{A \to B_2}
}
\hspace{3em}
\inferrule{
    \diffsto{B_1}{\delta}{B_2}
}{
    \diffsto{A \to B_1}{\minusdiff{A \to}{\delta}{}}{B_2}
}
\hspace{3em}
\inferrule{
}{
    \diffsto{T}{\replacediff{T}{T'}}{T'}
}
\end{align*}

Let's once again revisit our \CC{map} example, where the user inserts a one hole context
$\fun{\code{f}}{\cBool \to \cBool}{\clasp{}}$.
This can be given the following diff between two types:

\begin{align*}
\begin{array}{l}
\diffsto
    {T_1 }
    {\plusdiff{(\cBool \to \cBool) \to}{\cList\cBool \to \cList\cBool}{}}
    {T_2} 
\\[1em]
\begin{array}{rl}
    \text{where} &
        T_1 = \cList\cBool \to \cList\cBool \\ &
        T_2 = (\cBool \to \cBool) \to \cList\cBool \to \cList\cBool
\end{array}
\end{array}
\end{align*}


\noindent
When the user inserts a \code{\kw{\textlambda}} expression around the body of \CC{map}, Pantograph automatically makes the edits necessary to keep the program well typed. The system
adds an application to a hole at the two call sites, and alters the type signature, as shown in Figure~\ref{fig:insert_parameter_to_map_bignot}.
In the next section, we will explain how our typed editing system uses diffs to calculate and perform these changes. But first, we will further demonstrate the system's capabilities with a few more examples. 

\input{figures/insert_parameter_to_map_bigstep}

The inductive structure of diffs allows us to represent edits to the higher-order structure of functions. For example, suppose that the user wants to generalize \CC{map} so that the mapped function takes an index.
A type diff can describe the resulting change to the type of \CC{map}:
\begin{align*}
    \code{\plusdiff{\cInt \to}{\cBool \to \cBool}{} \to \cList\cBool \to \cList\cBool}.
\end{align*}

\noindent
Again, when the user edits the type of \CC{f}, Pantograph makes various edits to the rest of the code to maintain its well-typedness. This time, Pantograph wraps a \code{\kw{\textlambda}} around the \CC{not} function at the call site to account for the change in its type (Figure~\ref{fig:typed-editing-example-insert-ho-parameter-not}).

\input{figures/insert_higher_order_parameter_to_map_bigstep}

Of course,  it is not always desirable for an editor to fix typing issues automatically. Sometimes, Pantograph leaves errors in the program for the user to fix later.
For example, suppose that the user deletes the \code{f} parameter from a finished map function.
This amounts to removing the one hole context
$\fun{\code{f}}{\cBool \to \cBool}{\clasp{}}$ from the program.
This time, the edit gets a diff which subtracts a function argument:

\begin{align*}
    \minusdiff{(\cBool \to \cBool) \to }{\cList \cBool \to \cList \cBool}{}
\end{align*}

When the user makes the deletion, it leaves a couple of errors in the program. There is an unbound call to \CC{f}, and an out of place argument at the two call sites to \CC{map}.
While the system could simply replace the former with a hole and remove the latter two from the program, this would likely erase valuable work that the programmer wanted to keep. 




\begin{figure}[H]
    \centering
    \codeblockeditsequencetwosteps
        {\includepantographgraphics[0.2]{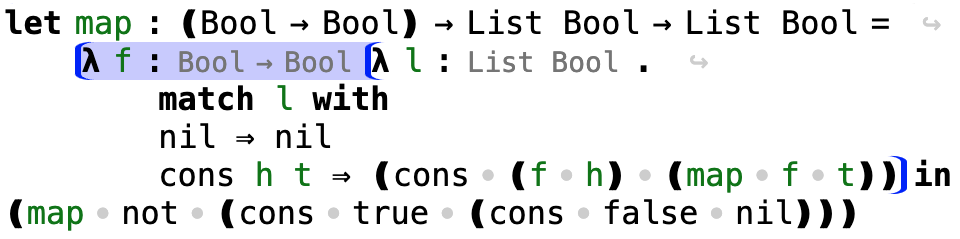}}
        {\includepantographgraphics[0.2]{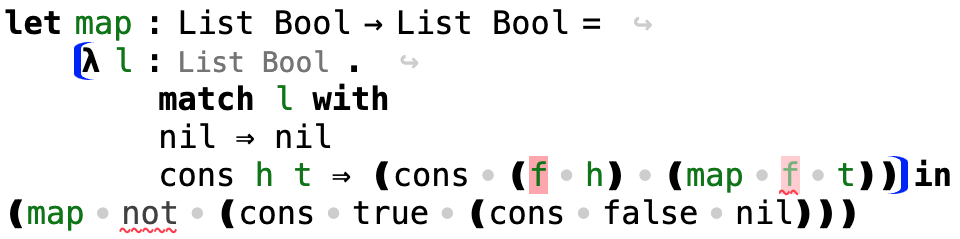}}
    \caption{Typed editing example 3: deleting a parameter. This deletion requires updating the rest the program according to the corresponding type diff, which also describes a change in the context in the body of \CC{map} where \CC{f} is no longer bound.\vspace*{-1em}}
    \label{fig:typed-editing-example-deleting-a-parameter}
\end{figure}

To allow such errors to exist in an otherwise well typed program, 
Pantograph has three final constructions.
The first two of these forms are: \emph{free variables}, which are variables whose binders have been removed, like \CC{f} in the example; and \emph{commented applications}, which are function arguments which no longer fit in place, like the first argument to \CC{map}.

\begin{align*}
    \inferrule{
        \Gamma \vdash f : A \\
        \Gamma \vdash a : B
    }{
        \Gamma \vdash \ghostapp{f}{a} : A
    }
    \;
    \textrm{\sc{Commented Application}}
    \hspace{3em}
    \inferrule{
    }{
        \Gamma \vdash \ghostvar{x}{T} : T
    }
    \;
    \textrm{\sc{Free Variable}}
\end{align*}

Any errors resulting from the user deleting a selection can be accounted for with these two constructs. However, there is one final way for a user to create an error that can't be resolved in this way: the user can simply replace one type annotation with another. Suppose that the user decides to delete the \CC{Bool} type in the output of the higher order argument to \CC{map}, and replace it with \CC{Int}. This is represented with a diff that has a replacement:

\begin{align*}
    (\cBool \to \replacediff{\cBool}{\cInt}) \to \cList\cBool \to \cList\cBool
\end{align*}




\begin{figure}[H]
    \centering
    \codeblockeditsequencetwosteps
        {\includepantographgraphics[0.2]{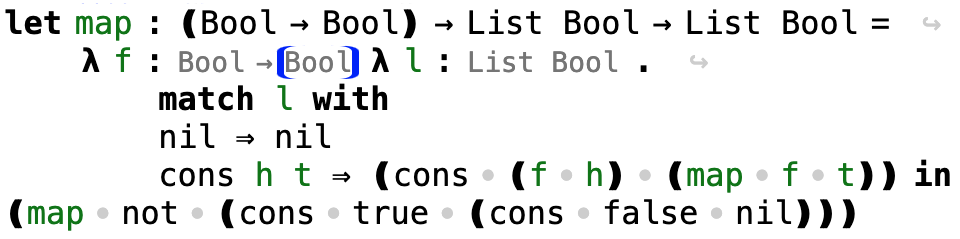}}
        {\includepantographgraphics[0.2]{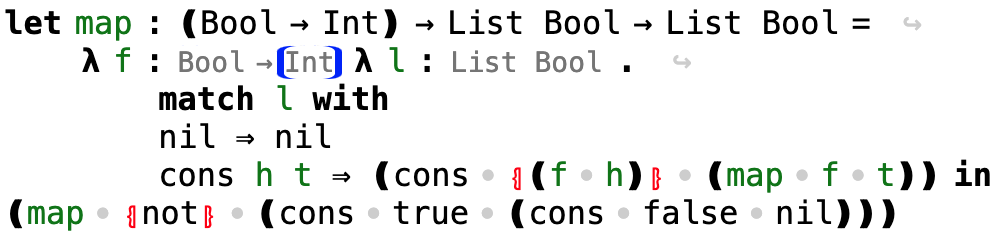}}
    \caption{Typed editing example 4: replacing one type with another. The corresponding type diff for this edit leaves some expressions ill-typed, so they must be placed in type error boundaries.
    \vspace*{-1em}}
    \label{fig:example-create-type-boundary}
\end{figure}

Pantograph is able to update the annotation of \CC{map}, but it can't resolve the type error at the call to $\code{f}$ so it inserts a \emph{type error boundary}. This is the third and final form that Pantograph uses to represent errors in a program. Its typing rule reflects it's meaning: it has a single child term, which can have a different type than the surrounding term.

\begin{align*}
    \inferrule{
        \Gamma \vdash t : T_1
    }{
        \Gamma \vdash \errorboundary{T_1}{T_2}{t} : T_2
    }
    \;
    \textrm{\sc{Type Error Boundary}}
\end{align*}


Type error boundaries are similar to the type errors that are placed by usual type checkers, except that they are a first-class term within the program that the user can interact with.
Such first-class errors are characteristic of structure editors that rigidly operate on typed terms (such as early iterations of Hazel~\cite{hazelnut}) because they allow errors to exist without resorting to fully untyped syntax.
However, Pantograph's ability to respond to an edit non-locally makes them especially useful.
The user can select and delete it just like any other form in the language --- and Pantograph will alter the surrounding code to fit the type inside!



\begin{figure}[H]
    \centering
    \codeblockeditsequencetwosteps
        {\includepantographgraphics[0.2]{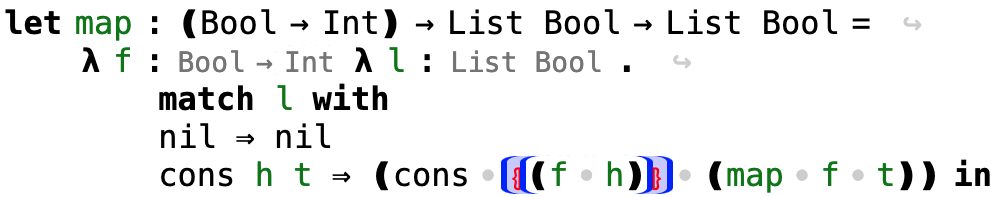}}
        {\includepantographgraphics[0.2]{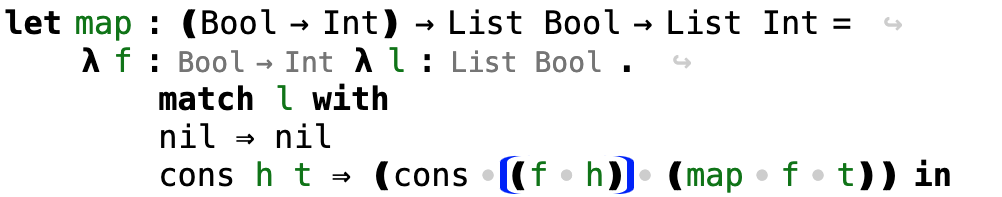}}
    \caption{Typed editing example 5: deleting a type error boundary. When the user deletes the type boundary, Pantograph automatically fixes the output type of \CC{map} to be $\cInt$.}
    \label{fig:example-delete-type-error-boundary}
\end{figure}

For instance, suppose that the programmer decides that the output type of \CC{Int} is actually the correct type for the output list of \CC{map}. The programmer can delete the type error boundary around $\code{f}$, resulting in the diff $\replacediff{\cBool}{\cInt}$ (Figure~\ref{fig:example-delete-type-error-boundary}).
Deleting a type error boundary tells the system that the term inside actually has the desired type, and the system will make the program surrounding it conform to that choice. Crucially, this isn't a special feature of error boundaries; deleting any one hole context will have the same effect.


%% file: figures/insert_parameter_to_map_bigstep.tex
\begin{figure}[H]
\centering



\codeblockeditsequencetwosteps
    {\includepantographgraphics[0.2]{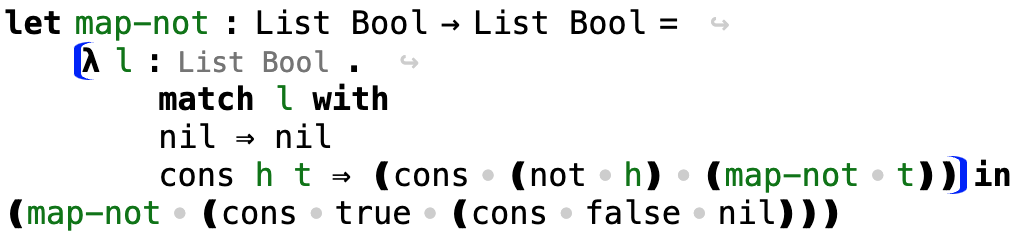}}
    {\includepantographgraphics[0.2]{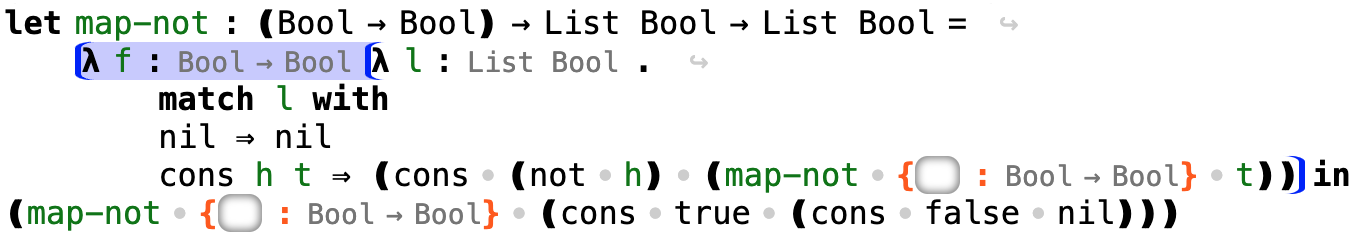}}

\caption{Typed editing example 1: adding a parameter to \CC{map-not}.}
\label{fig:insert_parameter_to_map_bignot}
\end{figure}

%% file: figures/insert_higher_order_parameter_to_map_bigstep.tex
\begin{figure}[H]
\centering
\codeblockeditsequencetwosteps
    {\includepantographgraphics[0.2]{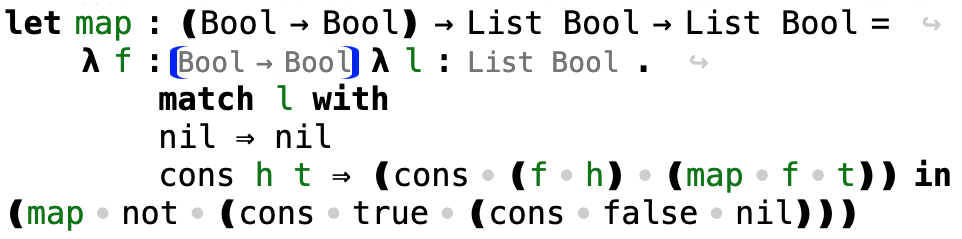}}
    {\includepantographgraphics[0.2]{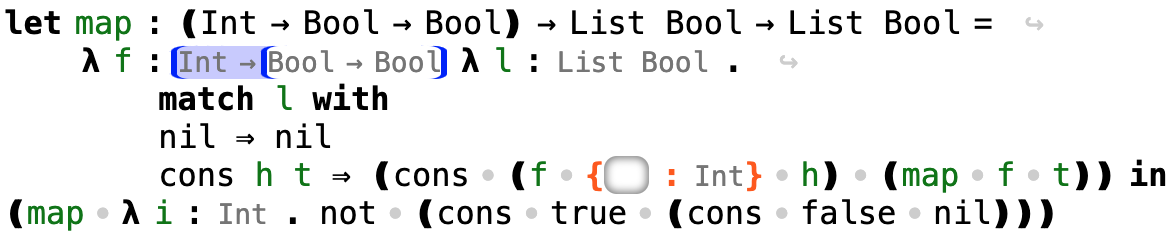}}

\caption{Typed editing example 2: adding a higher-order parameter to \CC{map}.}
\label{fig:typed-editing-example-insert-ho-parameter-not}
\end{figure}

%% file: simplediffmath.tex
\section{Computing Refactorings with diffs, by example}
\label{sec:simplemath}

In the previous section, we showed that type diffs precisely encode how an edit changes the type of a term, and gave several examples of how Pantograph changes a program in response.
In this section, we will describe at a high level how our system computes these changes automatically.
For now, we will elide some technical details in order to facilitate an end-to-end example; then, in the next section, we will formally develop our theory of typed edits in its entirety.

Consider the following simple example of an identity function applied to the constant 10, where the user inserts a $\lambda y : \cBool . \clasp{}$ one-hole context around the term \CC{10}, and Pantograph creates an application around the call to \CC{x} in response:

\vspace{-0.5em}
\begin{align*}
(\lambda x . x) [10] \xrightarrow{\mbox{\normalsize $\textrm{Insert } \lambda y : \cBool . \clasp{}$}} (\lambda x . x~\holety{\cBool}) (\lambda y . 10)
\end{align*}
\vspace{-1em}

To represent intermediate states of a computation, we introduce two forms $\up{d}{t}$ and $\down{d}{t}$ of \emph{diff boundaries}, which we call \emph{up} and \emph{down} boundaries respectively, into our grammar of terms.
Our system inserts an up boundary at the location of the user's edit to represent the change in types induced by the edit:

\vspace{-1em}
\begin{align*}
(\lambda x . x) \up{\plusdiffs{\cBool \to}{ \cInt}{}}{\lambda y . 10}
\tag{step 1}
\end{align*}

Next, Pantograph propagates this boundary through the surrounding program one language form at a time, making use of the corresponding typing rule at each step.
In our example the up boundary surrounds the argument to an application, so Pantograph looks at that typing rule, and in particular it's second premise:

\vspace{-0.5em}
$$
    \begin{array}{c}
        \overbrace{\Gamma \vdash t_1 : A \to B}^{\textrm{Output}} \hspace{2em}
        \overbrace{\Gamma \vdash t_2 : A}^{\textrm{Unify input diff with $A$}} 
        \\ \hline 
        \Gamma \vdash t_1~t_2 : B
    \end{array}
$$
Pantograph, then, needs to relate the type $A$ of this premise with the diff on the boundary $\plusdiff{\cBool \to }{\cInt}{}$. In general this is done using a unification procedure described
in the following section; in this case, it discovers that $A$ maps to this diff. Next, this new information needs to be propagated `through' the form, placing new boundaries around the form or its other children to account for the changes to $A$.
In this case, $A$ appears in the type of the first premise $A \to B$,
and so our algorithm creates a down boundary around the left child of the application:

\vspace{-0.3em}
\begin{align*}
\down{\plusdiffs{\cBool \to}{ \cInt}{} \to \cInt}{\lambda x . x} (\lambda y . 10)
\tag{step 2}
\end{align*}

\noindent
\begin{minipage}{0.75\textwidth}
At this point, the down boundary surrounds a $\lambda$ abstraction so our algorithm looks at the corresponding typing rule (on the right).
To proceed, Pantograph must unify the incoming diff with the type of the conclusion of the rule, $A \to B$.
The result is that $B$ is mapped to an identity diff $\cInt$, so our system ignores it.
On the other hand, $A$ maps to the interesting part of the type diff. The other appearance of $A$ in the rule is in the type of $x$ in the context of the premise, which means we need to propagate a change
to a {\em context}.
\end{minipage}
\begin{minipage}{0.25\textwidth}
\begin{align*}
\inferrule{
\overbrace{\Gamma, x : A \vdash t : B}^{\textrm{Output}}
}{
\underbrace{\Gamma \vdash \lambda x . t : A \to B}_{\textrm{Unify input diff with $A \to B$}}
}
\end{align*}
\end{minipage}

\vspace{1em}
In order to express such a change, we introduce a notion of \emph{context diffs}, which describe how a context changes, either by altering a variable by a given type diff, inserting a new variable, or removing a variable. We will discuss these further in Section~\ref{sec:general-diffs}.

\vspace{-0.9em}
\begin{align*}
    \Delta ~::=~        \varnothing \pipe
        \Delta, x : \delta \pipe
        \plusdiff{}{\Delta}{, x : T} \pipe
        \minusdiff{}{\Delta}{, x : T}
\end{align*}

Using a context diff, our system can propagate the knowledge of how the type of $x$ changes into the body of the abstraction. So far we have elided context diffs on the diff boundaries to facilitate a simpler introduction. However, the full form of diff boundaries that we need to use when there is a nontrivial context diff is $\Delta \vdash \delta$, encoding a change to both the context and type. Propagating such a diff results in the following term:

\vspace{-0.5em}
\begin{align*}
(\lambda x : \down{\varnothing, x : \plusdiffs{\cBool \to}{ \cInt}{} \vdash Int}{x}) (\lambda y . 10)
\tag{step 3}
\end{align*}

\noindent
The boundary now surrounds a variable and Pantograph looks up the corresponding rule:

\vspace{-1em}
\begin{align*}
\inferrule{
}{
\Gamma, x : T \vdash x : T
}
\end{align*}

There are two occurrences of $T$ in the conclusion of the rule, one in the context and one in the type. The existence of such nonlinearities, where the same variable appears twice in the same part of a typing rule, will inform some of the details of our exposition later on. In this case, the algorithm 
unifies the diff with the first occurrence of the $T$ in the context, creating a new boundary in the opposite direction with a diff corresponding to the other occurrence of $T$ in the type:

\vspace{-1em}
\begin{align*}
(\lambda x : \up{\plusdiffs{\cBool \to}{ \cInt}{}}{x}) (\lambda y . 10)
\tag{step 4}
\end{align*}

Finally, the system has arrived at the location where the edit must be made. Until now, propagation of the diffs though the syntax was determined by the structure of the typing rules. However, at this final step, the choice of what the system does with the diff is a more subjective design decision---we could chose to simply leave an error boundary. Pantograph automatically inserts arguments to functions when necessary, according to the following rule:

\vspace{-0.5em}
\begin{align*}
\upa{\Delta \vdash \plusdiffs{A \to}{\delta}{}}{t}
\step
\upa{\Delta \vdash \delta}{t \; \holety{A}}
\end{align*}

Pantograph has 8 such \emph{alteration} rules hard-coded in to determine how it makes edits, which we describe fully in Section~\ref{sec:alterationanddiffrules}. But in this case,
applying this rule gives us the final result of the edit:

\vspace{-0.5em}
\[
(\lambda x . x~\holety{\cBool}) (\lambda y . 10)
\tag{step 5}
\]

%% file: fullmath.tex
\section{Diffs and Diff Propagation In Detail}
\label{sec:fullmath}


So far we have shown examples of the automatic changes that Pantograph makes in response to a user's edit.
In this section, we will finally describe the underlying mechanism in detail.
First, we will unify the treatment of type and context diffs, introducing a category of diffs over arbitrary trees.
Then, we will describe an algorithm to propagate these boundaries through the program as a small-step operational semantics, and explore the metatheory of the entire system.

\subsection{Terms and Types as Trees}
\label{sec:syntaxastrees}
To operate on our types, contexts, and typing rules algorithmically, we will represent them in a 
standard intrinsic style: 
whereas in extrinsic typing rules the term is included in the typing judgement, in intrinsic ones the premises are implicitly understood to refer to the children of the form.
For example, below is the typing rule for an application $t_1~t_2$ rewritten into intrinsic style; the two premises are derivations for $t_1$ and $t_2$.

\vspace{-0.5em}
\begin{align*}
\inferrule{
\Gamma \vdash t_1 : A \to B
\\
\Gamma \vdash t_2 : A
}{
\Gamma \vdash t_1 \; t_2 : B
}
\hspace{2em}
\xrightarrow{becomes}
\hspace{2em}
\inferrule{
\Gamma \vdash A \to B
\\
\Gamma \vdash A
}{
\Gamma \vdash B
}
\end{align*}

Types and contexts both have particular grammars,
but
almost all of these details are irrelevant to the propagation algorithm. As a result, we can take inspiration from S-expressions~\cite{sexp}, and instead consider both of them as trees over a set of labels $l$
which we represent with $s$.
Each tree consists of a label and a list of children:

\vspace{-0.8em}
\begin{align*}
    s ::= l \; \overline{s}
\end{align*}
\vspace{-1.4em}

We collect all of the type constructors $\s{l_{\to}, l_{\ttt{Int}}, l_{\ttt{Bool}}, l_{\ttt{List}}}$ and context constructors $\s{l_{\varnothing}, l_{\square,\square}}$ into a set of labels $l$. We also make a label $l_x$ for each variable $x$, since these appear in typing rules too. In addition, we have a set of metavariables $\alpha$ and for each a label $l_\alpha$. Finally, we include a label for our typing judgement $l_\vdash$.
A typing rule can then be understood as $\overline{s} \times s$, a list of trees for each premise, and a tree for the conclusion.
For example, we can represent the context and type of the first premise of the \CC{let} typing rule ($\Gamma, x : A \vdash A$)
with a tree:
\begin{align*}
    (l_{\vdash} \; (l_{\square, \square} \; l_\Gamma \; l_x \; l_A) \; l_A)
\end{align*}
With this representation, all one hole contexts are simply trees with a child missing; we will use $c$ to represent single-label steps of a one-hole context, and $C$ to represent a list of such steps.

\begin{align*}
    c ::= l \; \overline{s} \; \clasp{} \; \overline{s}
    \hspace{5em}
    C ::= \overline{c}
\end{align*}

\subsection{The Category of Diffs}
\label{sec:general-diffs}
Now that we have abstracted away any specific notion of types, contexts, or judgements into the single abstraction that a tree of labels provides, we can define diffs between arbitrary trees.
We denote these with $d$, generalizing both the type diffs $\delta$ and the context diffs $\Delta$.
There are only four constructors, in direct correspondence with the diffs presented in Section~\ref{sec:typed-editing}: the first leaves a top level label unchanged, and changes each child by a diff. The second and third constructors add or remove a single label step $c$, and then apply another diff inside the hole of $c$. The final constructor simply replaces one tree with another.

$$
d ::= l \; \overline {d}
    \pipe \plusdiff{c}{d}{}
    \pipe \minusdiff{c}{d}{}
    \pipe \replacediff{s}{s}
$$

\vspace{0.5em}
We can also generalize our judgement $\diffsto{s_1}{d}{s_2}$ to range over these general purpose diffs:

\begin{align*}
\inferrule{
    \forall i, \diffsto{s_i} {d_i} {s_i'}
}{
    \diffsto{l \; \overline{s_i}} {l \; \overline{d_i}} {l \; \overline{s_i'}}
}
\hspace{3em}
\inferrule{
    \diffsto{s}{d} {s'}
}{
    \diffsto{s} {\plusdiff{c}{d}{}} {c \ohchole{s'}}
}
\hspace{3em}
\inferrule{
    \diffsto{s}{d} {s'}
}{
    \diffsto{c \ohchole{s}}{\minusdiff{c}{d}{}}{s'}
}
\hspace{3em}
\inferrule{
}{
    \diffsto{s_1}{\replacediff{s_1}{s_2}}{s_2}
}
\end{align*}
These diffs admit an identity and a composition operator, which will be useful to define Pantograph's algorithm.
\vspace{-1em}

\paragraph{Endpoints}
If $\diffsto{s_1}{d}{s_2}$, then we define $d.1$ and $d.2$ as the endpoints $s_1$ and $s_2$ respectively.

\vspace{-1em}


\paragraph{Identity diffs}
Given any tree $s$, there is an identity diff $\diffsto{s}{id_s}{s}$ given by 

\vspace{-1em}

\begin{align*}
    id_{l \; s_1 ... s_n} &= l \; id_{s_1} ... id_{s_n}
\end{align*}

We will often write just $id$ when the tree is unambiguous.

\vspace{-1em}

\paragraph{Composition}
Given two diffs sharing an endpoint $\diffsto{t_1}{d_1}{t_2}$ and $\diffsto{t_2}{d_2}{t_3}$, we can compose them together: $\diffsto{t_1}{d_1 \circ d_2}{t_3}$.
Composition is defined recursively over the structure of the diffs.
Each diff represents a change to a tree, and their composition is a single diff which performs both changes. Therefore, the resulting diff generally contains all of the `+' and `-' constructors from the two inputs combined.
The exception is if we compose two diffs which add and then remove the same one-hole context --- then the two cancel out.
In the following definition, we write $\overline{d}$ for a list of diffs, and $\overline{d}.1$ (or $\overline{d}.2$) for the list of left (or right) endpoints.


\begin{align*}
    (l \; d_1 ~\dots~ d_n) \circ (l \; d_1' ~\dots~ d_n') & = l \; (d_1 \circ d_1') ~\dots~ (d_n \circ d_n')
    \\
    \plusdiff{c}{d}{} \circ \minusdiff{c}{d'}{} & = d \circ d'
    \\
    d \circ \plusdiff{c}{d'}{} & = \plusdiff{c}{d \circ d'}{}
    \\
    \minusdiff{c}{d}{} \circ d' & = \minusdiff{c}{d \circ d'}{}
    \\
    %
    \plusdiff{\overline{a}.1 ~}{d}{~ \overline{b}.1} \circ (l \; \overline{a} \; d' \; \overline{b}) & = \plusdiff{\overline{a}.2 ~}{d \circ d'}{~ \overline{b}.2}
    \\
    (l \; \overline{a} \; d' \; \overline{b}) \circ \minusdiff{\overline{a}.2 ~}{d}{~ \overline{b}.2} & = \minusdiff{\overline{a}.1 ~}{d \circ d'}{~ \overline{b}.1}
    \\
    d_1 \circ d_2 & = \replacediff{d_1.1}{d_2.2}
    \hspace{3em} \textrm{otherwise}
\end{align*}

Finally, the next two theorems state that identities and compositions behave as expected, forming a category where the objects are trees and the morphisms are diffs:

\begin{theorem}[Identity-Compose]
    $$id \circ d = d \circ id = d$$   
\end{theorem}

\begin{theorem} [Associativity of composition]

Given any three diffs
$s_1 \xrightarrow{d_1} s_2 \xrightarrow{d_2} s_3 \xrightarrow{d_3} s_4$,

$$(d_1 \circ d_2) \circ d_3 = d_1 \circ (d_2 \circ d_3)$$
\end{theorem}

The proofs of these two theorems proceed by induction over the size of the diffs and cases over the possible constructors that form the diffs. The full proofs can be found in Appendix~\ref{sec:diffappendix}.





\subsection{Typing One-Hole Contexts with Diffs}
\label{sec:diffsetup}

As we have shown, Pantograph performs automatic edits when the user inserts or removes a one-hole context.
Each automatic edit begins by placing diff boundaries around the location of the edit.
We associate each one-hole context with a diff that describes how it changes the type and context of the program where it is inserted.
%
%
For a one-hole context {\em step} $c$, we write $\Delta \vdash c : \delta$ to represent that from the inside to the outside, the context and type of $c$ change by diffs $\Delta$ and $\delta$. 
For most language constructs, the types and contexts inside and outside their one-hole context $c$ are either exactly the same (which gives rise to an identity diff), or unrelated (which gives rise to a replace diff). Only for a few constructs is the relationship between inside and outside types and contexts more involved: $\lambda$ abstractions, applications, and forms which introduce variables.

\begin{align*}
    \minusdiff{}{\Delta}{,x : A} &\vdash \fun{x}{A}{\clasp{}} : \plusdiff{A \to}{B}{}
    \\
    \Delta &\vdash \clasp{} \; t : \minusdiff{A \to}{B}{}
    \\
    \minusdiff{}{\Delta}{,x : T_1} &\vdash \llet{x}{T_1}{t}{\clasp{}} : T_2
    \\
    \minusdiff{}{\Delta}{,x : T_1} &\vdash \llet{x}{T_1}{\clasp{}}{t} : \replacediff{T_1}{T_2}
    \\
    \minusdiff{}{\minusdiff{}{\Delta}{, h : T}}{, t : \cList~T} & \vdash \matchTwo{a}{\cnil}{c}{\ccons~h~t}{\clasp{}} : \cList~T
\end{align*}

To lift diffs from one-hole context steps to entire one-hole contexts, we can simply compose 
the diffs along the individual steps:

\begin{align*}
    \inferrule{
    }{
        id \vdash \clasp{} : id 
    }
    \hspace{3em}
    \inferrule{
        \Delta_2 \vdash c : \delta_2
        \\
        \Delta_1 \vdash C : \delta_1
    }{
        \Delta_1 \circ \Delta_2 \vdash c[C] : \delta_1 \circ \delta_2
    }
\end{align*}

This allows us to get a diff for any one hole context which goes from the context and type inside to the context and type outside.
In order to get the diff going the other way from the outside to the inside, we define a function which flips a diff and swaps its endpoints:

\begin{align*}
    \metatheorystyle{flip}~(l \; d_1 ~ ~\dots~ ~ d_n) &= l \; (\metatheorystyle{flip}~d_1) ~ ~\dots~ ~ (\metatheorystyle{flip}~d_n) \\
    \metatheorystyle{flip}~\plusdiff{c}{d}{} & = \minusdiff{c}{\metatheorystyle{flip}~d}{} \\
    \metatheorystyle{flip}~\minusdiff{c}{d}{} & = \plusdiff{c}{\metatheorystyle{flip}~d}{} \\
    \metatheorystyle{flip}~(\replacediff{s_1}{s_2}) & = \replacediff{s_2}{s_1}
\end{align*}




When the user makes an edit, Pantograph sets up diff boundaries into the program around the location of the edit. The type boundaries introduced in Section~\ref{sec:simplemath} have unsurprising typing rules:

\begin{align*}
    \inferrule{
        \Delta.1 \vdash t : \delta.1
    }{
        \Delta.2 \vdash \down{\Delta \vdash \delta}{t} : \delta.2
    }
    \hspace{3em}
    \inferrule{
        \Delta.2 \vdash t : \delta.2
    }{
        \Delta.1 \vdash \up{\Delta \vdash \delta}{t} : \delta.1
    }
\end{align*}
There are three ways to make such an edit: inserting a one-hole context, removing one, or directly editing a type annotation.

\paragraph{Inserting a one-hole context}

If the context and type of $C'$ 
change by diffs $\Delta$ and $\delta$ such that $\Delta \vdash C' : \delta$, then
the user may insert it if the term at the cursor has context $\Delta.2$ and type $\delta.1$.
Pantograph then sets up the configuration $C[\up{id \vdash \delta}{C'[\down{(\metatheorystyle{flip}~\Delta) \vdash id}{t}]}]$.
For example, if the user adds a parameter to a map function as in Figure~\ref{fig:insert_parameter_to_map_bignot}, the system will set up the following configuration:

\begin{center}\begin{tabular}{c}\begin{lstlisting}
let map : List Bool $\to$ List Bool =
    $\{$ $\lambda f : \cBool \to \cBool .$
        $\{$ ... $\}_{\plusdiffs{}{id}{, f : \cBool \to \cBool} \vdash id}^\downarrow$
    $\}_{id \vdash \plusdiffs{(Bool \to Bool) \to}{List ~ Bool \to List ~ Bool}{}}^\uparrow$
in ...
\end{lstlisting}\end{tabular}\end{center}

\paragraph{Deleting a one-hole context}
Given a selection $C_1[C_2[t]]$, the user can delete $C_2$, as in Figure~\ref{fig:typed-editing-example-deleting-a-parameter}.
If $\Delta \vdash C_2 : \delta$, then
Pantograph will set up $C_1[\up{id \vdash (\metatheorystyle{flip}~\delta)}{\down{\Delta \vdash id}{t}}]$.

\paragraph{Editing a type annotation}
When the user edits the annotation of a $\lambda$ abstraction as in Figure~\ref{fig:typed-editing-example-insert-ho-parameter-not} by a type diff $\delta$, Pantograph sets up the configuration

\begin{align*}
\up{id \vdash \delta \to id}{\fun{x}{\delta.2}{\down{id, x : \delta \vdash id}{t}}}
\end{align*}

Similarly, when the user edits the annotation of a  \CC{let} expression as in Figure~\ref{fig:example-create-type-boundary} by a type diff $\delta$, Pantograph sets up the configuration
$$
\llet{x}{\delta.2}{\down{id, x : \delta \vdash \delta}{t_1}}{\down{id, x : \delta \vdash id}{t_2}}
$$

\subsection{Diff Propagation}
\label{sec:diffpropagation}

Recall the example in Section~\ref{sec:simplemath}, and in particular the steps 1-3 where a diff boundary passed `through' an application and a $\lambda$ abstraction. 
Making use of our understanding of syntax as trees, we can define the algorithm which performs these steps.

\begin{wrapfigure}[3]{r}{0.3\textwidth}
\begin{minipage}{0.3\textwidth}
\vspace*{-1em}
  \begin{small}
$$\inferrule{\Gamma \vdash A \to B\\
\Gamma \vdash A}{\Gamma \vdash B}
$$
\end{small}
\end{minipage}
\end{wrapfigure}

Recall the intrinsic typing rule for applications from earlier (on the right).
It has three metavariables: $A$, $B$, and $\Gamma$. For each application in the program, each of these metavariables is instantiated as some specific tree.
When we propagate a diff boundary through a construct of our language, the construct remains the same, but the values of the metavariables from the typing rule change.
Therefore, we introduce the notion of a \emph{diff substitution} $\sigma$,
a mapping from the metavariables in a given typing rule to diffs. Whenever a diff boundary is propagated through a form, our algorithm finds a  diff substitution on the metavariables in its typing rule which describes how the instantiations of the metavariables change.
We write $\sigma \; s$ to represent the substitution of the metavariables in $s$ by $\sigma$; any metavariable not in the domain of $\sigma$ is mapped to an identity diff.

Using this notion of a diff substiution, we can attempt to propagate diff boundaries through any language construct. Suppose that we have some language construct \CC{r} with $n$ premises $s_1 \dots s_n$ and conclusion $s$ which may all refer to a shared set of metavariables in its intrinsic typing rule:

\vspace*{-1em}
\begin{small}
\begin{align*}
    \inferrule{
        s_1 \dots s_n
    }{
        s
    }
\end{align*}
\end{small}

Then, in order to propagate a diff boundary down into such a form, we need only find a diff substitution $\sigma$, with domain of the metavariables in $s$, such that the incoming diff has the form $\sigma \; s$. We may then apply that same substitution to each $s_i$ to get the diff which must be propagated down into the $i$th child:

\vspace{-0.7em}
\begin{align*}
    \down{\sigma \; s}{r \; t_1 \dots t_n} 
    \step
    r \; \down{\sigma \; s_1}{t_1} \dots \down{\sigma \; s_n}{t_n}
\end{align*}

\noindent
We can write a similar rule for dealing with an upwards boundary into a form:

\begin{align*}
    r \; t_1 \dots \up{\sigma \; s_i}{t_i} \dots t_n
    \step
    \up{\sigma \; s}{r \;
        \down{\sigma \; s_1}{t_1}
        \dots t_i \dots
        \; \down{\sigma \; s_n}{t_n}}
\end{align*}

%
%


\noindent
Specialized to applications in particular, we get the following three propagation rules:

\begin{align*}
    \upa{\Delta \vdash \delta_1 \to \delta_2}{t_1} \; t_2
    \step
    \upa{\Delta \vdash \delta_2}{t_1 \; \downa{\Delta \vdash \delta_1}{t_2}}
    \\
    t_1 \; \upa{\Delta \vdash \delta}{t_2}
    \step
    \upa{\Delta \vdash id}{\downa{\Delta \vdash \delta \to id}{t_1} \; t_2}
    \\
    \downa{\Delta \vdash \delta}{t_1 \; t_2}
    \step
    \downa{\Delta \vdash id \to \delta}{t_1} \; \downa{\Delta \vdash id}{t_2}
\end{align*}

\noindent
The second of these application rules recovers the behavior from step (2) in Section~\ref{sec:simplemath}.

We can similarly specialize the two generic rules above to any form in Pantograph.
However, they are not quite sufficient in the case of \emph{nonlinearities}, or typing rules in which the same metavariable appears multiple times in a premise or the conclusion. For example, consider the intrinsic typing rule for a recursive \CC{let} construct:

\vspace*{-1em}
\begin{small}
\begin{align*}
\inferrule{
    \Gamma, x : A \vdash A
    \\
    \Gamma, x : A \vdash B
}{
    \Gamma \vdash B
}
\end{align*}
\end{small}

The metavariable $A$ appears twice in the first premise.
If one were to apply a substitution to that first premise $\sigma (\Gamma, x : A \vdash A)$, then both $A$s would necessarily map to the same diff.
However, we would like to have the following propagation rule for recursive \CC{let} constructs, in which a type diff propagates up from the definition, causing a context diff to be sent back down.

\begin{align*}
\llet{x}{A.1}{\upa{\Delta, x:id \vdash \delta}{t_1}}{t_2}
\step
\upa{\Delta \vdash id}{\llet{x}{\delta.2}{\downa{id, x:\delta \vdash id}{t_1}}{\downa{\Delta, x:\delta \vdash id}{t_2}}}
\end{align*}

Our propagation rule from before won't help, because there is no substitution of metavariables such that $\Gamma, x : A \vdash A$ can be made equal to $\Delta, x : id \vdash \delta$, since $id \ne \delta$.
The solution will be to focus onto only the part of the incoming diff which is not the identity.
In particular, we can decompose the diff into a one-hole context consisting only of identity diffs, and a non-identity diff:

\begin{align*}
\Delta, x : id \vdash \delta = (\Delta, x : id \vdash \clasp{})[\delta]
\end{align*}

We can now finally write the propagation rules in their fully general form. Any language construct \CC{r} with a typing rule of the form above
%
gives rise to the following propagation rules (where $dom(\sigma)$ is the domain of $\sigma$, and $FV(s)$ is the set of metavariables in $s$):
$$
\inferrule{
s = c[s']
\\
dom(\sigma) = FV(s')
}{
\down{(\sigma_{id} \; C)[\sigma \; s']}{r \; t_1 ~\dots~ t_n}
    \;\; \leadsto  \;\;
\up{(\sigma \; C)[\sigma_{id} \; s']}{r \; \down{\sigma \; s_1}{t_1} ~\dots~ \down{\sigma \; s_n}{t_n}}
}
\textrm{{\sc Propagate}}\downarrow
$$

$$
\inferrule{
s_i = C[s_i']
\\
dom(\sigma) = FV(s_i')
}{
r \; t_1 ... \up{(\sigma_{id} \; C)[\sigma\; s_i']}{t_i} ... t_n
    \;\; \leadsto  \;\;
\up{\sigma \; s}{r \; \down{\sigma \; s_1}{t_1} ~\dots~ \down{(\sigma \; C)[\sigma_{id} \; s_i']}{t_i} ~\dots~ \down{\sigma \; s_n}{t_n}}
}
\textrm{{\sc Propagate}}\uparrow
$$

\noindent
This final version of the propagation rules is capable of propagating diff boundaries through every form in every situation that arises in Pantograph, even in the presence of nonlinearities.

\subsection{Pantograph's Automatic Edits}
\label{sec:alterationanddiffrules}

Once diff boundaries are placed after an edit, Pantograph propagates the diff boundaries through the program and alters the program according to a small-step operational semantics, which we describe in this section. We also reproduce all of the rules in a compact table in Appendix~\ref{sec:rule-small-app}.



\paragraph{Propagation rules} The first two rules are the {\sc Propagate} $\downarrow$ and {\sc Propagate} $\uparrow$ rules from the previous section which propagate diff boundaries through every form in our language.
The only exception is variables, whose propagation rules we list below \footnote
{In our implementation, using a typed representation of de Bruijn indices~\cite{deBruijn} the two variable propagation rules are also actually derived from the generic propagation rules as well.}
:

\vspace{0.4em}
\begin{tabular}{R C L R }
    \downa{C \ohchole{ \Delta ,~ x : \delta } \vdash id}{x}
    & \step &
    \upa{id \vdash \delta}{x}
    & \textrm{\sc{Propagate-Var}} \downarrow \!\! 1
    \\
    \downa{C \ohchole{ \Delta ,~ x : id } \vdash \delta}{x}
    & \step & 
    \upa{C \ohchole{ \Delta ,~ x : \delta } \vdash id}{x}
    & \textrm{\sc{Propagate-Var}} \downarrow \!\! 2
\end{tabular}

\paragraph{Alteration rules}
Next, Pantograph has a set of rules which enable the automatic edits shown by example in Section~\ref{sec:typed-editing}.
First, Pantograph can automatically insert and remove $\lambda$ abstractions (Figure~\ref{fig:typed-editing-example-insert-ho-parameter-not}).
The following three rules enable this behavior:

\vspace{0.4em}
\begin{tabular}{R C L R L}
\downa{\Delta \vdash \plusdiffs{A \to}{\delta}{}}{t}
& \step &
\fun{x}{A}{\downa{\plusdiffs{}{\Delta}{,~ x : A} \vdash \delta}{t}}
& \textrm{\sc{Insert-Abs}} & \downarrow
\\
\downa{\Delta \vdash \minusdiffs{A \to}{\delta}{}}{\fun{x}{A}{t}}
& \step &
\downa{\minusdiffs{}{\Delta}{,~ x : A} \vdash \delta}{t}
& \textrm{\sc{Delete-Abs}} & \downarrow
\\
\fun{x}{A}{\upa{\Delta, x : id \vdash \plusdiffs{A \to}{\delta}{}}{t}}
& \step &
\upa{\Delta \vdash A \to \delta}{\downa{\minusdiffs{}{\Delta}{,x : A} \vdash id}{t}}
& \textrm{\sc{Delete-Abs}} & \uparrow
\\[0.5em]
\end{tabular}

\vspace*{0.5em}
\noindent
Pantograph can also automatically insert applications, or replace them with commented ones as the type of a function changes (Figure~\ref{fig:typed-editing-example-insert-ho-parameter-not}). 
This is governed by the following rules:

\begin{tabular}{R C L R L}
\upa{\Delta \vdash \plusdiffs{A \to}{\delta}{}}{t}
& \step &
\upa{\Delta \vdash \delta}{t \; \holety{A}}
& \textrm{\sc{Insert-App}} & \uparrow
\\
\upa{\Delta \vdash \minusdiffs{A \to}{\delta}{}}{t_1} \; t_2
& \step &
\upa{\Delta \vdash \delta}{\ghostapp{t_1}{\; \downa{\Delta \vdash id}{t_2}}}
& \textrm{\sc{Displace-App}} & \uparrow
\\
\downa{\Delta \vdash \plusdiffs{A \to}{\delta}{}}{t_1 \; \holety{A}}
& \step &
\downa{\Delta \vdash id \to \delta}{t_1}
& \textrm{\sc{Delete-App}} & \downarrow
\\[0.5em]
\end{tabular}

\vspace*{0.5em}
\noindent
Finally, Pantograph will automatically replace bound variables with free variables when necessary and vice versa (Figure~\ref{fig:typed-editing-example-deleting-a-parameter}), according to the following rules:

\begin{tabular}{R C L R L}
%
\downa{C \ohchole{\minusdiffs{}{\Delta}{,~ x : T}} \vdash id}{x}
& \step &
\ghostvar{x}{T}
& \textrm{\sc{Local-To-Free}} &
\\[0.5em]
%
\downa{C \ohchole{ \plusdiffs{}{\Delta}{,~ x : T} } \vdash T}{\ghostvar{x}{T}}
& \step &
x
& \textrm{\sc{Free-To-Local}} &
\end{tabular}

All together, these eight rules describe how Pantograph automatically changes a program in response to a user's edit. The alteration rules take precedence over the propagation rules when both can apply. Additionally, {\sc Delete-App $\downarrow$} takes precedence over {\sc Insert-Abs $\downarrow$}.

\paragraph{Diff Boundary Rules}

Finally, Pantograph has eight rules used to deal with special cases involving diff boundaries themselves.
First, two rules eliminate diff boundaries containing only identity diffs.

\vspace{0.5em}
\begin{tabular}{R C L R L}
    \downa{id \vdash id}{t}
    & \step &
    t
    & \textrm{\sc{Identity}} & \downarrow
    \\
    \upa{id \vdash id}{t}
    & \step &
    t
    & \textrm{\sc{Identity}} & \uparrow
\end{tabular}

\vspace*{0.5em}
\noindent
Next, two rules enable diff boundaries to cross each other when the meet, so long as they don't both modify the same side of the typing judgement:

\vspace{.5em}
\begin{tabular}{R C L R L}
    %
    %
    \downa{\Delta \vdash id}{\upa{id \vdash \delta}{t}}
    & \step &
    \upa{id \vdash \delta}{\downa{\Delta \vdash id}{t}}
    & \textrm{\sc Interchange} & 1
    \\
    \downa{id \vdash T}{\upa{\Delta \vdash id}{t}}
    & \step &
    \upa{\Delta \vdash id}{\downa{id \vdash T}{t}}
    & \textrm{\sc Interchange} & 2 
    \\[1em]
\end{tabular}

\noindent
Two more rules stop type diffs from propagating in or out of function applications. For example, in Figure~\ref{fig:example-create-type-boundary}, a type diff is propagated up from the \CC{f} variable. The following two rules stop it around the function application, keeping the effect of the edit local to the definition it was made on and giving the user the choice of how to proceed.
We define a \emph{neutral form} as either a variable, a built-in function (such as \CC{cons}), or a neutral form applied to an argument.
These rules take precedence over the {\sc Propagate} but not the alteration rules.

\vspace{-.5em}
\begin{tabular}{R C L R L}
    \\
    \downa{\Delta \vdash \delta}{t}
    & \step &
    \multirow{2}{*}{
        \ensuremath{\begin{rcases*}
            \errorboundary{\delta.1}{\delta.2}{\downa{\Delta \vdash id}{t}} \\
            C \ohchole{\upa{\Delta \vdash id}{\errorboundary{\delta.2}{\delta.1}{t}}}
        \end{rcases*}}
        \begin{tabular}{l}
            if $t$ is a neutral form, \\
            and $C \ne \clasp{} \; t_2$
        \end{tabular}
    }
    & \textrm{\sc Neutral-Error} & \downarrow
    \\
    C \ohchole{\upa{\Delta \vdash \delta}{t}}
    & \step &
    & \textrm{\sc Neutral-Error} & \uparrow
    \\[1.3em]
\end{tabular}

\vspace*{0.5em}
\noindent
Finally, if no other rules apply, diff boundaries are converted into type error boundaries in the program:

\begin{tabular}{R C L R L}
    %
    %
    \downa{\Delta \vdash \delta}{t}
    & \step &
    \multirow{2}{*}{
        \ensuremath{\begin{rcases*}
            \errorboundary{\delta.1}{\delta.2}{\down{\Delta \vdash id}{t}} \\
            \up{\Delta \vdash id}{\errorboundary{\delta.2}{\delta.1}{t}}
        \end{rcases*}}
        \begin{tabular}{l}
            if no other rules apply \\
            and $\delta \ne id$
        \end{tabular}
    }
& \textrm{\sc Fallthrough-Error} & \downarrow
\\
\upa{\Delta \vdash \delta}{t}
& \step &
& \textrm{\sc Fallthrough-Error} & \uparrow
\\\\
\end{tabular}




%% file: proofs.tex
\subsection{Metatheory of the Propagation Rules}
\label{sec:proofs}

The system we have presented in this paper is quite intricate. In this section, we will state and sketch the proofs of the correctness properties of our automatic edit system --- full proofs can be found in Appendix~\ref{app:metatheory}.

\begin{theorem} [Progress]
\label{thm:progress}
If $t$ resulting from an edit has a diff boundary (other than an up boundary at the top), then for some $t'$, $t \step t'$
\end{theorem}

If at the end of propagation the program has the form $\up{\Delta \vdash \delta}{t}$, we prove an invariant on the contexts in up boundaries that shows that $\Delta = id$. In this case, Pantograph removes the boundary and the program has a new type.

Otherwise, given a boundary with diff $\Delta \vdash \delta$, if $\delta \ne id$, then by definition either the {\sc Fallthrough-Error} rule or some other rule applies. On the other hand, if $\delta = id$, then we must show that a {\sc Propagate} rule always applies. For most of the typing rules in Figure~\ref{fig:typing-rules}, this is straightforward, as the context in most conclusions and premises is a single metavariable $\Gamma$ with which any diff can unify. However, a few typing rules have a non-trivial context, such as that of a $\lambda$ abstraction. In the full proof in Appendix~\ref{app:progress} we use our invariant
to deal with these cases.

\begin{theorem} [Type Preservation]
\label{thm:preservation}
For any $t \leadsto t'$, if $\; \Gamma \vdash t : T$ then $\Gamma \vdash t' : T$
\end{theorem}

All of the diff boundary rules preserve the type of the term, as well as the type of sub-terms.
The proof proceeds by case analysis over the small-step propagation rules. Full details can be found in Appendix~\ref{app:preservation}.


\begin{theorem} [Termination]
\label{thm:terminationtheorem}
For any program with diff boundaries resulting from an edit in Pantograph, there is no infinite sequence of step rules that can be applied.
\end{theorem}

Proving termination of the boundary propagation is more involved. Taking a look at the rules above, we can observe that up boundaries can turn into down boundaries, but with the exception of the {\sc Variable}$*$ rules, no down boundary can ever turn into an up boundary.
Also, with the exception of the {\sc Insert}-$*$ rules, the upwards boundaries make progress to the top of the program and the downwards boundaries make progress to the leaves.
Therefore, the path of a boundary through the program will be to first go up, and then go down, and then disappear. 
We formalize this intuition in Appendix~\ref{app:termination}, providing a decreasing metric over terms with diff boundaries.

\paragraph{Nondeterminism}

In most cases, there is only one rule that can apply to a given diff boundary. However, the order in which multiple boundaries are propagated is left unspecified, and in some situations it is not obvious that the order does not matter. For example, the following situation arises during diff propagation at the second call to \CC{f} in Figure~\ref{fig:typed-editing-example-insert-ho-parameter-not}: 


$$
\down{\plusdiffs{\cInt \to}{\cBool \to \cBool}{}}{\up{\plusdiffs{\cInt \to}{\cBool \to \cBool}{}}{f}}
$$

\vspace*{0.5em}
\noindent
Either the down boundary can be stepped first, applying {\sc Insert-Abs} $\downarrow$ and then {\sc Delete-Abs} $\uparrow$, or the up boundary can be stepped first, applying {\sc Insert-App} $\uparrow$ and then {\sc Delete-App} $\downarrow$. But either way, the result is the same term \CC{f}.

More generally, we have proven the following confluence result.
Using $t_1 \leadsto^* t_2$ to mean that $t_1$ steps to $t_2$ after zero or more steps,

\begin{theorem} [Confluence]
\label{thm:confluence}
For any program with diff boundaries $t$ resulting from an edit in Pantograph,
if $t \leadsto^* t_1$ and $t \leadsto^* t_2$, then there exists $t'$ such that $t_1 \leadsto^* t'$ and $t_2 \leadsto^* t'$
\end{theorem}

A proof of confluence is given in Appendix~\ref{app:confluence}.
The core idea of the confluence proof is to consider
all of the pairs of rules which could step the same term, and show that the same term can be reached regardless of which rule was chosen first.
The difficult cases are those like the example above, where an up diff boundary is inside a down diff boundary, because in these cases two different rules can alter the same part of the program.
However, such situations are rare.
We again observe that up boundaries can only turn to down boundaries at the {\sc Propagate-Variable} rules. Therefore, with an exception in neutral forms, the property that no up boundary is a descendant of a down boundary is preserved by the rules.
In the full proof, we formalize this intuition by proving two invariants that are preserved by all of Pantograph's rules.
These invariants reduce the number of possible cases to only eight, all of which we discuss in the full proof.


%% file: userstudy-v2.tex
\section{User Study}
\label{sec:userstudy}

To convincingly establish the effectiveness of a particular editor, one would need to undertake a thorough empirical comparison of its effectiveness compared to other editors---a herculean task way beyond the scope of this work.
Instead of comparing our editor to other state-of-the-art structure editors (such as Hazel or MPS),
in this section we empirically investigate the feasibility of our approach---whether programmers can learn to think and program in the paradigm presented in this paper.
Furthermore, we investigate qualitatively which aspects of Pantograph worked well or badly, and compare and contrast how participants use Pantograph and the text editor.
We describe a user study in which participants were first given a short introductory tutorial on using Pantograph, before being asked to complete a series of basic programming tasks using either Pantograph or a text editor.

\paragraph{Participants}

We sought participants with prior experience with strongly-typed and functional programming languages, so
we recruited 13 students (11 male, 1 female, 1 unstated; ages 19-25) from CMSC 433, an upper level functional programming course at the University of Maryland, and offered participants a \$30 compensation.
Recruitment materials included a link to Pantograph, and one participant used it recreationally before the study.

\paragraph{Editors and Tasks}

We designed a version of Pantograph \footnote{Available \href{https://pantographeditor.github.io/pantograph-2024-user-study-tutorial/}{here}.} and a text editing environment, both using a simple ML-style programming language, for the study.
The text editor was based on Monaco~\cite{monaco}, the library behind the popular editor Visual Studio Code~\cite{vscode}. We augmented Monaco with syntax highlighting, automatic type-checking, and inline typing diagnostics.
Further, we designed software to conduct the study, available
\href{https://pantographeditor.github.io/pantograph-2024-user-study/}{here}.

Each participant solved the same sequence of 12 tasks (T1-T12). The tasks were designed in six pairs and each participant solved one task from each pair in each editor.
Four tasks (T1, T2, T7, T8) were taken from a prior study on Tylr~\cite{tylr}, asking participants to transcribe a given program and then edit it into a second given one. 
Another pair of tasks (T5, T11) was inspired by a prior study of MPS~\cite{mpsstudy}, in which participants were instructed to fix definitions of a common mathematical law given a correct version to copy.
Because we also wanted to test each participant's ability to program using our typed editing paradigm rather than merely copy given code, we designed six programming tasks: two tasks to implement a function using a given fold function (T3, T9), two tasks to generalize a given function implementation in a specified way (T4, T10), and two tasks to implement a simple recursive function over a list (T6, T12).


\paragraph{Procedure}

The participants were given a 1-hour tutorial on Pantograph consisting of interactive examples and exercise which explained typed tree editing, zipper editing and selections, and type boundaries.
During the tutorial, we actively answered questions and assisted participants.
Additionally, each participant was given 5 minutes to familiarize themselves with the text editor.

After the tutorial, the participants were randomly divided into two groups. 
Group A solved the first six tasks in Pantograph and the last six tasks in the text editor, while group B used the editors in the reverse order.
While solving tasks, participants were allowed to ask questions about the language or editor but not about how to solve the task.
When a participant's code passed a task's test suite, the software indicated that they could continue to the next task.
Participants were informed that they should skip any task that they thought they couldn't finish or was taking them too long.

\subsection{Quantitative Analysis}

The user study yielded roughly 10 hours of screen recordings.
In all, 156 tasks were attempted by the participants.
Two participants (one in Group A and one in Group B) skipped at least six tasks, including all of the programming tasks in both editors---we chose to omit these two participants' data from the following analysis, although we discuss this issue further in the limitations section.
Further, we discarded the data points (4 in Pantograph and 2 in text) where our test suite mistakenly indicated to a participant that their solution was correct when it was actually incorrect.
This left us with 124 data points on tasks, 121 of which were solved and 3 of which were skipped.

\begin{figure}
    \centering
    \includegraphics[width=\textwidth]{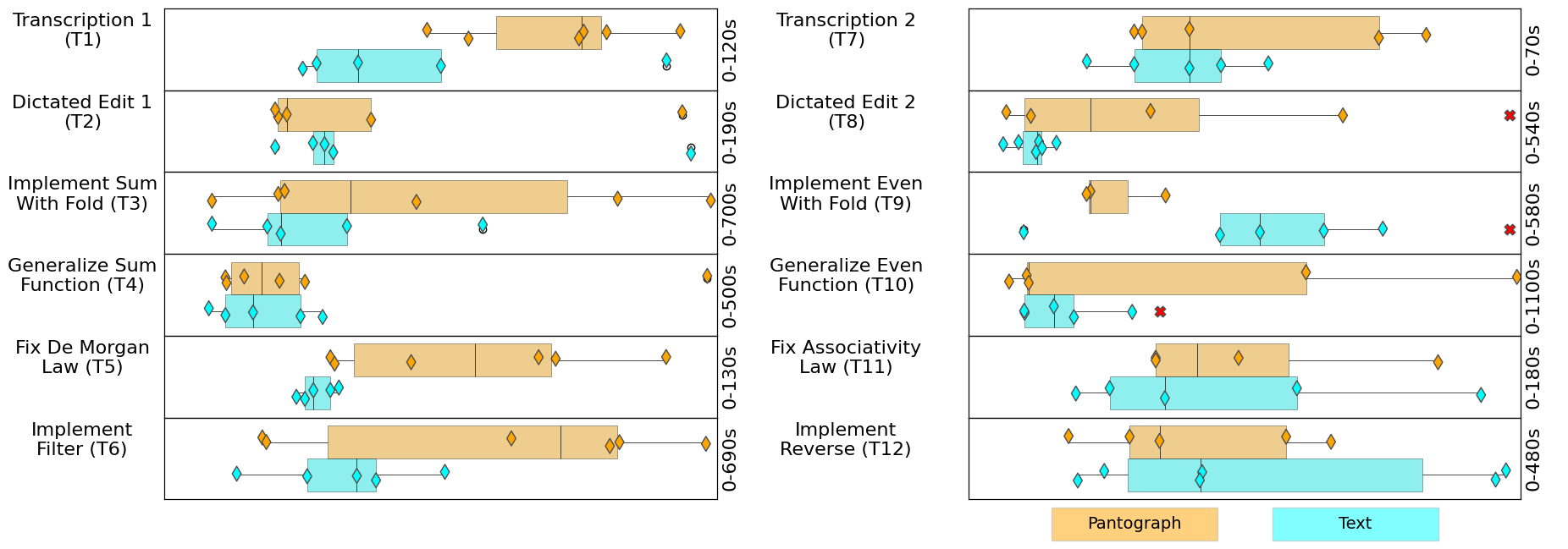}
    \caption{Task durations (in seconds) split by editor. For each task, Pantograph times are on the top in orange, and text times are on the bottom in blue. Tasks that participants chose to skip are marked as a red `X'. }
    \label{fig:alldata}
\end{figure}

\begin{figure}
    \centering
    \includegraphics[width=0.5\linewidth]{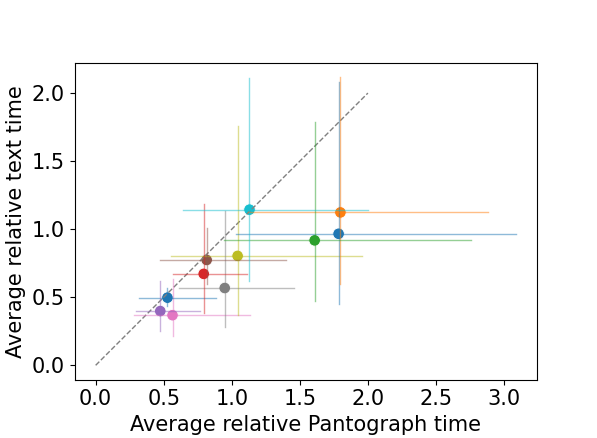}
    \caption{
    %
    %
    %
    %
    %
    %
    %
    Geometric mean and variance of relative task times for each participant (denoted as a dot), and for each editor. The dashed line shows parity between the editors --- participants below the line solved tasks slower in Pantograph then the text editor. One standard deviation is shown as horizontal and vertical lines.
    }
    \label{fig:duration-ratios}
\end{figure}


Figure~\ref{fig:alldata} shows the duration of each task for each editor.
Overall, Pantograph was slower than the text editor for most tasks.
It was $1.4\times$ slower on average overall, measured as the geometric mean of the ratio between the average duration of each editor on each task.

In order to view these task durations by participant, we need a way to compare times across different tasks. We define a participant's \emph{relative task time} as the ratio of their time to the average time taken on that task.
We plotted the average relative task time in each editor for each participant in Figure~\ref{fig:duration-ratios}.
To account for skipped tasks in this calculation, we used the maximum time taken for that task by any participant.
Some of the participants performed similarly in both editors, while some performed significantly more slowly in Pantograph than the text editor.
Overall, most of the participants in the study were able to effectively solve the tasks using Pantograph.

\input{userstudy-qa}

 \subsection{Limitations}
As discussed above, no other structure editor was used in this study, so our results offer little insight into how our system compares to other structure editors.
Additionally, while the tasks in our study were intended to represent realistic programming scenarios, the results may not generalize to real world programming in a fully developed system.
Finally, the participants had no experience with Pantograph before the study (with one exception) but had an hour to practice with it during the study; at the same time, they had plenty of prior experience using a text editor for similar languages, but only five minutes to practice with it during the study.
It is not clear how the results might change if participants had the same amount of practice in both editors.

Unlike prior studies on structure editors such as~\citet{tylr, teentylr, mpsstudy} which focus on transcription tasks, we also included programming tasks.
Although we intended the tasks to be easy enough to complete in the time allotted for the study, two participants were still not able to complete any of these coding tasks in either editor, getting stuck on an early problem and then quickly skipping later problems.
As we had not planned task time limits beforehand we decided on our problem skipping protocol during the study, and it is possible that this design flaw influenced the participant's decisions to skip problems.

Finally, there was a large potential for bias and non-reproducibility:
we facilitated the user study in person, and the participants were aware that the we created Pantograph, which has been shown to cause bias \cite{biasstudy}.
Additionally, the Pantograph tutorial involved us actively teaching participants in a one-on-one manner, which could lead to different results with different teachers.
Our decision to allow participants to ask questions about the editor and language while solving the tasks may have additionally affected the results of the study, as our decisions in answering these questions were subjective.

Taking into account all of these limitations, we re-emphasize that our goal was to demonstrate our approach's feasibility rather than its effectiveness relative to state of the art structure editors
and explore the advantages and disadvantages of Pantograph compared to a traditional
text editor.
Through this user study, we learned that students were able to learn to use Pantograph relatively effectively in a short amount of time, but we also discovered many opportunities for improvement 
in our ongoing efforts to develop an intrinsically typed structure editor.
We leave a thorough empirical evaluation of different structure editors for future work.



%% file: userstudy-qa.tex
\subsection{Qualitative Analysis}


While the quantitative data demonstrate that the participants were generally able to solve tasks in Pantograph, albeit somewhat slower than in the text editor, a qualitative analysis is necessary to answer more specific questions. In order to discover which aspects of Pantograph helped or hindered the participants, and to compare the way that participants solved problems in the two editors, we performed a reflexive thematic analysis following the six step process described in \citet{reflexivethematicanalysis}.

First, we coded the screen recordings, over several viewings, by noting common behavioral patterns as participants solved tasks.
While in many cases we were unable to interpret what a participant may have been thinking from a screen recording, we focused on patterns that could be identified unambiguously.
Next, we organized the codes into themes.
Finally, we developed these themes into the final analysis that is the rest of this section.
We emphasize that reflexive thematic analysis is inherently subjective \cite{reflectingreflexive}, particularly in our choices of which codes and themes to focus on.


\paragraph{Attempts to use Pantograph's interface like a text editor}
There was a wide variety in participants' understanding of Pantograph's basic interface.
One common theme that emerged in some participants were attempted actions with Pantograph that would make sense in a text editor but do not work in Pantograph's interface. Two codes in particular seemed to unambiguously show this principle.
The first is that some participants tried to input an entire expression consisting of multiple forms into a query such as \CC{1 + 1}, despite that in Pantograph forms can only be input one at a time and so these queries are not valid.
The second is that some participants, in a situation where the cursor was around a term that they intended to apply to an argument, attempted to immediately query the argument without first inserting an application form.
These particular patterns only occurred for the 4 participants who had the longest relative Pantograph times, and one participant who skipped all of the coding problems.

Conversely, one code represented an understanding that Pantograph should not be used like a text editor. Four participants opted to skip writing type annotations while transcribing programs in T1 and T7, and instead allow Pantograph to infer them automatically --- a behavior with no analog in text editing. These were the 4 participants with the fastest relative Pantograph times.

\paragraph{Diff propagation working well} 
Two tasks (T4 and T10) had a simple intended solution in Pantograph which would make use of diff propagation to add an argument to a function.
All but two participants made use of the diff propagation to solve the problem - one manually rewrote a function call instead, and the other skipped the problem.
Additionally, while no other tasks necessarily required diff propagation, most participants made use of it to at some point to fix a mistake.

\paragraph{Diff propagation getting in the way}
On the other hand, there were two common patterns in which diff propagation obstructed a participant from programming in the way they wanted to. These were not characteristic of only participants who had longer relative Pantograph times, but rather there was no clear correlation - we believe these to represent flaws with Pantograph.
Most of these cases involved creation of type boundaries.

The first pattern was that participants accidentally changed the type of a variable by making an edit to one of its arguments.
We can be certain that these alterations were unintentional because they occurred in problems where a type annotation was given and should not be changed.
Most participants (7 of 13) had this problem at some point.
Typically the edits were locally sensible, but the edit unintentionally caused a type diff to propagate to a variable definition.

The second pattern of confusion arose when a participant wanted to insert the child of a form at their cursor \textit{before} inserting the parent form.
In some cases Pantograph supports with, like wrapping $\clasp{} + \hole$ around $1$.
But in particular on task T12, participants needed to input lambda abstractions surrounding a match expression. 
A few participants decided to first input the match before the lambda abstractions (1 in text and 3 in Pantograph), even though directly inputting the lambda abstractions afterwards alters the type of the surrounding function by diff propagation.

\paragraph{Syntax Errors}
In both editors, participants tried to use invalid syntax.
In the text editor the invalid syntax was marked by the editor with a syntax error, and in Pantograph the query is invalid and cannot be inserted into the program in the first place.
Syntax errors were much more common in the text editor --- all but 1 participant input syntax errors in text at some point, while only 3 participants ever input invalid syntax into a Pantograph query, excepting the earlier mentioned pattern of inputting multiple forms in one query. 
One benefit of a structure editor like Pantograph is that it requires less memorization of syntax, although it is possible that more practice with the text editor could have prevented many of the syntax errors made in text.

\paragraph{Type errors}
Again in both editors, participants tried to input ill-typed forms---in the text editor this leads to a type error after the fact, while in Pantograph either a query is not allowed, or a type boundary appears elsewhere in the program.
Type errors appeared in 24 attempted tasks in the text editor, while participants attempted to input ill-typed queries in 14 attempted tasks in Pantograph.
Visible types on holes in Pantograph may have helped participants avoid trying to directly input an ill-typed term more often than in the text editor, although as mentioned above there were also many situations where type boundaries appeared due to diff propagation, not counted here as ill-typed queries.


\paragraph{Selections} Two tasks (T2 and T11) had a simple intended solution involving a selection.
Most participants used selections, except a few who rewrote the expressions entirely (1 time in Pantograph and 6 times in text).
Additionally, almost all edits made using text selections (excepting small selections in a single word) corresponded to structured edits: 
5 selection based text edits were directly equivalent to a zipper edit with a one-hole context; 
7 were not quite equivalent to a zipper edit but were immediately followed by a small additional edit that made it equivalent to one;
5 selection based text edits amounted to swapping two entire terms;
and 2 more amounted to moving a term. On the other hand, only one selection based text edit did not correspond to a structured edit - a participant wrote part of an expression, and then deleted it with a selection.
%
Finally, in Pantograph, participants were often unsure about how to make a selection, as evidenced by behavior making several experimental selections before deciding what to do next---which occurred 15 times during tasks attempted in Pantograph, compared to only 6 times in those attempted in the text editor.

%% file: limitations.tex
\section{Discussion}
\label{sec:limitations}

\paragraph{Limitations of Zipper Editing}
In Section~\ref{sec:zipper-editing}, we showed that in our simply-typed ML-like language, most common edits take the form of adding or removing a one-hole context in the program. However, in some other settings this form of edit doesn't turn out to be as useful. For example, consider the following edit to an imperative c-style program:

\vspace*{-1em}
\begin{figure}[H]
    \centering
    \begin{tikzpicture}[scale=0.9]
        \node[draw,anchor=west] (step0) at (0,0) {\lstinline[language=java]
            |$s_1$; $\text{\textselection{while(c)\{}}$ $s_2$; $s_3$; $\text{\textselection{\}}}$ $s_4$; $s_4$;|};
        \node[draw,anchor=west] (step1) at ($(step0.east) + (1,0)$) {\lstinline[language=java]
            |$s_1$; $s_2$; $s_3$; $s_4$; $s_4$;|};
        \node[draw,anchor=west] (step2) at ($(step1.east) + (1,0)$) {\lstinline[language=java]
            |$s_1$; $s_2$; $\text{\textselection{while(c)\{}}$ $s_3$; $s_4$; $\text{\textselection{\}}}$ $s_4$;|};
        \draw[->,thick] (step0.east) to node[above,yshift=1em]{cut selection?} (step1.west);
        \draw[->,thick] (step1.east) to node[above,yshift=1em]{paste selection?} (step2.west);
    \end{tikzpicture}
\end{figure}


Visually, this sequence of edits appears like a zipper cut and paste. However, it doesn't seem possible to understand it mathematically as a zipper selection. If `$;$' is a binary operator, then we may be able to understand the initial cut as the deletion of a one-hole context. However, this would leave the `$;$' operators associated in the wrong way for the paste to be possible.

This situation is not limited to imperative programming. For instance, consider a similar example that might occur in a markup language like HTML:


\vspace*{-1em}
\begin{figure}[H]
    \centering
    \begin{tikzpicture}[scale=0.9]
        \node[draw,anchor=west] (step0) at (0,0) {
        \lstinline[language=xml]
            !<$a$>$\text{\textselection{<div>}}$<$b$><$c$>$\text{\textselection{</div>}}$<$d$><$e$>!};
        \node[draw,anchor=west] (step1) at ($(step0.east) + (1,0)$) {\lstinline[language=xml]
            !<$a$><$b$><$c$><$d$><$e$>!};
        \node[draw,anchor=west] (step2) at ($(step1.east) + (1,0)$) {\lstinline[language=xml]
            !<$a$><$b$>$\text{\textselection{<div>}}$<$c$><$d$>$\text{\textselection{</div>}}$<$e$>!};
        \draw[->,thick] (step0.east) to node[above,yshift=1em]{cut selection?} (step1.west);
        \draw[->,thick] (step1.east) to node[above,yshift=1em]{paste selection?} (step2.west);
    \end{tikzpicture}
\end{figure}


\paragraph{Limitations of Typed Editing}
Our goal in creating Pantograph was to build an editor which only allows edits which are locally well-typed, and yet the user doesn't need to edit typing annotations. Instead, the user may edit terms in locally sensible ways, and the corresponding annotations are automatically updated. However, we found that certain trade-offs are required.

For example, consider the following program in which the user inserts a lambda at a hole. The hole is already at a function type, so there are two options for what might happen: the lambda can either fill the hole, therefore leaving the type of the term unchanged. Or, the lambda can go around the hole, thus adding another argument to the type of the term.

\begin{figure}[H]
    \centering
    \begin{tikzpicture}[scale=0.9]
        \node[draw,anchor=west] (step0) at (0,0) {\includepantographgraphics[0.22]{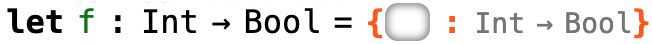}};
        \node[draw,anchor=west] (step10) at (6.7,0.6) {\includepantographgraphics[0.22]{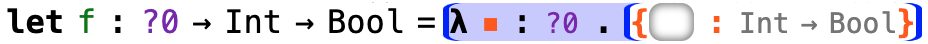}};
        \node[draw,anchor=west] (step11) at (6.7,-0.6) {\includepantographgraphics[0.22]{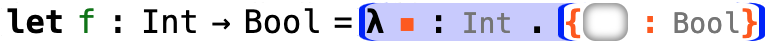}};
        \draw[-{[flex,sep]>},thick]
            (step0.east)
            .. controls ($(step0.east) + (0.8,0)$) and ($(step10.west) + (-0.8,0)$) ..
            node[left,yshift=1em]{outside}
            (step10.west);
        \draw[-{[flex,sep]>},thick]
            (step0.east)
            .. controls ($(step0.east) + (0.8,0)$) and ($(step11.west) + (-0.8,0)$) ..
            node[left,yshift=-1em]{inside}
            (step11.west);
    \end{tikzpicture}
\end{figure}

As another example, consider the following program in which the user inserts an application around $x$. Here, $x$ is already a function so there are two possible outcomes.
Either
$x$ gets a new argument, leaving the type of $g$ unchanged.
Or,
$x$ is applied to its existing argument, changing the type of $g$.

\begin{figure}[H]
    \centering
    \begin{tikzpicture}[scale=0.9]
        \node[draw,anchor=west] (step0) at (0,0) {\includepantographgraphics[0.23]{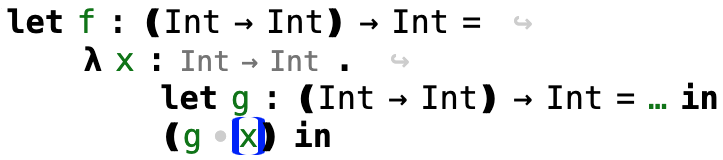}};
        \node[draw,anchor=west] (step10) at (7.5,1.2) {\includepantographgraphics[0.23]{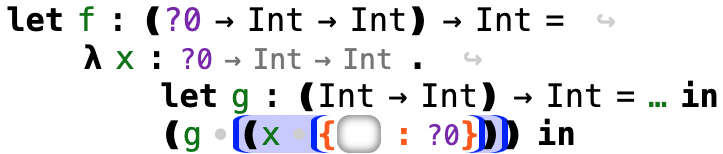}};
        \node[draw,anchor=west] (step11) at (7.5,-1.2) {\includepantographgraphics[0.23]{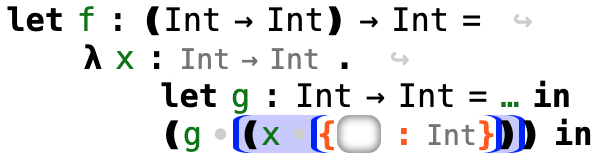}};
        \draw[-{[flex,sep]>},thick]
            (step0.east)
            .. controls ($(step0.east) + (0.8,0)$) and ($(step10.west) + (-0.8,0)$) ..
            node[left,yshift=2em]{update \CC{x} and \CC{f}}
            (step10.west);
        \draw[-{[flex,sep]>},thick]
            (step0.east)
            .. controls ($(step0.east) + (0.8,0)$) and ($(step11.west) + (-0.8,0)$) ..
            node[left,yshift=-2em]{update \CC{g}}
            (step11.west);
    \end{tikzpicture}
\end{figure}

Of course, many less rigid systems exist which re-typecheck the program after an edit. Our typed refactoring system could be integrated with such tools while retaining many of its benefits.
%
Instead, we opted to explore a more opinionated point in the design space of editors.
In intrinsically typed editors,
as the above examples show,
either the programmer must sometimes be required to edit the type annotations directly, or the programmer must sometimes be required to input extra information with an edit which can be used to distinguish between the multiple possible outcomes.

In Pantograph, we chose a mixture of the two options. We solve the first example in the user interface by making two distinct cursor positions: inside a hole, and around a hole. Inputting the $\lambda$ expression inside the hole gives a function of the existing type, while wrapping it around the hole adds a new parameter. We solve the second example by requiring the user to edit the type annotation in order to achieve the first outcome.

%% file: related_work.tex
\section{Related Work}
\label{sec:relatedwork}

\paragraph{Traditional Tree-Based Structure Editors}
Most existing structure editors share a common core design. As described in the introduction, these editors represent the program as a tree with holes. The user may insert forms into holes, delete nodes, and copy/paste entire nodes. Often, there are additional list structures in which sub-lists may be manipulated.

A wide variety of tree-based structure editors have been developed by both academia and industry with a wide variety of design goals.
Some are designed for advanced type theories, such as
ALF~\cite{alf} for Martin-L{\"o}f type theory, and
Alfa~\cite{alfa} for the Agda dependently-typed programming language.
Others represent the tree as nested blocks, such as
Scratch~\cite{scratch}, a well-known structure editor designed for teaching programming to children.
Others have been designed for general purpose programming, such as MPS~\cite{mps} which allows the user to edit languaged defined by custom DSLs, or Lamdu~\cite{lamdu} which edits a Haskell-like functional language.

While these editors offer a variety of innovative designs,
two problems often arise in tree editors. First, as we described in the introduction, it is difficult to edit code by only moving entire subtrees, a problem often referred to as viscosity \cite{BLACKWELL2003103}.
Second, while the approach straightforwardly maintains well-typedness for a simple type system like that in Scratch, most syntactic structure editors with a more complex type system resort to type checkers which give errors in the same way as on a text editor, leading to semantically meaningless partial programs.

While many tree editors re-typecheck the program after each edit, others are intrinsically typed, like Pantograph.
Intrinsically typed editors place errors \emph{during} an edit, giving them access to more information than a checking algorithm which runs \emph{after} the edit.
%
%
For example, suppose that a programmer deletes an argument from a function which inputs three integers in Pantograph:

\begin{tikzpicture}
    \node (step0) at (0,0) {\includegraphics[scale=0.25]{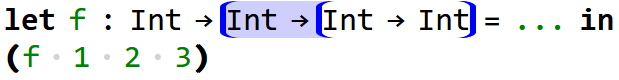}};
    \node (step1) at (8,0) {\includegraphics[scale=0.25]{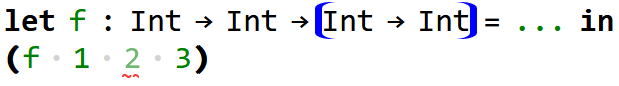}};
    \draw[->,thick] (step0.east) to node[above,yshift=0.5em]{delete} (step1.west);
\end{tikzpicture}

\noindent
Pantograph is able to place a commented application at the corresponding argument.
In contrast, a type checker only has access to the (untyped) program state after the edit, and so can not distinguish which argument of the function was deleted.



\paragraph{Hazel}

The closest related work that addresses the viscosity problem—and one of the most successful
structure editors in recent history in general, is Hazel \cite{hazelwebsite}, a structured editing
system with a typed language and an edit semantics.

Early iterations of the Hazel structure editor ensure that every state is statically meaningful, by defining statics \cite{hazelnut} and dynamics ~\cite{hazel-live-lits} for incomplete programs in a tree-based structure editor.
In particular, Hazel leverages a gradual type system \cite{gradualhazel2, gradualhazel1}, which allows ill-typed (according to a strict type system) programs to be considered well-typed by converting the type errors into dynamic type checks.
It can also run programs with holes, giving users feedback about partial programs.

To address viscosity, {\em Tylr} \cite{tylr,teentylr} introduces gradual structure editing,
allowing the user to locally break the tree structure of the program and get hints
about how to fix the ill-formed fragment.
Tylr introduced a concept of ``structural obligations'' that, given an ill-formed program, encode where certain syntactic delimiters must be inserted in order to yield a well-formed program. 
In essence, gradual structured editing allows the user to edit their program as if it was text, get structured support for the fragment of the program that is well-structured, and hints about how to fix the fragment that is ill-formed.
The authors conducted two user studies, which demonstrate that this approach can improve fluidity in comparison to a more traditional rigidly tree-structured editor.

\citet{teentylr} also identify several specific barriers to fluidity with strict tree editors, including \emph{multiplicity}, or the lack of the ability to place terms temporarily anywhere in the program during edits. While our zipper editing paradigm improves the fluidity of a tree editor, multiplicity remains a problem in Pantograph --- in particular, swapping two terms is difficult.

Recent versions of Hazel incorporate Tylr, replacing its older tree-based interface. 
Furthermore, they re-check the program after each edit using a novel algorithm.
\citet{typeerrorlocalization, polymorphichazel} replace a standard type checking algorithm with a marking algorithm, which inputs untyped preterms and marks them with various marks indicating typing problems, including ones similar to our type error boundaries. The algorithm is superior at placing type errors to many prior type checking algorithms because of its focus on neutrality, or not making unwarranted assumptions about error placements.
In addition, the algorithm can assign static meaning to many programs on which a traditional type checker would give up, and can give suggestions for automatic fixes for type errors.
However, as we discussed earlier, approaches like this which re-check the program after an edit contrast with Pantograph's approach of placing errors during an edit.

\paragraph{Other Editors}

Other ways of manipulating structured syntax have been devised.
Paredit~\cite{paredit} is a tree editor which provides structured operations on S-expressions~\cite{sexp}.
Beyond tree editors, graph editors, like the system design environment LabVIEW~\cite{labview}, represent the program as nodes which the programmer can connect with edges.

In order to compactly represent and contrast the diverse designs of existing structure editors, we chose 8 representative editors and compare them on four criteria. The first two criteria pertain to how strictly structured the editor is.
The last two criteria pertain to fluidity; they are not intended as a benchmark of fluidity, but rather are chosen to highlight the differences between different approaches.

\begin{center}
\begin{tabular}{ |p{2.0cm} p{1.9cm} p{1.8cm} p{1.9cm} p{2.6cm} | } 
    \hline 
     Editor
        & \raggedright All states are well-formed trees
        & \raggedright Intrinsically typed edits or re-check
        & \raggedright Swap terms conveniently (multiplicity)
        & Can re-order e.g. \newline $C[((x * 2) + 1)]$ and \newline $C[((x + 1) * 2)]$
    \\
    \hline \hline
    Text & No & Re-check & Yes & Yes, 1 cut-paste
    \\
    Paredit & Yes & (Untyped) & Yes & Yes, many actions
    \\
    MPS & Yes & Re-check & No & No
    \\
    Scratch & Yes & Intrinsic & Yes & Yes, 6 moves
    \\
    Early Hazel & Yes & Intrinsic & No & No
    \\
    Tylr & Partially & (Untyped) & Yes & Yes, 2 cut-paste
    \\
    Current Hazel & Partially & Re-check & Yes & Yes, 2 cut-paste
    \\
    Pantograph & Yes & Intrinsic & No & Yes, 1 cut-paste
    \\
    \hline
\end{tabular}
\end{center}

\paragraph{Other related works}



Many approaches have been developed for locating (and fixing) errors in existing programs, contrasting with Pantographs intrinsically typed approach as we discussed earlier. There is a large body of work on improving a type checker's ability to locate type errors, including \citet{typeerrorlocationwand}, which improves on a Hindley Millner type inference algorithm to track the reasons for failed unification; and \citet{typeerrorlocationtip} and \citet{typeerrorlocationschilling}, which report a slice of all program locations contributing to an error. Bidirectional type checking \cite{bidirectional-type-checking}, prescribes specific locations in programs where types are inferred or checked.

Automated program repair \cite{summaryOfProgramRepair} uses various techniques to automatically find and fix bugs in a program.
For example, GenProg \cite{classicProgramRepeairPaper} uses a genetic algorithm to randomly vary a program until it passes test cases.
In contrast to Pantograph, these algorithms operate on existing code rather than as part of the editing process.

Ornaments \cite{originalOrnamentsPaper} encode relationships between data types. 
\citet{ornamentRefactoring} use ornaments to facilitate automatic refactoring of code.
Ornaments can encode more possible relations between types than our diffs (although only at data types), but require a user-written definition rather than being derived from the shape of any edit as diffs are in Pantograph.


%% file: conclusion.tex
\section{Conclusion}
\label{sec:conclusion}

In this paper we presented a zipper editing scheme which generalizes text editing in a structured setting, allowing 
users to edit a program while maintaining syntactic well-formedness. In this way, zipper editing is conceptually analogous to text editing while providing the advantages of structure editing. 
We also introduced a type editing system which extends zipper editing. Without requiring the user to learn any new interactions beyond zipper editing, it allows edits to maintain the well-typedness of the program. The user may make any edit which locally can be reconciled to be well-typed, and our diff propagation system will keep the rest of the program aligned to whatever changes were made.

The modern text editing interface has stood as the dominant code editing interface for decades, and programmers have correspondingly internalized its concepts as second nature.
This familiarity creates a barrier to entry for alternative editing systems that ask users to learn new concepts. 
However, the widespread adoption of complex IDEs and editor plugins suggests a strong desire for functionality that is difficult to integrate with text editing.
Since programs are fundamentally tree-structured, structure editing offers innate advantages conceptually organizing basic and advanced edits alike. 
In the introduction, we posed the possibility of designing a general fluid structure editor that never resorts to operating over untyped or ill-formed syntax.
Our contributions with Pantograph show that this goal is achievable,
by placing a powerful typed editing system into an interface
as simple, expressive, as language-generic as, and in direct correspondence with, those in text editing.

%% file: smallsteprules2.tex
\newpage
\section{Small-Step Diff Rules}
\label{sec:rule-small-app}

\begingroup
\renewcommand{\arraystretch}{1.3} 

\begin{tabular*}{\linewidth}{R C L @{\extracolsep{\fill}} R @{\extracolsep{0.2em}} L}
%
%
\multicolumn{5}{c}{(a) Propagation Rules}
\\
\multicolumn{5}{c}{\textrm{if form $r$ has the typing rule \inferrule{s_1 \dots s_n}{s}},}
\\
\multicolumn{3}{l}{
    $$
    \inferrule{
    s = C[s']
    \\
    dom(\sigma) = FV(s')
    }{
    \down{(\sigma_{id} \; C)[\sigma \; s']}{r \; t_1 ~\dots~ t_n}
        \;\; \leadsto  \;\;
    \up{(\sigma \; C)[\sigma_{id} \; s']}{r \; \down{\sigma \; s_1}{t_1} ~\dots~ \down{\sigma \; s_n}{t_n}}
    }
    $$
} & \textrm{\sc{Propagate}} & \downarrow
\\[1em]
\\
\multicolumn{3}{l}{
    $$
    \inferrule{
    s_i = C[s_i']
    \\
    dom(\sigma) = FV(s_i')
    }{
    r \; t_1 ... \up{(\sigma_{id} \; C)[\sigma\; s_i']}{t_i} ... t_n
        \; \leadsto  \;\;
    \up{\sigma \; s}{r \; \down{\sigma \; s_1}{t_1} ~\dots~ \down{(\sigma \; C)[\sigma_{id} \; s_i']}{t_i} ~\dots~ \down{\sigma \; s_n}{t_n}}
    }
    $$
} & \textrm{\sc{Propagate}} & \uparrow
%
\\[1em]
\\
\downa{C \ohchole{ \Delta ,~ x : \delta } \vdash id}{x}
& \step &
\upa{id \vdash \delta}{x}
& \textrm{\sc{Propagate-Var}} & \downarrow \!\! 1
\\
\downa{C \ohchole{ \Delta ,~ x : id } \vdash \delta}{x}
& \step & 
\upa{C \ohchole{ \Delta ,~ x : \delta } \vdash id}{x}
& \textrm{\sc{Propagate-Var}} & \downarrow \!\! 2
%
%
\\[1em]
\multicolumn{5}{c}{(b) Alteration Rules} 
\\[0.5em]
\downa{\Delta \vdash \plusdiffs{A \to}{\delta}{}}{t}
& \step &
\fun{x}{A}{\downa{\plusdiffs{}{\Delta}{,~ x : A} \vdash \delta}{t}}
& \textrm{\sc{Insert-Abs}} & \downarrow
\\
\downa{\Delta \vdash \minusdiffs{A \to}{\delta}{}}{\fun{x}{A}{t}}
& \step &
\downa{\minusdiffs{}{\Delta}{,~ x : A} \vdash \delta}{t}
& \textrm{\sc{Delete-Abs}} & \downarrow
\\
\fun{x}{A}{\upa{\Delta, x : id \vdash \plusdiffs{A \to}{\delta}{}}{t}}
& \step &
\upa{\Delta \vdash A \to \delta}{\downa{\minusdiffs{}{\Delta}{,x : A} \vdash id}{t}}
& \textrm{\sc{Delete-Abs}} & \uparrow
\\[0.5em]
%
%
\upa{\Delta \vdash \plusdiffs{A \to}{\delta}{}}{t}
& \step &
\upa{\Delta \vdash \delta}{t \; \holety{A}}
& \textrm{\sc{Insert-App}} & \uparrow
\\
\upa{\Delta \vdash \minusdiffs{A \to}{\delta}{}}{t_1} \; t_2
& \step &
\upa{\Delta \vdash \delta}{\ghostapp{t_1}{\; \downa{\Delta \vdash id}{t_2}}}
& \textrm{\sc{Displace-App}} & \uparrow
\\
\downa{\Delta \vdash \plusdiffs{A \to}{\delta}{}}{t_1 \; \holety{A}}
& \step &
\downa{\Delta \vdash id \to \delta}{t_1}
& \textrm{\sc{Delete-App}} & \downarrow
\\[0.5em]
%
\downa{C \ohchole{\minusdiffs{}{\Delta}{,~ x : T}} \vdash id}{x}
& \step &
\ghostvar{x}{T}
& \textrm{\sc{Local-To-Free}} &
\\[0.5em]
%
\downa{C \ohchole{ \plusdiffs{}{\Delta}{,~ x : T} } \vdash T}{\ghostvar{x}{T}}
& \step &
x
& \textrm{\sc{Free-To-Local}} &
\\[1em]
\multicolumn{5}{c}{(c) Diff Boundary Rules} 
\\[0.5em]
C[t] & \step & C[t'] \hspace{2em}
\textrm{if}~~~ t \step t'
& \textrm{\sc{Step-Inside}} &
\\
\upa{id \vdash id}{t}
& \step &
t
& \textrm{\sc{Identity}} & \uparrow
\\
\downa{id \vdash id}{t}
& \step &
t
& \textrm{\sc{Identity}} & \downarrow
\\
\upa{id \vdash id}{t}
& \step &
t
& \textrm{\sc{Identity}} & \uparrow
\\[0.5em]
%
%
\downa{\Delta \vdash \delta}{t}
& \step &
\multirow{2}{*}{
    \ensuremath{\begin{rcases*}
        \errorboundary{A}{B}{\downa{\Delta \vdash id}{t}} \\
        C \ohchole{\upa{\Delta \vdash id}{\errorboundary{B}{A}{t}}}
    \end{rcases*}}
    \begin{tabular}{l}
        if $\diffsto{A}{\delta}{B}$, \\[-0.8em]
        $t$ is a neutral form, \\[-0.6em]
        and $C \ne \clasp{} \; t_2$
    \end{tabular}
}
& \textrm{\sc Neutral-Error} & \downarrow
\\
C \ohchole{\upa{\Delta \vdash \delta}{t}}
& \step &
& \textrm{\sc Neutral-Error} & \uparrow
\\[1em]
%
%
\downa{\Delta \vdash id}{\upa{id \vdash \delta}{t}}
& \step &
\upa{id \vdash \delta}{\downa{\Delta \vdash id}{t}}
& \textrm{\sc Interchange} & 1
\\
\downa{id \vdash T}{\upa{\Delta \vdash id}{t}}
& \step &
\upa{\Delta \vdash id}{\downa{id \vdash T}{t}}
& \textrm{\sc Interchange} & 2
\\[1em]
%
%
\downa{\Delta \vdash \delta}{t}
& \step &
\multirow{2}{*}{
    \ensuremath{\begin{rcases*}
        \errorboundary{A}{B}{\down{\Delta \vdash id}{t}} \\
        \up{\Delta \vdash id}{\errorboundary{B}{A}{t}}
    \end{rcases*}}
    \begin{tabular}{l}
        if $\diffsto{A}{\delta}{B}$, \\[-0.8em]
        and no other rules apply
    \end{tabular}
}
& \textrm{\sc Fallthrough-Error} & \downarrow
\\
\upa{\Delta \vdash \delta}{t}
& \step &
& \textrm{\sc Fallthrough-Error} & \uparrow
\end{tabular*}

\endgroup

%% file: metatheoryAppendix.tex
\section{Proofs about the Diff Propagation System}
\label{app:metatheory}

In this appendix we write the full proofs of the theorems shown in section~\ref{sec:proofs}.
The proof of termination introduces a few lemmas which are referenced by the other proofs, so we present it first.

\subsection{Termination of Small-Step Rules}
\label{app:termination}

\begin{customthm}{\ref{thm:terminationtheorem}} [Termination]
\label{thm:termination}
For any program with diff boundaries resulting from an edit in Pantograph, there is no infinite sequence of step rules that can be applied.
\end{customthm}

Looking at the rules defined in Appendix~\ref{sec:rule-small-app}, we can observe that while up boundaries can turn into down boundaries, 
no down boundary can ever turn into an up boundary with the exception of the {\sc Propagate} and {\sc Propagate-Var} rules. 
Also, with the exception of the {\sc Insert}-$*$ rules, the upwards boundaries make progress to the top of the program and the downwards boundaries make progress to the leaves.
Therefore, the path of a boundary through the program will generally be to first go up, and then go down, and then disappear. 
In the following proof, we formalize this intuition.

While we conjecture that any setup of diff boundaries will terminate, we only prove that configurations of diff boundaries that can occur in Pantograph terminate. Towards this end, we prove two lemmas about properties that hold for all states in Pantograph.

\begin{lemma}
    \label{lemma:uplemma}
    All up boundaries have a diff either with only a type diff $id \vdash \delta$ or only a change to one variable in context $id[x : \delta, id]\vdash id$.
\end{lemma}

\paragraph{Proof:} The diff boundaries initially set up, as described in Section~\ref{sec:diffsetup}, satisfy this property because the up diff has an identity context. Most of the step rules in Section~\ref{sec:alterationanddiffrules} can directly be observed to hold this property. To see that The {\sc Propagate} rules preserve this property, we need to consider all of the typing rules in Figure~\ref{fig:typing-rules}.

First, observe that {\sc Propagate} $\downarrow$ can only create an up boundary when there is a nonlinearity in the conclusion of a typing rule. However, no typing rules have an appearance of a nonlinear variable in the context of the conclusion, so output up diff boundary can only have non-identity diffs in the type.

Next we consider {\sc Propagate} $\uparrow$.
In most typing rules, the context on all premises and the conclusion is simply a metavariable $\Gamma$, so the property holds trivially. The remaining typing rules are $\lambda$ abstraction, recursive \CC{let} forms, and \CC{match} expressions.

\begin{itemize}
    \item For $\lambda$ abstractions, the input up diff may by assumption modify only one of $\Gamma$, $T_1$, or $T_2$. Therefore, the property is preserved.
    \item For \CC{let} expressions, from either premise only one of $\Gamma$, $T_1$, or $T_2$ may be modified. Therefore, the property is preserved.
    \item For \CC{match} expressions, for any premise only one of $\Gamma, T_1$ or $T_2$ may be modified. Therefore, the property is preserved.
\end{itemize}

To state the next lemma, we clarify the definition of neutral forms and boundaries in a neutral form.

\begin{definition} [Neutral form]
\label{def:neutral}
    A neutral form is a term t such that either
    \begin{itemize}
        \item $t = x$
        \item $t$ is a form that has a nonlinearity in its conclusion, such as \CC{cons}
        \item $u$ is a neutral form, and $t = u~t'$
        \item $u$ is a neutral form, and $t = \ghostapp{u}{t'}$
        \item $u$ is neutral form, and $t = \up{d}{t}$ or $t = \down{d}{t}$
    \end{itemize}
\end{definition}

\begin{definition}[Maximal Neutral Form]
    A maximal neutral form is a subexpression $u$ within a program $t = C[u]$ such that $u$ is a neutral form, and the innermost step of $C$ is not of the form $\clasp{}~t'$.
\end{definition}

\begin{definition}[Boundary in a Neutral Form]
    We say that a boundary is in a neutral form if one of:
    \begin{itemize}
        \item $\up{d}{u}$ and $u$ is a neutral form
        \item $\down{d}{u}$ and $u$ is a non-maximal neutral form
    \end{itemize}
\end{definition}

\begin{lemma}
    \label{lemma:neutralboundaries}
    The following property is preserved by the step rules:
    \begin{itemize}
        \item All boundaries in a neutral form have a diff $\Delta \vdash \delta$ where either $\Delta = id$ or $\delta = id$
    \end{itemize}
\end{lemma}

We refer to these as type-only and context-only boundaries respectively.

\paragraph{Proof:}
We prove that boundaries already in a neutral form maintain this property, and that boundaries entering a neutral form must have this property.

The rules that can operate on a boundary within a neutral form are {\sc Propagate $\updownarrow$}, {\sc Propagate-Var $\updownarrow$}, {\sc Insert-App}, {\sc Displace-App} and {\sc Delete-App}, and {\sc Fallthrough-Error}. Each of these preserves the property.


Next, we consider all of the possible ways that a boundary can enter a neutral form.
Boundaries resulting directly from an edit inside of a neutral form satisfy this property because as described in Section~\ref{sec:diffsetup} the initial diffs after an edit satisfy this property.
Boundaries entering a neutral form from the top must be context-only because of the  {\sc Neutral} $\downarrow$ rule.
Finally, a boundary which enters a neutral form through an argument $u \; \up{d}{t}$ (or a commented argument) must also have this property by Lemma~\ref{lemma:uplemma}.

Next, we make a definition to help deal with the rules that can reflect down boundaries up.
These rules only occur inside of neutral forms.
The variable rules are two such rules. Also, some other propagation rules can have this property. Specifically, any rule which has two of the same metavariable in the conclusion, like the \CC{cons} rule in Figure~\ref{fig:typing-rules}, can reflect down boundaries upwards. For example, the following \CC{cons} rule is a special case of the general propagation rule in Appendix~\ref{sec:rule-small-app}.
This rule is used in the final example of section \ref{sec:typed-editing} where the user deletes a type boundary.

$$
    \down{id \vdash \delta \to \cList id \to \cList id}{\ccons}
    \step
    \up{id \vdash id \to \cList \delta \to \cList \delta}{\ccons}
$$

No such construct in our language with multiple appearances of the same metavariable in the conclusion has any premises.
Therefore, neutral forms are exceptional in terms of diff propagation for two reasons. The first is that it is inside of a neutral form, at the variable or \CC{cons} constuctor, that downward boundaries can be reflected upwards. The second is that two of the propagation rules, namely the {\sc Neutral} $\uparrow$ and {\sc Neutral} $\downarrow$ rules, interact with them; specifically, they prevent any type diff from entering or exiting a neutral form, while allowing context diffs to pass.

\begin{definition} [up-like and down-like]
    We label each boundary as up-like or down-like.
    For most boundaries, up boundaries are up-like and down boundaries are down-like.
    The exception is for a type-only boundary in a neutral form: up type-only boundaries are down-like, and down type-only boundaries are up-like.
\end{definition}

No rule allows a down-like boundary to turn into an up-like boundary.
To understand these definitions, consider a type-only down boundary on a neutral form consisting of a single variable applied to an argument. It will propagate as follows:

$$
\down{id \vdash \delta_1 \to \delta_2}{x \; t}
\step
\down{id \vdash \delta_1}{x} \; \down{id \vdash \delta_2}{t}
\step
\up{... x : \delta_1 ... \vdash id}{x} \; \down{id \vdash \delta_2}{t}
$$

The type-only down boundary was up-like. When it turned into two down boundaries, the boundary around $x$ is still up-like, while the boundary around $t$ is now down-like. Finally, when the boundary around the variable reflects upwards, it still retains its up-like nature. If this context-only boundary eventually reaches a recursive let, it may be reflected back down, turning into a down-like boundary.







The proof proceeds by describing a well-founded ordering on terms, which must always decrease by each step rule.

The first component of the ordering is an ordering on diff boundaries called $\le_{ud}$. It is given by the up-like or down-like property of diff boundaries. In this ordering, an up-like boundary is greater than a down-like boundary.

The next component of the ordering on diff boundaries is based on their position, called $\le_{pos}$.

First, we define a notion of the length of a one-hole context. This is almost the number of steps in the one-hole context, except for a special case at neutral forms to account for how boundaries move up through neutral forms:

\begin{align*}\begin{array}{l}
\metatheorystyle{length} : C \to \mathbb{N}
\\
\metatheorystyle{length}~C[x~t_1 \dots t_n~ \clasp{}] = 2n + \metatheorystyle{length}~C
\\
\metatheorystyle{length}~C[c[\clasp{}]] = 1 + \metatheorystyle{length}~C
\\
\metatheorystyle{length}~\clasp{} = 0
\end{array}\end{align*}

We then define the following function which gives the distance that the boundary has left to travel:

\begin{align*}\begin{array}{l}
    \mdistance \; : \textrm{diff boundary} \to Int
    \\ 
    \begin{rcases}
        \mdistance \; C_1 \ohchole{\down{d_1}{t_1}} = \textrm{ the maximum depth of $t_1$} \\
        \mdistance \; C_1 \ohchole{\up{d_1}{t_1}} = \textrm{ the length of $C_1$}
    \end{rcases}
    \text{not a type-only boundary in neutral form} 
    \\
    \begin{rcases}
        \mdistance \; C_1 \ohchole{\down{d_1}{x \; t_1 ... t_n}} = 2 n + \textrm{the length of $C_1$} \\
        \mdistance \; C_1\ohchole{\up{d_1}{t_1} \; arg_1 ~ \dots ~ arg_n} = n + \textrm{ the maximum depth of $arg_1 ... arg_n$}
    \end{rcases}
    \text{else}
\end{array}\end{align*}

Using this function, we define an ordering $\le_{pos}$ on two diff boundaries within two programs. 

\begin{align*}
    C_1 \ohchole{
        \updownboundary{\Gamma_1 \vdash T_1}{t_1}
    } >_{rc} 
    C_2 \ohchole{
        \updownboundary{\Gamma_2 \vdash T_2}{t_2}
    } 
    \textrm{ if } \mdistance \; T_1 > \mdistance \; T_2
\end{align*}

All of the rules are non-increasing along this ordering. The only rules which don't decrease are the {\sc Insert-Abs} $\downarrow$ and {\sc Insert-App} $\uparrow$ rules. To deal with these two exceptions, we will need one final component to the ordering.

The final component of the ordering is given by counting the $+$s and $-$s in the diffs. For a diff $d$, let $\mcount \; d$ be the number of $\plusdiff{A \to}{ B}{}$ and $\minusdiff{A \to}{B}{}$ forms appearing in the diff.

\begin{align*}
    C_1 \ohchole{
        \updownboundary{\Gamma_1 \vdash T_1}{t_1}
    }
    >_{c}
    C_2 \ohchole{
        \updownboundary{\Gamma_2 \vdash T_2}{t_2}
    }
    \textrm{ if } \mcount \; T_1 > \mcount \; T_2
\end{align*}

In the rules on which the $\le_{pos}$ ordering doesn't decrease, the $\le_{c}$ decreases.
Therefore, on the lexicographical ordering formed by the combination $(\le_{ud}, \le_{pos}, \le_{c})$, the output boundaries of every rule are strictly less than the input boundary.

Of course, as the system is stepping the boundaries, there may be many boundaries in the program. We need an ordering over entire programs with many boundaries, not just single boundaries, in order to prove termination of our algorithm.
We construct an ordering over the set of all of the diff boundaries within the program. We use a Dershowitz-Manna ordering \cite{setordering}. In this ordering, if $S_1$ and $S_2$ are sets of boundaries and $b$ is a particular boundary,

\begin{align*}
    \textrm{if} \; \forall x \in S_2, b > x, \;\textrm{then}\; S_1 \cup \s{b} > S_1 \cup S_2
\end{align*}   

Under this order, for every step rule (besides the variable rules) if $S_1$ is the set of boundaries in $t_1$ and $S_2$ the set for $t_2$, then for any $t_1 \leadsto t_2$, $S_2 < S_1$. Because the ordering is well-founded, this proves the theorem.

\subsection{Type preservation}
\label{app:preservation}

Each small-step rule in Appendix~\ref{sec:rule-small-app} preserves the well-typedness of the program.

\begin{customthm}{\ref{thm:preservation}}
If $\;\Gamma \vdash t : T$, and $t \leadsto t'$, then $\Gamma \vdash t' : T$.
\end{customthm}

The proof proceeds by cases over the rules.
Each step rule converts and input term to an output, with some common subterms preserved. Therefore, our task is to show that for each case, the type of the term overall is preserved, and the type of each subterm is preserved. We list all of the cases below. The two {\sc Propagate} rules require some reasoning about diff substitutions, and all of the other cases follow directly from the definitions.

\begin{itemize}
    \item
    {\sc Propagate} $\downarrow$

    Suppose that a construct $r$ has an intrinsic typing rule
    $$
    \inferrule{
        s_1 \; \dots \; s_n
    }{
        s
    }
    $$
    with $s = C[s']$,
    And the form $\down{(\sigma_{id} \; C)[\sigma \; s']}{r \; t_1 \dots t_n}$ is well-typed.

    Let $\sigma.1$ and $\sigma.2$ be endpoints of $\sigma$, that is for all metavariables $\alpha$, $(\sigma.1~\alpha) = (\sigma~\alpha).1$ and $(\sigma.2~\alpha) = (\sigma~\alpha).2$.
    Then, the metavariables in this instantiation of $r$'s typing rule is given by $\sigma.1$, and each $t_i$ has type and context $\sigma.1 \; s_i$. The type and context of the term overall is $(\sigma.1 \; C)[\sigma.2 \; s']$.
    
    Then the output of the rule is

    $$
    \up{(\sigma \; C)[\sigma_{id} \; s']}{r \; \down{\sigma \; s_1}{t_1} ~\dots~ \down{\sigma \; s_n}{t_n}}
    $$

    Now, the type and context of the form overall is still $(\sigma.1 \; C) [\sigma.2 \; s']$, and the type and context of each $t_i$ is still $\sigma.1 \; s_i$.
    \item
    {\sc Propagate} $\uparrow$

    Suppose that a construct $r$ has an intrinsic typing rule
    $$
    \inferrule{
        s_1 \; \dots \; s_n
    }{
        s
    }
    $$
    with $s_i = C[s_i']$,
    And the form
    $ r \; t_1 ... \up{(\sigma_{id} \; C)[\sigma\; s_i']}{t_i} ... t_n $
    is well-typed.
    
    Like the previous case, let $\sigma.1$ and $\sigma.2$ be endpoints of $\sigma$.
    Then, the metavariables in this instantiation of $r$'s typing rule is given by $\sigma.1$, and each $t_j$ with $j \ne i$ has type and context $\sigma.1 \; s_i$. $t_i$ has type and context $(\sigma.1 \; C) [\sigma.2 \; s_i']$. The type and context of the term overall is $\sigma.1 \; s$.

    Then the output of the rule is
    $$
    \up{\sigma \; s}{r \; \down{\sigma \; s_1}{t_1} ~\dots~ \down{(\sigma \; C)[\sigma_{id} \; s_i']}{t_i} ~\dots~ \down{\sigma \; s_n}{t_n}}
    $$

    Now, the type and context of the form overall is still $\sigma.1~s$. The type and context of each $t_j$ with $j \ne i$ is still $\sigma.1~s_j$. Finally, the type and context of $t_i$ is still $(\sigma.1 \; C) [\sigma.2~ s_i']$.
    \item {\sc Propagate-Var} $\downarrow 1$
    Suppose that
    $
    \downa{C \ohchole{ \Delta ,~ x : \delta } \vdash id}{x}
    $
    is well typed. Then, the whole expression is at a context $C.2[\Delta.2, x : \delta]$ and type $\delta.1$. Then the output
    $
    \upa{id \vdash \delta}{x}
    $
    is still well-typed at the same context and type.
    \item {\sc Propagate-Var} $\downarrow 2$
    Suppose that
    $
    \downa{C \ohchole{ \Delta ,~ x : id } \vdash \delta}{x}
    $
    is well typed. Then, the whole expression is at a context $C.2[\Delta.2, x : \delta.1]$ and type $\delta.2$. Then the output
    $
    \upa{C \ohchole{ \Delta ,~ x : \delta } \vdash id}{x}
    $
    is still well-typed at the same context and type.
    \item {\sc Insert-Abs $\downarrow$}

    Suppose that $\down{\Delta \vdash \plusdiffs{A \to}{\delta}{}}{t}$
    is well typed.

    Then, $\Delta.1 \vdash t : \delta.1$, and the whole term has context $\Delta.2$ and type $A \to \delta.2$.

    Then, the output $\fun{x}{A}{\down{\plusdiffs{}{\Delta}{, x : A} \vdash \delta}{t}}$
    is still well-typed at the same context and type.
    
    \item
    {\sc Delete-Abs} $\downarrow$

    Suppose that
    $
    \down{\Delta \vdash \minusdiffs{A' \to }{\delta}{}}{\lambda x : A. t}
    $
    Is well typed.
    Then, $A' = A$, $\Delta.1, x : A \vdash t : \delta.2$, and the context and type of the whole term are $\Delta.2$ and $\delta.2$.
    
    Then the output
    $
    \down{\plusdiffs{}{\Delta}{, x : A} \vdash \delta}{t}
    $
    Is also well-typed, and the whole term as well as $t$ have the same context and type as before.

    \item
    {\sc Delete-Abs} $\uparrow$

    Suppose that $\fun{x}{A}{\up{\Delta, x : A \vdash \plusdiffs{A \to}{\delta}{}}{t}}$
    is well typed. Then, $\Delta.2, x : A \vdash t : \delta.2$, and the whole term has context $\Delta.1$ and type $A \to \delta.1$.

    Then the output
    $\up{\Delta \vdash A \to \delta}{\down{\minusdiffs{}{\Delta}{, x : A} \vdash id}{t}}$
    is still well typed at the same context and type.

    \item
    {\sc Insert-App} $\uparrow$

    Suppose that $\up{\Delta \vdash \plusdiffs{A \to}{\delta}{}}{t}$ is well typed.
    Then $\Delta.2 \vdash t : A \to \delta.2$, and the whole term has context $\Delta.1$ and type $\delta.1$.

    Then the output
    $\up{\Delta \vdash \delta}{t_1 ~ \holety{A}}$
    is still well typed at the same context and type.
    
    \item
    {\sc Displace-App} $\uparrow$

    Suppose that $\up{\Delta \vdash \minusdiffs{A \to}{\delta}{}}{t}$ is well typed.
    Then $\Delta.2 \vdash t : \delta.2$, and the whole term has context $\Delta.1$ and type $A \to \delta.1$.

    Then the output
    $\up{\Delta \vdash \delta}{\ghostapp{t_1}{\down{\Delta \vdash id}{t_2}}}$
    is still well typed at the same context and type.

    \item
    {\sc Delete-App} $\downarrow$

    Suppose that $\down{\Delta \vdash \plusdiffs{A \to}{\delta}{}}{t_1~\holety{A}}$. Then $\Delta.1 \vdash t_1 : \delta.1$, and the whole term has context $\Delta.2$ and type $\delta.2$.

    Then the output
    $\down{\Delta \vdash id \to \delta}{t_1}$
    is still well typed at the same context and type.

    \item
    {\sc Local-To-Free}

    Suppose that $\down{C[\minusdiffs{}{\Delta}{, x : T}] \vdash id}{x}$
    is well typed.
    Then the term has type $T$.

    Then the output
    $\ghostvar{x}{T}$
    is well typed at the same type $T$ and any context.

    \item
    {\sc Free-To-Local}
    
    Suppose that $\down{C[\plusdiffs{}{\Delta}{, x : T}] \vdash T}{x}$ is well typed.
    Then the term has type $T$ in a context where $x : T$.

    Then the output
    $x$
    is well typed at the same type $T$.

    \item
    {\sc Identity} $\downarrow$

    Suppose that $\down{id \vdash id}{t}$ is well typed at a context and type.

    Then the output $t$ is still well-typed at the same context and type.
    
    \item
    {\sc Identity} $\uparrow$

    Suppose that $\up{id \vdash id}{t}$ is well typed at a context and type.

    Then the output $t$ is still well-typed at the same context and type.

    \item
    {\sc Interchange} $1$
    Suppose that
    $ \downa{\Delta \vdash id}{\upa{id \vdash \delta}{t}} $
    is well typed. Then, $\Delta.1 \vdash t : \delta.2$, and the whole term is at the context $\Delta.2$ and type $\delta.1$.
    
    Then the output
    $ \upa{id \vdash \delta}{\downa{\Delta \vdash id}{t}} $
    is still well-typed at the same context and type.
    
    \item
    {\sc Interchange} $2$
    Suppose that
    $ \downa{id \vdash \delta}{\upa{\Delta \vdash id}{t}} $
    is well typed. Then, $\Delta.2 \vdash t : \delta.1$, and the whole term is at the context $\Delta.1$ and type $\delta.2$.
    
    Then the output
    $ \upa{\Delta \vdash id}{\downa{id \vdash \delta}{t}} $
    is still well-typed at the same context and type.

    \item
    {\sc Neutral-Error} $\downarrow$ and {\sc Fallthrough-Error} $\downarrow$

    Suppose that $\down{\Delta \vdash \delta}{t}$
    is well typed.
    Then, $\Delta.1 \vdash t : \delta.1$
    and the whole term has context $\Delta.2$ and type $\Delta.2$.

    Then the output $\errorboundary{\delta.1}{\delta.2}{\down{\Delta \vdash id}{t}}$
    is well typed at the same context and type.
    
    \item
    {\sc Neutral-Error} $\uparrow$ and {\sc Fallthrough-Error} $\uparrow$

    Suppose that $\up{\Delta \vdash \delta}{t}$
    is well typed.
    Then, $\Delta.2 \vdash t : \delta.2$
    and the whole term has context $\Delta.1$ and type $\Delta.1$.

    Then the output $\up{\Delta \vdash id}{\errorboundary{\delta.1}{\delta.2}{t}}$
    is well typed at the same context and type.

\end{itemize}

\subsection{Confluence}
\label{app:confluence}

We will motivate some aspects of the confluence proof through two examples. 
First, we demonstrate that although confluence holds of all states that occur in Pantograph, it does not hold of all configurations of diff boundaries.

\begin{example}[Confluence Does not Hold of All Configurations]
\end{example}
In the program
$$
\down{id \vdash d_1 \to id}{\up{id \vdash d_2}{t_1}~t_2}
$$

after one step, the two boundaries will point in to each other, and the {\sc Interchange} rules will not apply. Therefore, the {\sc Fallthrough-Error} rules will create an error boundary. But the location of the error will differ depending on which boundary is stepped first.

Luckily, this situation can not occur in Pantograph, as we will prove in a moment. However, the following situation gives an example of nondeterminism that can occur, but is confluent.

\begin{example} [An Edit that Leads to Nondeterminism]
    
\end{example}
If the programmer alters the annotation on the $\lambda$ abstraction by a diff, for example
$\cInt \to \replacediff{\cInt}{\cBool}$
,in the program:

$$
\llet{f}{(\cInt \to \cInt) \to \cInt}{\fun{x}{\cInt \to \cInt}{x~10}}{\dots}
$$

Then, the system will set up:

$$
\llet{f}{(\cInt \to \cInt) \to \cInt}{\up{id \vdash (\cInt \to \replacediffs{\cInt}{\cBool}) \to \cInt{}}{\fun{x}{\cInt \to \cBool}{\down{f : id, x : \cInt \to (\replacediffs{\cInt}{\cBool}) \vdash id}{x~10}}}}{\dots}
$$

Then, after propagating the resulting boundaries for some period, the algorithm could arrive at the term

$$
\llet{f}{(\cInt \to \cBool) \to \cInt}{\fun{x}{\cInt \to \cBool}{\down{f : (\replacediffs{\cInt}{\cBool}) \to \cInt, x : id \vdash id}{\up{id \vdash \cInt \to (\replacediffs{\cInt}{\cBool})}{x}~10}}}{\dots}
$$

In this state, again, either boundary can be stepped next. But in this case the {\sc Interchange} rules will apply, the boundaries can pass through each other, and the same result will be reached either way.

\begin{customthm}{\ref{thm:confluence}} [Confluence]
\label{thm:appendixconfluence}
If $t \leadsto^* t_1$ and $t \leadsto^* t_2$, then there exists $t'$ such that $t_1 \leadsto^* t'$ and $t_2 \leadsto^* t'$
\end{customthm}

Newman's lemma \cite{newmanlemma} states that if a rewrite system is terminating and locally confluent, then it is confluent. Therefore given Theorem~\ref{thm:termination} we only need to prove local confluence, which states that if a term can step by a single step to two different terms, then by any number of steps those two terms step to the same term:

\begin{definition} [Local Confluence]
If $t \leadsto t_1$ and $t \leadsto t_2$, then there exists $t'$ such that $t_1 \leadsto^* t'$ and $t_2 \leadsto^* t'$
\end{definition}

To prove local confluence, we will need to consider all possible pairs of rules that can step from the same term.
One major source of such pairs are when a down boundary and an up boundary propagate into the same term from opposite sides, like the two examples above.
First, we can rule out many of these cases by proving that two properties apply to all configurations in Pantograph.

\begin{property}
    \label{prop:oneup}
    There is at most one up-like boundary
\end{property}

\begin{lemma}
    \label{lemma:oneup}
    Property~\ref{prop:oneup} is true of the initial states in Pantograph, and is preserved by all of the step rules.
\end{lemma}

\paragraph{Proof:}
Each of the initial states in Section~\ref{sec:diffsetup} has an up type-only boundary and a down context-only boundary. If the edit is not in a netural form, it therefore has one up-like boundary. If it is in a neutral form, it has none.

As for the preservation of the property, we will show that every step rule that inputs an up-like boundary will output at most one, and every step rule that inputs none will output none.

First, we consider the case that the up-like boundary is not in a neutral form, and is therefore an up boundary.
It can be seen at a glance that all of the rules in Section~\ref{sec:alterationanddiffrules} which input an up boundary output at most one.
Similarly, most of the rules which input a down boundary do not output an up boundary. The only exception is the {\sc Propagate $\downarrow$} rule, which can output an up boundary if the conclusion of the typing rule of the form that it propagates into is nonlinear. However, by definition, such forms can only be in a neutral form.

Next, we consider the case inside of a neutral form. Here, down type-only boundaries are up-like by definition. By Lemma~\ref{lemma:neutralboundaries}, every boundary propagating within a neutral form is either type-only or context-only.
If a type-only or context-only form steps by {\sc Delete-App $\uparrow$} or {\sc Insert-App $\uparrow$}, the property is preserved.
Considering each of the forms that exist in neural forms (Definition~\ref{def:neutral}),
the propagation rules preserved the property.
The only remaining possibility is a diff boundary propagating up into the argument of a neutral form;
consider a neutral form $u$, propagating a boundary up into an argument $u~\up{\Delta \vdash \delta}{t} \step \up{\Delta \vdash id}{\down{\Delta \vdash \delta \to id \vdash id}{u}~t}$. However, by Lemma~\ref{lemma:uplemma}, this can create only one up-like boundary.

\begin{property}
    \label{prop:noaround}
    If there is one up-like boundary, then there are no other boundaries above it. Specifically,
    \begin{itemize}
        \item If the up-like boundary is not a type-only boundary in a neutral form, then the program has the form $C[\up{d}{t}]$ and there are no diff boundaries in $C$
        \item if the up-like boundary is a type-only boundary in a neutral form, then the program has the form $C[\down{id \vdash d}{u} \; t_1 \dots t_n]$ where $\down{id \vdash d}{u} \; t_1 \dots t_n$ is a maximal neutral form in the program, and there are no diff boundaries in $C$, and no other boundaries in the neutral form.
    \end{itemize}
\end{property}

\begin{lemma}
    \label{lemma:noaround}
    Property~\ref{prop:noaround} is true of the initial states in Pantograph, and is preserved by all of the step rules.
\end{lemma}

\paragraph{Proof:}
The initial state in Pantograph contains an up type-only boundary and a down context-only boundary, as described in Section~\ref{sec:alterationanddiffrules}. When the edit is not in a neutral form, this satisfies the property. The exception is if the edit is made in a neutral form. But then, the up type boundary is down-like, and the property holds vacuously as there are no up-like boundaries.

Next, we consider all of the ways that the two configurations of up-like boundaries can be created by the step rules, and show that in all cases if the property is true of the input, then it is true of the output.
\begin{itemize}
    \item The rules that can create an up-like boundary in the program of the form $C[\up{d}{t}]$ are {\sc Propagate} $\uparrow$, {\sc Insert-App}, and {\sc Displace-App}. In each of these rules, if there are no boundaries in the one-hole context $C$ beforehand, then there will also be none in the one-hole context around the up boundary in the output of the rule.
    \item There are only two rules that can create a type-only down boundary in a neutral form of the form $C[\down{id\vdash d}{t} \; t_1 \dots t_n]$. The first is {\sc Propagate $\downarrow$}, given an input either $C[\down{id \vdash t}{u \; t_1} \; t_2 \dots t_n]$ or $C[u \; \up{id \vdash d}{t_1} \dots t_n]$. The other possibility is the {\sc Delete-App $\downarrow$} rule. All of these possible cases preserve the property.
\end{itemize}

With these two properties in mind, we consider all of the possible pairs of step rules that act on the same term $t$.

Each step rule inputs a term with a particular boundary and acts on that boundary, along with possibly the forms directly above and below the boundary. If there are multiple step rules which act on the same term, then that must be because there are multiple boundaries.

However, if the parts of the terms which are modified by the rule are separate, then local confluence will trivially hold. This is described by the following two cases:

\begin{itemize}
    \item
    If $t_i \step t_i'$ and $t_j \step t_j'$, then:
    
    \begin{tikzcd}
                       & C[r \dots t_i \dots t_j \dots ] \arrow[ld, "\textrm{{\sc Step-Inside}}"'] \arrow[rd, "\textrm{\sc Step-Inside}"] &                   \\
    C[r \dots t_i' \dots t_j \dots ] \arrow[rd, "\textrm{\sc Step-Inside}"', dashed] &                                    & C[r \dots t_i \dots t_j' \dots ] \arrow[ld, "\textrm{\sc Step-Inside}", dashed] \\
                       & C[r \dots t_i' \dots t_j' \dots ]                                  &                  
    \end{tikzcd}
    
    \item 
    Many step rules preserve a child of the stepped term, and have the form $C[x] \step C'[x]$.
    If for any $x$, $C[x] \step C'[x] $, and $t_2 \step t_2'$, then:
    
    \begin{tikzcd}
                       & C[t] \arrow[ld, "\textrm{{\sc Step-Inside}}"'] \arrow[rd, "\textrm{\sc Step-Inside}"] &                   \\
    C'[t] \arrow[rd, "\textrm{\sc Step-Inside}"', dashed] &                                    & C[t'] \arrow[ld, "\textrm{\sc Step-Inside}", dashed] \\
                       & C'[t']                                  &                  
    \end{tikzcd}
\end{itemize}

This leaves cases where the two step rules act on the overlapping parts of the program.

A first straightforward case is that whenever the {\sc Identity} rules apply, a {\sc Propagate} rule also applies.

If a form $r$ has typing a typing rule

$$
\inferrule{s_1 \dots s_n}{s}
$$

Then:

\begin{tikzcd}
                   & \down{id}{r \; t_1 \dots t_n} \arrow[ld, "\textrm{\sc Identity $\downarrow$}"'] \arrow[rd, "\textrm{\sc Propagate $\downarrow$}"] &                   \\
r \; t_1 \dots t_n &                                    & r \; \down{id}{t_1} \dots \down{id}{t_n} \arrow[ll, "\textrm{$n$ steps {\sc Identity $\downarrow$}}", dashed]
\end{tikzcd}

The situation for {\sc Propagate $\uparrow$} and {\sc Identity $\uparrow$} is similar.

There are no other rules which input the same boundary. The remaining cases where multiple rules can apply to the same term are when a subexpression $\down{d_1}{r~t_1 \dots \up{d_2}{t_i} \dots t_n}$ or $\down{d_1}{\up{d_2}{t_i}}$ exists in the program. Some rule will apply to both the up and the down boundary, and in many cases both rules can affect the intermediate form $r$.
There are six rules which propagate down into a form, and six which propagate up. At first, this would seem to lead to 36 total cases. However, we know by Lemmas~\ref{lemma:oneup} and \ref{lemma:noaround} that this situation could only occur if the up boundary is a type-only boundary in a neutral form. Otherwise, the configuration would violate \ref{prop:noaround}. Therefore, we only need to consider the cases where either the two boundaries are directly touching and the up boundary is type-only in a neutral form, or where there is an application or commented application between the two boundaries. We list all of these cases below; the final one will require another lemma.
    
\begin{itemize}
    \item

    \begin{tikzcd}
                       & \down{\plusdiffs{A \to}{B}{}}{\up{\plusdiffs{A \to}{B}{}}{t}} \arrow[ld, "\textrm{{\sc Insert-Abs} $\downarrow$}"'] \arrow[rd, "\textrm{\sc Insert-App $\uparrow$}"] &                   \\
     \fun{x}{A}{\up{\plusdiffs{A \to}{B}{}}{t}} \arrow[rd, "\textrm{\sc Delete-Abs $\uparrow$}"', dashed] &                                    & \down{\plusdiffs{A \to}{B}{}}{t~\holety{A}} \arrow[ld, "\textrm{\sc Delete-App $\downarrow$}", dashed] \\
                       & t                                  &                  
    \end{tikzcd}

    \item 
    \begin{tikzcd}
                       & \down{\Delta \vdash id}{\ghostapp{\up{id \vdash \delta}{u}}{t}} \arrow[ld, "\textrm{\sc Propagate $\uparrow$}"'] \arrow[rd, "\textrm{\sc Propagate $\downarrow$}"] &                   \\
    \down{\Delta \vdash id}{\up{id \vdash B}{\ghostapp{u}{t}}} \arrow[d, "\textrm{\sc Interchange 1}"', dashed] &                                    & \ghostapp{\down{\Delta \vdash id}{\up{id \vdash \delta}{u}}}{{\down{\Delta \vdash id}{t}}}  \arrow[d, "\textrm{\sc Interchange 1}", dashed]
       \\
       \up{id \vdash B}{\down{\Delta \vdash id}{\ghostapp{u}{t}}} \arrow[rd, "\textrm{\sc Propagate $\downarrow$}"', dashed] && \ghostapp{\up{id \vdash \delta}{\down{\Delta \vdash id}{u}}}{{\down{\Delta \vdash id}{t}}}  \arrow[ld, "\textrm{\sc Propagate $\uparrow$}", dashed]
       \\
       & \up{id \vdash \delta}{\ghostapp{\down{\Delta \vdash id}{u}}{\down{\Delta \vdash id}{t}}} &                  
    \end{tikzcd}

    \item

    \begin{tikzcd}
                       & \down{\Delta \vdash id}{\up{id \vdash \plusdiffs{A \to}{B}{}}{t}} \arrow[ldd, "\textrm{\sc Interchange 1}"'] \arrow[rd, "\textrm{\sc Insert-App $\uparrow$}"] &                   \\
     &                                    & \down{\Delta \vdash id}{\up{id \vdash B}{t~\holety{A}}} \arrow[d, "\textrm{\sc Interchange 1}", dashed] \\
        {\up{id \vdash \plusdiffs{A \to}{B}{}}{\down{\Delta \vdash id}{t}}} \arrow[rdd, "\textrm{\sc Insert-App $\uparrow$}"', dashed] &  & \up{id \vdash B}{\down{\Delta \vdash id}{t~\holety{A}}} \arrow[d, "\textrm{\sc Propagate $\downarrow$}", dashed]
       \\
       && \up{id \vdash B}{\down{\Delta \vdash id}{t}~\down{\Delta \vdash id}{\holety{A}}} \arrow[ld, "\textrm{\sc Propagate $\downarrow$}", dashed]
       \\
       & \up{id \vdash B}{\down{\Delta \vdash id}{t}~\holety{A}}                                  &                  
    \end{tikzcd}

    \item

    \begin{tikzcd}
                       & \down{\Delta \vdash id}{\up{id \vdash \minusdiffs{A \to}{B}{}}{t_1}~t_2} \arrow[ld, "\textrm{\sc Propagate $\downarrow$}"'] \arrow[rd, "\textrm{\sc Displace-App $\uparrow$}"] &                    
        \\
        \down{\Delta \vdash id}{\up{id \vdash \minusdiffs{A \to}{B}{}}{t_1}}~\down{\Delta \vdash id}{t_2} \arrow[d, "\textrm{\sc Interchange 1}"', dashed]  &                                    & \down{\Delta \vdash id}{\up{id \vdash B}{\ghostapp{t_1}{t_2}}}\arrow[d, "\textrm{\sc Interchange 1}", dashed]
        \\
    \up{id \vdash \minusdiffs{A \to}{B}{}}{\down{\Delta \vdash id}{t_1}}~\down{\Delta \vdash id}{t_2} \arrow[rd, "\textrm{\sc Displace-App $\uparrow$}"', dashed] &                                    & \up{id \vdash B}{\down{\Delta \vdash id}{\ghostapp{t_1}{t_2}}} \arrow[ld, "\textrm{\sc Propagate $\downarrow$}", dashed]
        \\
                       & \up{id \vdash B}{\ghostapp{\down{\Delta \vdash id}{t_1}}{{\down{\Delta \vdash id}{t_2}}}}                                  &                   
    \end{tikzcd}

    \item

    \begin{tikzcd}
                       & \down{\Delta \vdash id}{\up{id \vdash A \to B}{u}~t} \arrow[ld, "\textrm{\sc Propagate $\uparrow$}"'] \arrow[rd, "\textrm{\sc Propagate $\downarrow$}"] &                   \\
    \down{\Delta \vdash id}{\up{id \vdash B}{u~\down{id \vdash A}{t}}} \arrow[d, "\textrm{\sc Interchange 1}"', dashed] &                                    & \down{\Delta \vdash id}{\up{id \vdash A \to B}{u}}~\down{\Delta \vdash id}{t} \arrow[d, "\textrm{\sc Interchange 1}", dashed]
       \\
       \up{id \vdash B}{\down{\Delta \vdash id}{u~\down{id \vdash A}{t}}} \arrow[d, "\textrm{\sc Propagate $\downarrow$}"', dashed] && \up{id \vdash A \to B}{\down{\Delta \vdash id}{u}}~\down{\Delta \vdash id}{t} \arrow[d, "\textrm{\sc Propagate $\uparrow$}", dashed]
       \\
       \up{id \vdash B}{\down{\Delta \vdash id}{u}~\down{\Delta \vdash id}{\down{id \vdash A}{t}}} \arrow[rd, "\textrm{Lemma~\ref{lemma:reordering}}"', dashed] && \up{id \vdash B}{\down{\Delta \vdash id}{u}~\down{id \vdash A}{\down{\Delta \vdash id}{t}}} \arrow[ld, "\textrm{Lemma~\ref{lemma:reordering}}", dashed]
       \\
       & \up{id \vdash B}{\down{\Delta \vdash id}{u}~t'} &                  
    \end{tikzcd}

\end{itemize}

In this final case, we have two terms $\down{\Delta \vdash id}{\down{id \vdash A}{t}}$ and $\down{id \vdash A}{\down{\Delta \vdash id}{t}}$. To show that these two terms step to the same result, we prove two lemmas by mutual induction which show that two diff boundaries propagated into the same term in different orders gives the same result.

\begin{lemma}[Reordering outside of neutral forms]
\label{lemma:reordering}
If $t$ is not a neutral form (unless maximal) and contains no boundaries,
then there exists $t'$ such that both

\begin{itemize}
    \item 
    $ \down{C.2[\Delta] \vdash id}{\down{C[id] \vdash \delta}{t}} \step t' $
    \item
    $ \down{C[id] \vdash \delta}{\down{C.1[\Delta] \vdash id}{t}} \step t' $
\end{itemize}
\end{lemma}

\paragraph{Proof:}
We proceed by cases. There must be some rule that can step $\down{C[id]  \vdash \delta}{t}$, and we give a case for each possibility:

\begin{itemize}
    \item If $\down{C[id]  \vdash \delta}{t}$ can step by the {\sc Propagate $\downarrow$} rule because no alterations rules apply, then we proceed generically. Suppose that $t = r~t_1 \dots t_n$, for a construct $r$ with typing rule
    $$
    \inferrule{s_1 \dots s_n}{s}
    $$
    Then, because $t$ is not a neutral form, by definition of neutral form $s$ must be linear and so have at most one instance of each metavariable. Therefore, propagation proceeds by the simple case. Suppose that $\delta = \sigma~s$. Then,

    \begin{itemize}
        \item 
        $ \down{C.2[\Delta] \vdash id}{\down{C[id] \vdash \sigma~s}{r~t_1 \dots t_n}} \leadsto^* r~\down{C.2[\Delta] \vdash id}{\down{C[id] \vdash \sigma~s_1}{t_1}} \dots \down{C.2[\Delta] \vdash id}{\down{C[id] \vdash \sigma~s_n}{t_n}}$
        \item
        $ \down{C[id] \vdash \sigma~s}{\down{C.1[\Delta] \vdash id}{r~t_1 \dots t_n}} \leadsto^* r~ \down{C[id] \vdash \sigma~s_1}{\down{C.1[\Delta] \vdash id}{t_1}} \dots \down{C[id] \vdash \sigma~s_n}{\down{C.1[\Delta] \vdash id}{t_n}}$
    \end{itemize}
    
    \item If $\down{C[id]  \vdash \delta}{t}$ can step by the {\sc Insert-Abs $\downarrow$} rule ,
    \begin{itemize}
        \item 
        $ \down{C.2[\Delta] \vdash id}{\down{C[id] \vdash \plusdiffs{A \to}{B}{}}{t}} \leadsto^* \fun{x}{A}{\down{C.2[\Delta] \vdash id}{\down{C[id] \vdash B}{t}}}$
        \item
        $ \down{C[id] \vdash \plusdiffs{A \to}{B}{}}{\down{C.1[\Delta] \vdash id}{t}} \leadsto^* \fun{x}{A}{\down{C[id] \vdash B}{\down{C.1[\Delta] \vdash id}{t}}} $
    \end{itemize}
    
    \item If $\down{C[id]  \vdash \delta}{t}$ can step by the {\sc Delete-Abs $\downarrow$} rule ,
    \begin{itemize}
        \item 
        $
        \down{C.2[\Delta] \vdash id}{\down{C[id] \vdash \minusdiffs{A \to}{B}{}}{\fun{x}{A}{t}}}
        \leadsto^*
        \down{C.2[\Delta] \vdash id}{\down{C[id] \vdash B}{t}}
        $
        \item
        $
        \down{C[id] \vdash \minusdiffs{A \to}{B}{}}{\down{C.1[\Delta] \vdash id}{\fun{x}{A}{t}}}
        \leadsto^*
        \down{C[id] \vdash B}{\down{C.1[\Delta] \vdash id}{t}}
        $
    \end{itemize}
    
    \item If $\down{C[id]  \vdash \delta}{t}$ can step by the {\sc Delete-App $\downarrow$} rule ,
    \begin{itemize}
        \item 
        $
        \down{C.2[\Delta] \vdash id}{\down{C[id] \vdash \plusdiffs{A \to}{B}{}}{t~\holety{A}}}
        \leadsto^*
        \down{C.2[\Delta] \vdash id}{\down{C[id] \vdash B}{t}}
        $
        \item
        $
        \down{C[id] \vdash \plusdiffs{A \to}{B}{}}{\down{C.1[\Delta] \vdash id}{{t~\holety{A}}}}
        \leadsto^*
        \down{C[id] \vdash B}{\down{C.1[\Delta] \vdash id}{t}}
        $
    \end{itemize}
    
    \item If $\down{C[id]  \vdash \delta}{t}$ can step by the {\sc Neutral-Error $\downarrow$} rule,
    \begin{itemize}
        \item 
        $ \down{C.2[\Delta] \vdash id}{\down{C[id] \vdash \delta}{u}} \leadsto^* \errorboundary{\delta.1}{\delta.2}{\down{C.2[\Delta] \vdash id}{\down{C[id] \vdash id}{u}}} $
        \item
        $ \down{C[id] \vdash \delta}{\down{C.1[\Delta] \vdash id}{u}} \leadsto^* \errorboundary{\delta.1}{\delta.2}{\down{C[id] \vdash id}{\down{C.1[\Delta] \vdash id}{u}}} $
    \end{itemize}
    
    \item If $\down{C[id]  \vdash \delta}{t}$ can step by the {\sc Fallthrough-Error $\downarrow$} rule,
    \begin{itemize}
        \item 
        $ \down{C.2[\Delta] \vdash id}{\down{C[id] \vdash \delta}{t}} \leadsto^* \errorboundary{\delta.1}{\delta.2}{\down{C.2[\Delta] \vdash id}{\down{C[id] \vdash id}{t}}} $
        \item
        $ \down{C[id] \vdash \delta}{\down{C.1[\Delta] \vdash id}{t}} \leadsto^* \errorboundary{\delta.1}{\delta.2}{\down{C[id] \vdash id}{\down{C.1[\Delta] \vdash id}{t}}} $
    \end{itemize}
\end{itemize}

\begin{lemma}[Reordering inside of neutral forms]
\label{lemma:reorderingneutral}
If $u$ is a neutral form and contains no boundaries,
then there exists $t'$ such that both

\begin{itemize}
    \item 
    $ \down{C.2[\Delta] \vdash id}{\down{C[id] \vdash id}{t}} \step t' $
    \item
    $ \down{C[id] \vdash id}{\down{C.1[\Delta] \vdash id}{t}} \step t' $
\end{itemize}
\end{lemma}

\paragraph{Proof:}
Because $u$ is a neutral form, there are only a few cases. In some cases, we use the induction hypothesis, and Lemma~\ref{lemma:reordering}

\begin{itemize}
    \item \begin{itemize}
        \item
        $\down{C.2[\Delta] \vdash id}{\down{C[id] \vdash id}{u~t}} ~ \leadsto^* ~  \down{C.2[\Delta] \vdash id}{\down{C[id] \vdash id}{u}}~\down{C.2[\Delta] \vdash id}{\down{C[id] \vdash id}{t}}$
        \item
        $ \down{C[id] \vdash id}{\down{C.1[\Delta] \vdash id}{u~t}} \leadsto^*  \down{C[id] \vdash id}{\down{C.1[\Delta] \vdash id}{u}}~\down{C[id] \vdash id}{\down{C.1[\Delta] \vdash id}{t}}$
    \end{itemize}
    \item \begin{itemize}
        \item
        $\down{C.2[\Delta] \vdash id}{\down{C[id] \vdash id}{\ghostapp{u}{t}}} ~ \leadsto^* ~  \ghostapp{\down{C.2[\Delta] \vdash id}{\down{C[id] \vdash id}{u}}}{\down{C.2[\Delta] \vdash id}{\down{C[id] \vdash id}{t}}}$
        \item
        $ \down{C[id] \vdash id}{\down{C.1[\Delta] \vdash id}{\ghostapp{u}{t}}} \leadsto^*  \ghostapp{\down{C[id] \vdash id}{\down{C.1[\Delta] \vdash id}{u}}}{\down{C[id] \vdash id}{\down{C.1[\Delta] \vdash id}{t}}}$
    \end{itemize}
    \item \begin{itemize}
        \item $ \down{\ldots x : \delta \ldots \vdash id}{\down{\ldots x : id \ldots \vdash id}{x}} \leadsto^* \up{id \vdash \delta}{x}$
        \item $ \down{\ldots x : id\ldots \vdash id}{\down{\ldots x : \delta \ldots \vdash id}{x}} \leadsto^* \up{id \vdash \delta}{x}$
    \end{itemize}
    \item \begin{itemize}
        \item $ \down{\ldots x : id \ldots \vdash id}{\down{\ldots x : id \ldots \vdash id}{x}} \leadsto^* x$
        \item $ \down{\ldots x : id\ldots \vdash id}{\down{\ldots x : id \ldots \vdash id}{x}} \leadsto^* x$
    \end{itemize}
    \item \begin{itemize}
        \item $ \down{\ldots \minusdiffs{\ldots}{, x : T}{} \ldots \vdash id}{\down{\ldots x : id \ldots \vdash id}{x}} \leadsto^* \ghostvar{x}{T}$
        \item $ \down{\ldots x : id\ldots \vdash id}{\down{\ldots \minusdiffs{\ldots}{, x : T}{} \ldots \vdash id}{x}} \leadsto^* \ghostvar{x}{T}$
    \end{itemize}
\end{itemize}

\subsection{Progress}
\label{app:progress}

\begin{customthm}{\ref{thm:progress}} [Progress]
For any term $t_0$ resulting from an edit as described in Section~\ref{sec:diffsetup}, if $t_0 \leadsto^* t$ and $t$ has a diff boundary other than an up boundary at the top, then for some $t'$, $t \step t'$.
\end{customthm}

At the end of propagation, the program may have the form $\up{\Delta \vdash \delta}{t}$. By Lemma~\ref{lemma:uplemma}, $\Delta$ may not add or remove any variables.
Further, by Lemma~\ref{lemma:noaround}, there may not be multiple up boundaries at the top of the program.
Therefore, as the program started in the empty context, it must also end in the empty context.

Given a diff boundary $\Delta \vdash \delta$, if $\delta \ne id$ then by definition either a {\sc Fallthrough-Error} rule applies or some other rule applies.

We claim that if $\delta = id$, then some propagation rule will apply.

\begin{itemize}
    \item If the boundary is a down boundary, then we have $\down{\Delta \vdash id}{t}$. If $t$ is a variable, then by definition the {\sc Propgate-Var} $\downarrow$ rule applies. Otherwise, looking at all of the typing rules in Figure~\ref{fig:typing-rules}, the context on the conclusion of all of the rules is a single metavariable $\Gamma$. Therefore, the {\sc Propgate} rule always applies, since trivially there is a substitution mapping $\Gamma$ to $\Delta$ such that $\sigma~\Gamma = \Delta$.
    \item If the boundary is an up boundary, then we have $C[\up{\Delta \vdash id}{t}]$. For most typing rules in Figure~\ref{fig:typing-rules}, most premises also have a context which is a single metavariable $\Gamma$. However, there are a few rules, specifically the rules for $\lambda$ abstractions, \CC{let} expressions, and \CC{match} expressions, where a premise has a context of a different form; either $\Gamma, x : T$ or $\Gamma, h : T_1, t : \cList~T_1$ as in the \CC{match} rule. However, by Lemma~\ref{lemma:uplemma}, an up boundary with $\delta = id$ must have a particular form that changes only one variable. Therefore, the {\sc Propagate} $\uparrow$ always applies.
\end{itemize}

%% file: diffappendix.tex
\section{Diff proofs}
\label{sec:diffappendix}

\begin{theorem}[Identity-Compose]
    $$id_{d.1} \circ d = d \circ id_{d.2} = d$$   
\end{theorem}

The proof proceeds by induction over the structure of $d$, with one case for each of the four diff constructors.

\begin{itemize}
    \item If $d = l~d_1 \dots d_n$, then $(l~id_{d_1.1} \dots id_{d_n.1}) \circ d = l~(id_{d_1.1} \circ d_1) \dots (id_{d_n.1} \circ d_n)$, and $ d \circ (l~id_{d_1.1} \dots id_{d_n.1}) = l~(d_1 \circ id_{d_1.1}) \dots (d_n \circ id_{d_n.2})$. These are equal to $d$ by the induction hypothesis.
    \item If $d = \plusdiff{c}{d'}{}$, then $d.1 = d'.1$ and $d.2 = c[d'.2]$.
    
        $id_{d.1} \circ d = id_{d'.1} \circ \plusdiff{c}{d'}{} = \plusdiff{c}{id_{d'.1} \circ d'}{}$, and $id_{d'.1} \circ d' = d$ by the induction hypothesis.
        
        $d \circ id_{d.1} = \plusdiff{c}{d'}{} \circ c[id_{d'.2}] = \plusdiff{c}{d' \circ id_{d'.1}}{}$, and $d' \circ id_{d'.2} = d$ by the induction hypothesis.
        
    \item If $d = \minusdiff{c}{d'}{}$, then $d.1 = c[d'.1]$ and $d.2 = d'.2$.
    
        $id_{d.1} \circ d = c[id_{d'.1}] \circ \minusdiff{c}{d'}{} = \minusdiff{c}{id_{d'.1} \circ d'}{}$, and $id_{d'.1} \circ d' = d$ by the induction hypothesis.
        
        $d \circ id_{d.1} = \minusdiff{c}{d'}{} \circ id_{d'.2} = \minusdiff{c}{d' \circ id_{d'.1}}{}$, and $d' \circ id_{d'.2} = d$ by the induction hypothesis.

    \item If $d = \replacediff{s_1}{s_2}$, then $id \circ d = d \circ id = \replacediff{s_1}{s_2}$ directly.
\end{itemize}

\begin{theorem} [Associativity of composition]
Given any three diffs
$s_1 \xrightarrow{d_1} s_2 \xrightarrow{d_2} s_3 \xrightarrow{d_3} s_4$,

then
$(d_1 \circ d_2) \circ d_3 = d_1 \circ (d_2 \circ d_3)$
\end{theorem}

The proof proceeds by induction over the size of the diffs, and cases over the possible constructors of the diffs.

There are four constructors of diffs, $l \; d_1 ... d_n$, $\plusdiff{c}{d}{}$, $\minusdiff{c}{d}{}$, and $\replacediff{s_1}{s_2}$. We proceed by cases; there are $4^3 = 64$ combinations. In each case, associativity either follows directly or after using the induction hypothesis. Luckily, many of the cases can be combined, and all 13 resulting of the cases are listed below.
For each case, the number of combinations that it accounts for is described, as well as a proof that associativity holds by computing both possible associations.

Given 
$$
s_1 \xrightarrow{d_1} s_2 \xrightarrow{d_2} s_3 \xrightarrow{d_3} s_4
$$

\begin{enumerate}
    \item If $d_3 = \plusdiff{c}{d_3'}{}$, then
        \begin{itemize}
            \item  $(d_1 \circ d_2) \circ \plusdiff{c}{d_3'}{} = \plusdiff{c}{(d_1 \circ d_2) \circ d_3'}{}$
            \item $d_1 \circ (d_2 \circ \plusdiff{c}{d_3'}{}) = d_1 \circ \plusdiff{c}{d_2 \circ d_3}{} = \plusdiff{c}{d_1 \circ (d_2 \circ d_3')}{}$
        \end{itemize}
        This case accounts for $16$ combinations.
    \item If $d_1 = \minusdiff{c}{d_1'}{}$, then
        \begin{itemize}
            \item  $(\minusdiff{c}{d_1'}{} \circ d_2) \circ d_3' = \minusdiff{c}{d_1' \circ d_2}{} \circ d_3 = \minusdiff{c}{(d_1' \circ d_2) \circ d)3}{}$
            \item $\minusdiff{c}{d_1'}{} \circ (d_2 \circ d_3) = \minusdiff{c}{d_1' \circ (d_2 \circ d_3)}{}$
        \end{itemize}
        This case accounts for 12 combination, because 4 of the possible 16 overlap with case (1).
    \item If $d_1 = \plusdiff{c_1}{d_1'}{}$ and $d_2 = \minusdiff{c_2}{d_2'}{}$, then
        \begin{itemize}
            \item If $c_1 = c_2 = c$, then
                \begin{itemize}
                    \item $(\plusdiff{c}{d_1'}{} \circ \minusdiff{c}{d_2'}{}) \circ d_3 = (d_1' \circ d_2') \circ d_3$
                    \item $\plusdiff{c}{d_1'}{} \circ (\minusdiff{c}{d_2'}{} \circ d_3) = \plusdiff{c}{d_1'}{} \circ \minusdiff{c}{d_2' \circ d_3}{} = d_1' \circ (d_2' \circ d_3)$
                \end{itemize}
            \item $If c_1 \ne c_2$, then
                \begin{itemize}
                    \item $(\plusdiff{c_1}{d_1'}{} \circ \minusdiff{c_2}{d_2'}{}) \circ d_3 = (\replacediff{s_1}{s_3}) \circ d_3 = \replacediff{s_1}{s_4}$
                    \item $\plusdiff{c_1}{d_1'}{} \circ (\minusdiff{c_2}{d_2'}{} \circ d_3) = \plusdiff{c_1}{d_1'}{} \circ \minusdiff{c_2}{d_2' \circ d_3}{} = \replacediff{s_1}{s_4}$
                \end{itemize}
        \end{itemize}
        This case accounts for 3 combinations, because 1 of the possible 4 overlaps with case (1).
    \item If $d_2 = \plusdiff{c}{d_2'}{}$ and $d_3 = \minusdiff{c}{d_3'}{}$
        \begin{itemize}
            \item If $c_1 = c_2 = c$, then
                \begin{itemize}
                    \item $(d_1 \circ \plusdiff{c}{d_2'}{}) \circ \minusdiff{c}{d_3'}{} \circ d_3 = \plusdiff{c}{d_1 \circ d_2'}{} \circ \minusdiff{c}{d_3'}{} = (d_1 \circ d_2') \circ d_3'$
                    \item $d_1 \circ (\plusdiff{c}{d_2'}{} \circ \minusdiff{c}{d_3'}{}) = d_1 \circ (d_2' \circ d_3')$
                \end{itemize}
            \item $If c_1 \ne c_2$, then
                \begin{itemize}
                    \item $(d_1 \circ \plusdiff{c_1}{d_2'}{}) \circ \minusdiff{c_2}{d_3'}{} = \plusdiff{c_1}{d_1 \circ d_2'}{} \circ \minusdiff{c_2}{d_3'}{} = \replacediff{s_1}{s_4}$
                    \item $d_1 \circ (\plusdiff{c_1}{d_2'}{} \circ \minusdiff{c_2}{d_3'}{}) = d_1 \circ (\replacediff{s_2}{s_4}) = \replacediff{s_1}{s_4}$
                \end{itemize}
        \end{itemize}
        This case accounts for 3 combinations, because 1 of the possible 4 overlaps with case (2).
    \item If $d_2 = \plusdiff{c.1}{d_2'}{}$ and $d_3 = c[d_3']$
        \begin{itemize}
            \item $(d_1 \circ \plusdiff{c.1}{d_2'}{}) \circ c[d_3'] = \plusdiff{c.1}{d_1 \circ d_2'}{} \circ c[d_3'] = \plusdiff{c.2}{(d_1 \circ d_2') \circ d_3'}{}$
            \item $d_1 \circ (\plusdiff{c.1}{d_2'}{} \circ c[d_3']) = d_1 \circ \plusdiff{c.2}{d_2' \circ d_3'}{} = \plusdiff{c.2}{d_1 \circ (d_2' \circ d_3'}{}$
        \end{itemize}
        This case accounts for 3 combinations, because 1 of the possible 4 overlaps with case (2).
    \item If $d_1 = c[d_1']$ and $d_2 = \minusdiff{c.2}{d_2'}{}$
        \begin{itemize}
            \item $(c[d_1'] \circ \minusdiff{c.2}{d_2'}{}) \circ d_3 = \minusdiff{c.1}{d_1' \circ d_2'}{} \circ d_3 = \minusdiff{c.1}{(d_1' \circ d_2') \circ d_3}{}$
            \item $c[d_1'] \circ (\minusdiff{c.2}{d_2'}{} \circ d_3) = c[d_1'] \circ \minusdiff{c.2}{d_2' \circ d_3}{} = \minusdiff{c.1}{d_1' \circ (\circ d_2' \circ d_3)}{}$
        \end{itemize}
        This case accounts for 3 combinations, because 1 of the possible 4 overlaps with case (1).
    \item If $d_1 = \plusdiff{c.1}{d_1'}{}$, $d_2 = c[d_2']$, and $d_3 = \minusdiff{c.2}{d_3'}{}$
        \begin{itemize}
            \item $(\plusdiff{c.1}{d_1'}{} \circ c[d_2']) \circ \minusdiff{c.2}{d_3'}{} = \plusdiff{c.2}{d_1' \circ d_2'}{} \circ \minusdiff{c.2}{d_3'}{} = (d_1' \circ d_2') \circ d_3'$
            \item $\plusdiff{c.1}{d_1'}{} \circ (c[d_2'] \circ \minusdiff{c.2}{d_3'}{}) = \plusdiff{c.1}{d_1'}{} \circ \minusdiff{c.1}{d_2' \circ d_3'}{} = d_1' \circ (d_2' \circ d_3')$
        \end{itemize}
    \item If $d_1 = \plusdiff{c.1}{d_1'}{}$, $d_2 = c[d_2']$, and $d_3 = c[d_3']$
        \begin{itemize}
            \item $(\plusdiff{c.1}{d_1'}{} \circ c[d_2']) \circ c[d_3'] = \plusdiff{c.1}{d_1' \circ d_2'}{} \circ c[d_3'] = \plusdiff{c.2}{(d_1' \circ d_2') \circ d_3'}{}$
            \item $\plusdiff{c.1}{d_1'}{} \circ (c[d_2'] \circ c[d_3']) = \plusdiff{c.1}{d_1'}{} \circ c[d_2' \circ d_3'] = \plusdiff{c.2}{d_1' \circ (d_2' \circ d_3')}{}$
        \end{itemize}
    \item If $d_1 = c_1[d_1']$, $d_2 = c_2[d_2']$, and $d_3 = \minusdiff{c_2.2}{d_3'}{}$
        \begin{itemize}
            \item $(c_1[d_1'] \circ c_2.2[d_2']) \circ \minusdiff{c.2}{d_3'}{} = (c_1 \circ c_2)[d_1' \circ d_2'] \circ \minusdiff{c_2.2}{d_3'}{} = \minusdiff{c_1.1}{(d_1' \circ d_2') \circ d_3'}{}$
            \item $c_1[d_1'] \circ (c_2.2[d_2'] \circ \minusdiff{c.2}{d_3'}{}) = c[d_1'] \circ \minusdiff{c_2.1}{d_1' \circ d_3'}{} = \minusdiff{c_1.1}{d_1' \circ (d_2' \circ d_3')}{} $
        \end{itemize}
    \item If $d_1 = l \; a_1 ... a_n$, $d_2 = l \; b_1 ... b_n$, and $d_3 = l \; c_1 ... c_n$
        \begin{itemize}
            \item $(l \; a_1 ... a_n \circ l \; b_1 ... b_n) \circ l \; c_1 ... c_n = (l \; (a_1 \circ b_1) ... (a_n \circ b_n)) \circ l \; c_1 ... c_n$
            
            $= l \; ((a_1 \circ b_1) \circ c_1) ...((a_n \circ b_n) \circ c_n)$
            \item $l \; a_1 ... a_n \circ (l \; b_1 ... b_n \circ l \; c_1 ... c_n) = l \; a_1 ... a_n \circ (l \; (b_1 \circ c_1) ... (b_n \circ c_n))$
            
            $= l \; (a_1 \circ (b_1 \circ c_1)) ...(a_n \circ (b_n \circ c_n))$
        \end{itemize}
    \item If $d_1 = \replacediff{s_1}{s_2}$, and cases (1) and (5) do not apply, then
    
        \begin{itemize}
            \item $(\replacediff{s_1}{s_2} \circ d_2) \circ d_3 = \replacediff{s_1}{s_3} \circ d_3 = \replacediff{s_1}{s_4}$
            \item $\replacediff{s_1}{s_2} \circ (d_2 \circ d_3) = \replacediff{s_1}{s_4}$
        \end{itemize}
        The extra condition that cases (1) and (5) do not apply ensures that all of the compositions result in replace diffs. If for example $d_3$ were a `+' diff, then the result would not be a replace diff. However, we already handled that case. Similarly, if case (5) held, then the composite $d_2 \circ d_3$ would be a `+' diff.

        This case accounts for 10 combinations, because 4 of the possible 16 overlap with case (1), one overlaps with case (5), and one overlaps with case (4).
    \item If $d_2 = \replacediff{s_2}{s_3}$ and cases (1) and (2) do not apply, then
        \begin{itemize}
            \item $(d_1 \circ \replacediff{s_2}{s_3}) \circ d_3 = \replacediff{s_1}{s_3} \circ d_3 = \replacediff{s_1}{s_4}$
            \item $d_1 \circ (\replacediff{s_2}{s_3} \circ d_3) = d_1 \circ \replacediff{s_2}{s_4} = \replacediff{s_1}{s_4}$
        \end{itemize}
        
        This case accounts for 6 combinations, because 4 of the possible 16 overlap with case (1), 3 overlap with case (2) (that didn't overlap with case 1), and 3 overlap with case (11) (that didn't overlap with case 1).
    \item If $d_3 = \replacediff{s_3}{s_4}$ and cases (2) and (6) do not apply, then
        \begin{itemize}
            \item $(d_1 \circ d_2) \circ \replacediff{s_3}{s_4} = \replacediff{s_1}{s_4}$
            \item $d_1 \circ (d_2 \circ \replacediff{s_3}{s_4}) = d_1 \circ \replacediff{s_2}{s_4} = \replacediff{s_1}{s_4}$
        \end{itemize}
        Similar to case (11), the extra condition ensures that all composites are replace diffs.
        
        This case accounts for 4 combinations, because 4 of the possible 16 overlap with case (2), 1 overlaps with case (3), 1 overlaps with case (6), 4 overlap with case (11), and 2 overlap with case (12) (that didn't overlap with case (1) or (11)).
\end{enumerate}